\documentclass[10pt]{iopart}

\usepackage{iopams}  
\usepackage{graphicx}
\usepackage{url}
\usepackage[tight]{subfigure}
\usepackage{multirow}

\usepackage{setspace}
\usepackage{placeins}

\begin{document}

\title[Divertor Heat Load in ASDEX Upgrade]{Divertor Heat Load in ASDEX Upgrade L-Mode in Presence of External Magnetic Perturbation}

\author{M.Faitsch$^{1}$, B.Sieglin$^{1}$, T.Eich$^{1}$, A.Herrmann$^1$, W.Suttrop$^{1}$ and the ASDEX Upgrade Team}
\address{$^1$Max-Planck-Institute for Plasma Physics, Boltzmannstr.~2, D-85748 Garching, Germany }
\ead{Michael.Faitsch@ipp.mpg.de}

\begin{abstract}
Power exhaust is one of the major challenges for a future fusion device.
Applying a non-axisymmetric external magnetic perturbation is one technique that is studied in order to mitigate or suppress large edge localized modes which accompany the high confinement regime in tokamaks.
The external magnetic perturbation brakes the axisymmetry of a tokamak and leads to a 2D heat flux pattern on the divertor target.
The 2D heat flux pattern at the outer divertor target is studied on ASDEX Upgrade in stationary L-Mode discharges.
The amplitude of the 2D characteristic of the heat flux depends on the alignment between the field lines at the edge and the vacuum response of the applied magnetic perturbation spectrum.
The 2D characteristic reduces with increasing density.
The increasing divertor broadening $S$ with increasing density is proposed as the main actuator.
This is supported by a generic model using field line tracing and the vacuum field approach that is in quantitative agreement with the measured heat flux.
The perturbed heat flux, averaged over a full toroidal rotation of the magnetic perturbation, is identical to the non-perturbed heat flux without magnetic perturbation.
The transport qualifiers, power fall-off length $\lambda_q$ and divertor broadening $S$, are the same within the uncertainty compared to the unperturbed reference.
No additional cross field transport is observed.

\end{abstract}

\section{Introduction}
Power exhaust is one of the major challenges for a future fusion device.
Future fusion devices are considered to operate in a regime with enhanced confinement, the so called H-Mode.
This regime is usually accompanied by edge localized modes (ELMs) which lead to periodic bursts of energy and particles from the confined plasma towards the plasma facing components.
Applying a non-axisymmetric external magnetic perturbation (MP) is one technique that is studied in order to mitigate or suppress large ELMs in next step fusion devices such as ITER~\cite{Lang2013, Loarte2014}.
The thermal load due to ELMs might limit the lifetime of the divertor of these devices~\cite{Loarte2003}.
Many of today's experiments are equipped with magnetic coils to study the physics and feasibility of ELM mitigation/suppression with an external magnetic perturbation, e.g. ASDEX Upgrade~\cite{Suttrop2009}, DIII-D~\cite{Evans2004}, EAST~\cite{SunEAST2016}, JET~\cite{Liang2007}, KSTAR~\cite{Jeon2012}, MAST~\cite{Kirk2015}, NSTX~\cite{Ahn2010}.
Most of the studies focus on changes of global plasma parameters, e.g. density, or the increase in ELM frequency~\cite{Kirk2015, Thornton2015, SunEAST2016}.\\
Applying external magnetic perturbation brakes the axisymmetry of a tokamak and leads to a 2D heat flux pattern on the divertor target~\cite{Muller2013,Jakubowski2009,Harting2012,Ahn2010,Thornton2014, Faitsch2016}.
It is reported that for ITER it might be necessary to rotate the magnetic perturbation in order to prevent local over-heating due to the toroidally asymmetric heat load~\cite{Loarte2014}.
In this paper the 2D heat flux is studied at ASDEX Upgrade in stationary L-Mode discharges.\\
In section~\ref{Experiment} the plasma parameters and measurements are introduced.
In section~\ref{ExperimentResults} the results obtained from the experiments are shown.
Section~\ref{Modelling} introduces a generic model for the target heat flux pattern using field line tracing in the vacuum field approach.
Section~\ref{ModellingResults} compares this model with experimental data.
Conclusions and a summary are given in section~\ref{Conclusion}.

\section{Experiment}\label{Experiment}
\begin{figure}[ht] 
	\centering

\includegraphics[width=0.5\textwidth]{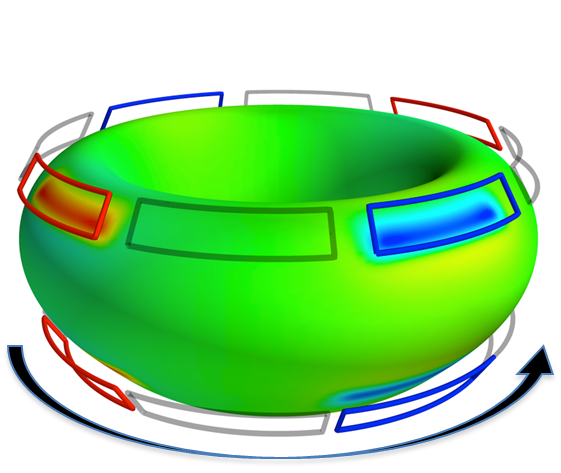}

\caption{ASDEX Upgrade magnetic perturbation coils.
Shown is an n\,=\,2 perturbation with a \textit{differential phase} $\Delta\phi\,=\,0$ between the upper row and the lower row.}
\label{fig:AUGCoils}
\end{figure}

ASDEX Upgrade~\cite{AUG2013} is equipped with two toroidal rows of 8 saddle coils, one below the outer midplane (lower coils) and one above the outer midplane (upper coils)~\cite{Suttrop2009, Suttrop2011}, shown in~\fref{fig:AUGCoils}.
The power supplies are able to produce a rigid rotation of a toroidal mode number n\,=\,2 perturbation with various poloidal mode spectra.
The poloidal spectrum is varied by the phase between the currents of the two sets of coils ({\textit{differential phase}, $\Delta \phi$)~\cite{Teschke2015}.
The magnetic perturbation is rotated with 1\,Hz in all discharges discussed in this paper.
This frequency corresponds to the current inside the coils.
With an n\,=\,2 perturbation the magnetic perturbation phase is rotated by $\pi$ within 1\,s.\\
The temperature evolution of the divertor target is measured using an infrared (IR) system~\cite{Sieglin2015}, measuring at a wavelength of $4.7\,\mu\mathrm{m}$ with a FWHM of 125\,nm and a frequency of 800\,Hz, optimized for L-Mode discharges.
The resolution on the outer divertor target is about 0.6\,mm/pixel.
The heat flux to the divertor target is calculated using THEODOR~\cite{Herrmann1995, Sieglin2015}. 
The IR system is observing a toroidal location at an angle of $\phi_{IR} = 213^\circ$ in the ASDEX Upgrade coordinate system.
This corresponds to a phase of $\phi_{IR} = 33^\circ$ in relation to the n\,=\,2 perturbation.
The global plasma parameters for the reference shot \#\,32212 are shown in \fref{fig:TimeTraceReferenceShot}.
The same parameters are used for the study of the \textit{differential phase} in section~\ref{diffPhase}.
In section~\ref{densSteps} discharges with higher stationary densities are discussed.
The scenario has a toroidal magnetic field of -2.5\,T and a plasma current of 0.8\,MA.
About 370\,kW of external ECR heating is applied to the center in order to increase the temperature rise at the divertor target and thus increasing the IR signal quality whilst still staying in L-Mode.
The plasma shape and current distribution is fully evolved at around 2.0\,s.
In the reference discharge a constant magnetic perturbation was applied between 4.5\,s and 5.0\,s.
The application of the magnetic perturbation does not change the global plasma parameters.
This allows to study the sole effect of the magnetic perturbation on the heat flux.
Also, the discharges performed with the rigid rotation have a reference phase without the magnetic perturbation in the beginning of the discharges before 2.5\,s.
This is done to have the possibility to check and hence ensure the similarity of the discharges as well as providing a comparison with the axisymmetric target profiles.
This results in about 3.5 rotations throughout every discharge, ensuring the reproducibility of the measurement.

\begin{figure}[ht] 
	\centering

\includegraphics[width=0.5\textwidth]{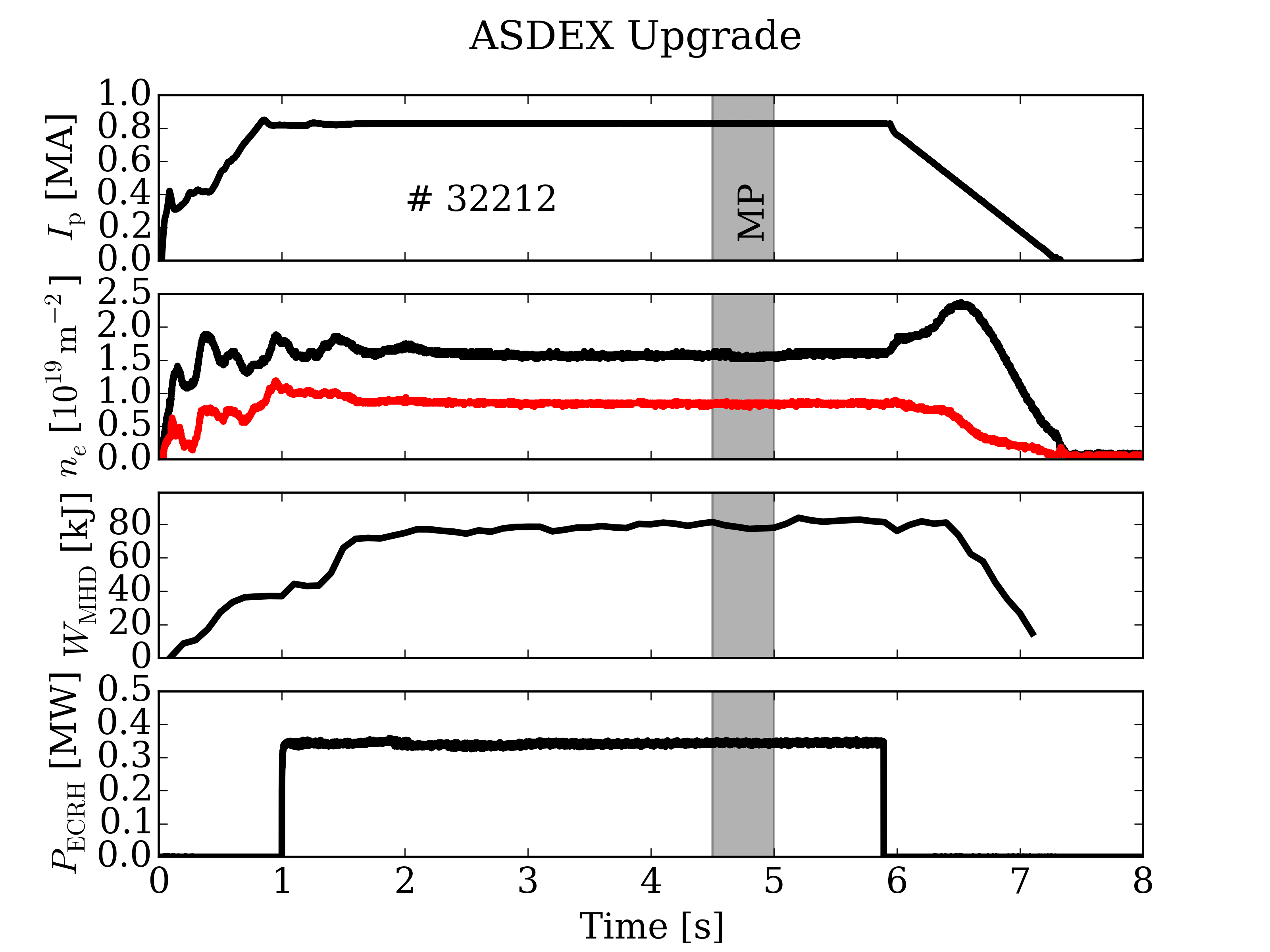}

\caption{Time traces of the global parameters for the reference shot \#\,32212 without magnetic perturbation until 4.5\,s and a steady magnetic perturbation phase until 5.0\,s.
The top panel shows the plasma current that is kept at 0.8\,MA.
The second panel shows the core and edge line integrated density in black and red, respectively.
The third panel shows the stored energy in the plasma.
The bottom panel shows the external heating.}
\label{fig:TimeTraceReferenceShot}
\end{figure}

The heat flux onto the outer divertor target for the discharge \#\,32217 is shown in \fref{fig:RefHeat}.
The heat flux profile obtained in the phase without the magnetic perturbation is shown in black.
The red curve shows the heat flux profile for the phase with the magnetic perturbation.
The heat flux profile without magnetic perturbation is described using the 1D diffusive model presented in~\cite{Eich2011}:
\begin{eqnarray} \label{eq:diffusiveModel}
  q(s) &= \frac{q_0}{2} \exp \left(\left(\frac{S}{2 \lambda_q}\right)^2 - \frac{s}{\lambda_q f_x} \right) \cdot \mathrm{erfc} \left(\frac{S}{2 \lambda_q} - \frac{s}{S f_x}\right) & \left[\frac{\mathrm{MW}}{\mathrm{m}^2}\right]
\end{eqnarray}
with s the target location, $S$ the divertor broadening, $\lambda_q$ the power fall-off length and $f_x$ the poloidal flux expansion.
The structure observed in the heat flux profile with magnetic perturbation is referred to as lobe structure, e.g.~\cite{Kirk2015, Ahn2010}, or strike line splitting, e.g.~\cite{Jakubowski2009, Harting2012}, with the characteristic of multiple distinguished peaks as a consequence of the non-axisymmetry.
A hot spot is present, marked in the figures with a grey area.
This hot spot exhibits a larger temperature increase which is not taken into account in the evaluation of the heat flux.
This leads to a too high estimated heat flux.
The area is thus excluded from any further discussions.

\begin{figure}[ht] 
	\centering

\includegraphics[width=0.5\textwidth]{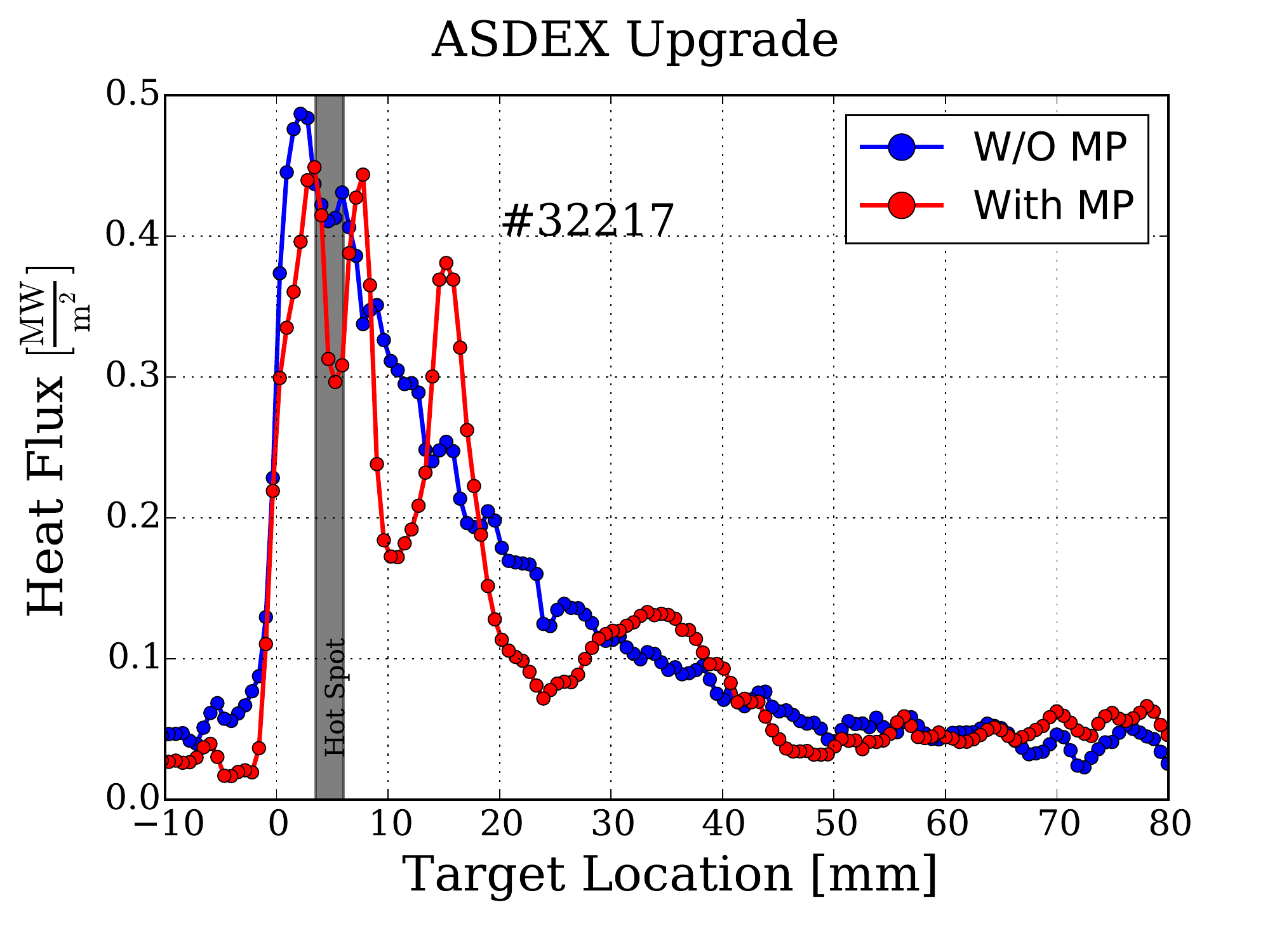}

\caption{1D heat flux profile for discharge \#\,32217 with (red) and without (blue) magnetic perturbation.
In the area in grey a hot spot is present.}
\label{fig:RefHeat}
\end{figure}

\subsection{Differential Phase}
The \textit{differential phase} $\Delta\phi$ between the upper and lower coil currents is changed in steps of $\frac{\pi}{2}$ with fixed maximum coil current amplitudes $I_{\mathrm{coil}}$\,=\,1\,kA.
Additional discharges are performed using only the upper (lower) coils.
The \textit{differential phase} of $\Delta\phi\,=\,-\frac{\pi}{2}$ is field line aligned at the edge (q\,=\,5 surface) and is therefore called the \textit{resonant} configuration~\cite{Suttrop2011}.
The \textit{differential phase} with $\Delta\phi\,=\,+\frac{\pi}{2}$ is called \textit{non-resonant} configuration accordingly.
The \textit{resonant} configuration with a reduced amplitude of the perturbation field $I_{\mathrm{coil,red}}\,=\,\frac{1}{\sqrt{2}}~I_{\mathrm{coil}}$ is performed to investigate the influence of the perturbation strength on the 2D structure and changes in the transport properties in the scrape-off layer.

\section{Divertor Heat Loads}\label{ExperimentResults}
In this section the results obtained with the IR thermography system are discussed.

\subsection{2D Heat Flux Profiles}\label{diffPhase}
The heat flux evolution on the outer divertor target for various \textit{differential phases} $\Delta \phi$ between lower and upper coil currents is shown in~\fref{fig:DifferentialPhasesIRTime}.
\begin{figure}[htb!]
    \centering

\subfigure[\label{fig:2DResonant} Resonant $\Delta\phi\,=\,-\frac{\pi}{2}$]{\includegraphics[width=0.3 \textwidth]{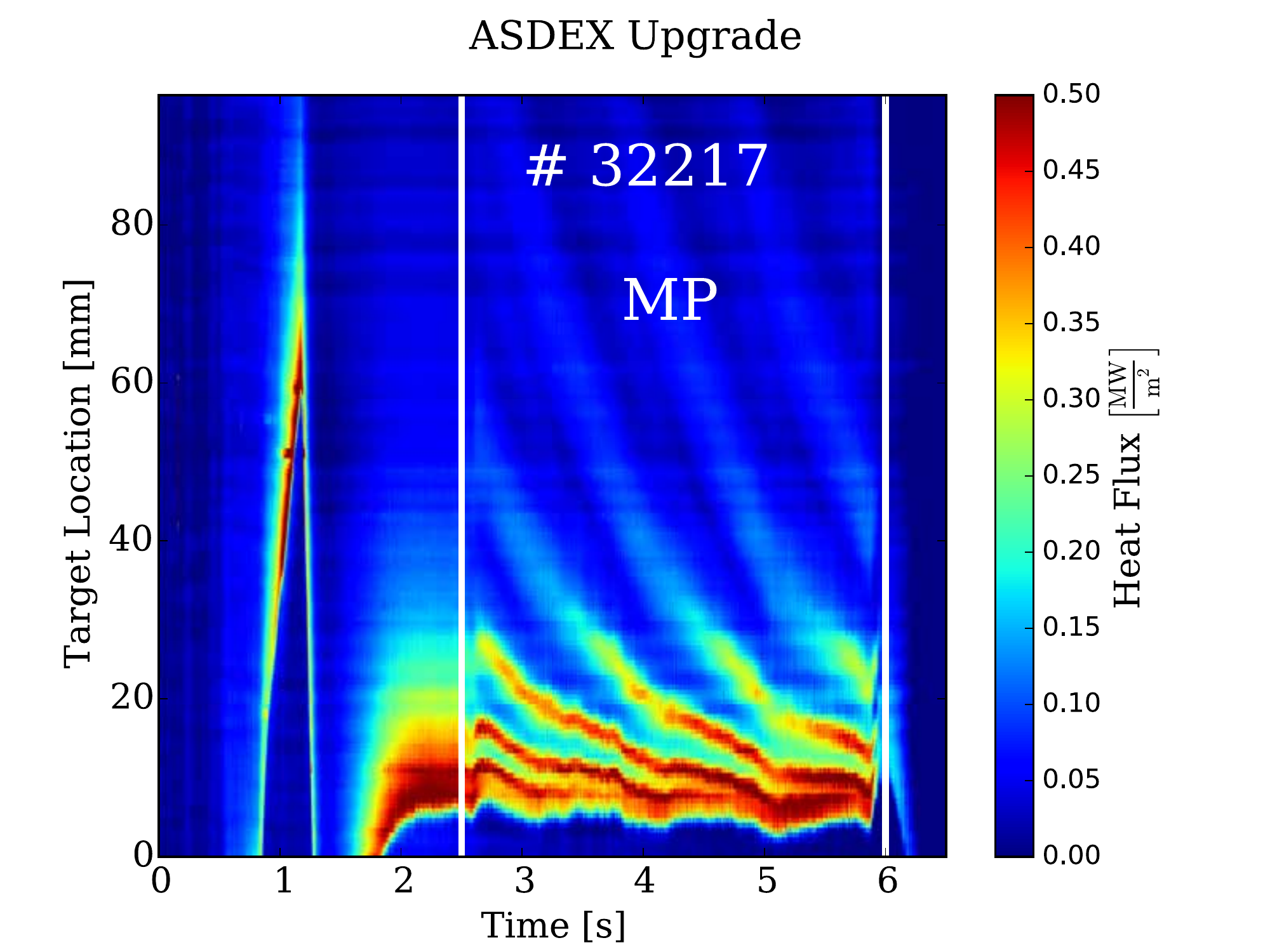}}
\subfigure[\label{fig:2DNonResonant} Non-Resonant $\Delta\phi\,=\,+\frac{\pi}{2}$]{\includegraphics[width=0.3 \textwidth]{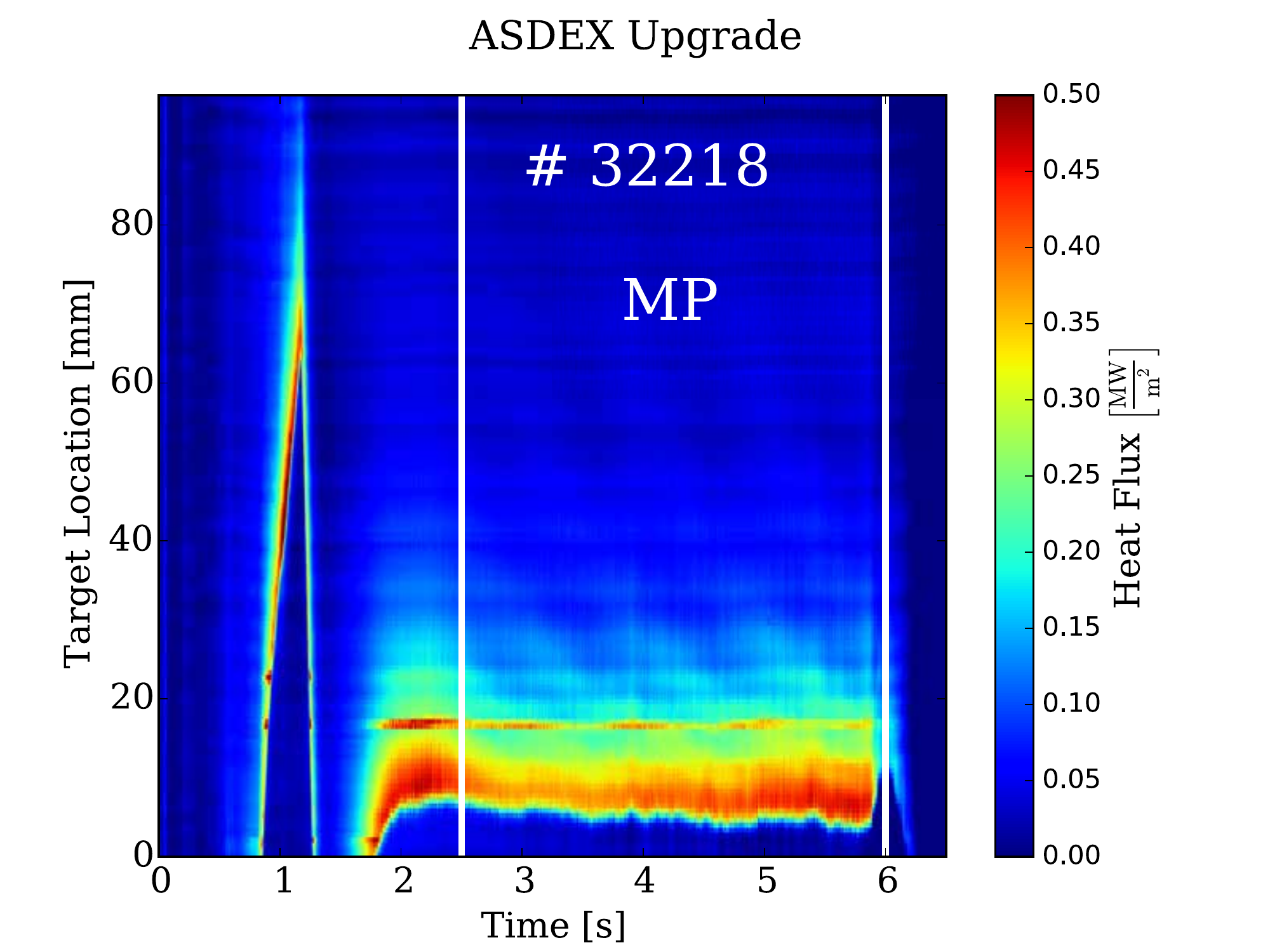}}
\subfigure[\label{fig:2DUpper} Upper Only]{\includegraphics[width=0.3 \textwidth]{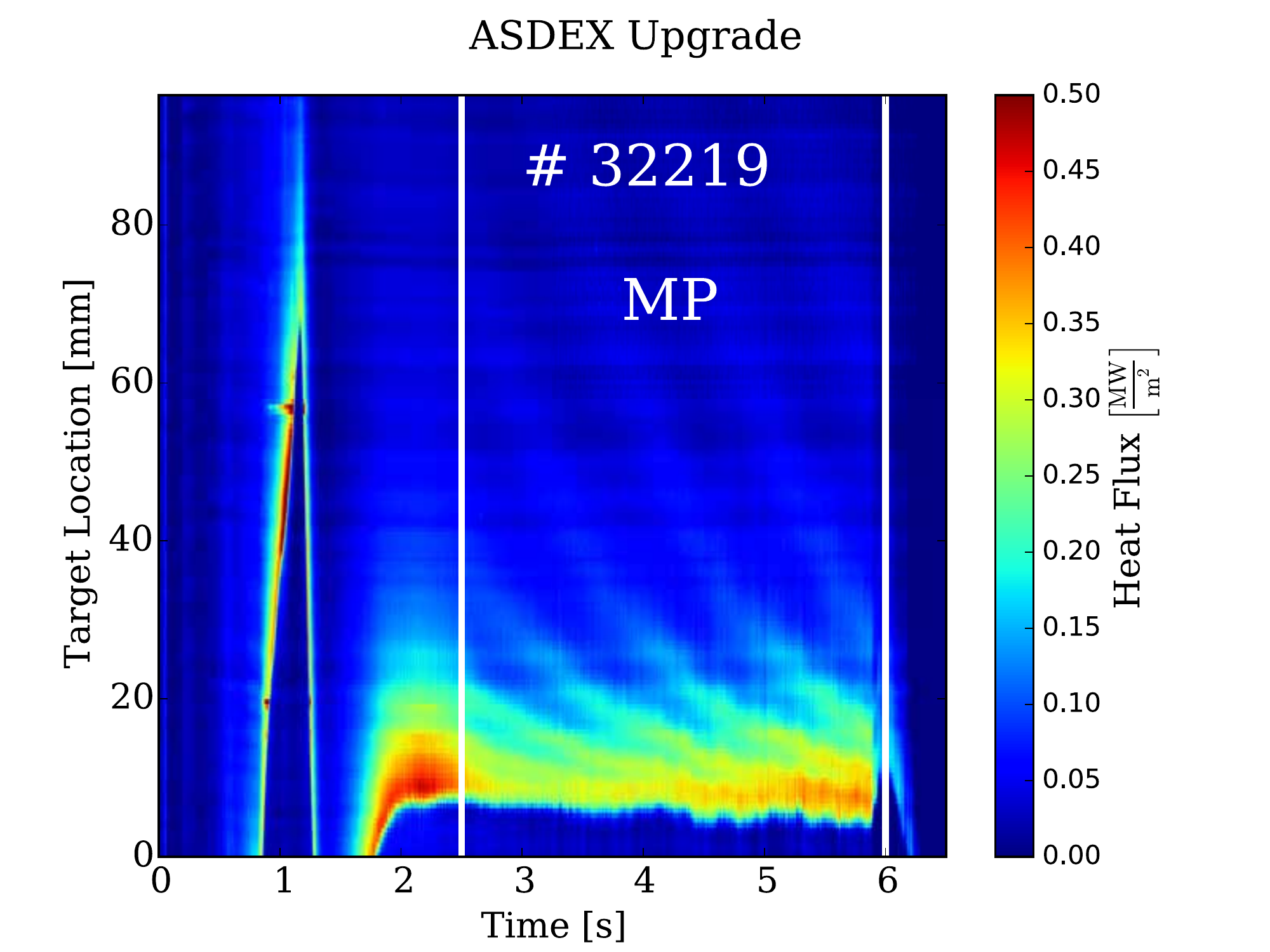}}\\
\subfigure[\label{fig:2DLower} Lower Only]{\includegraphics[width=0.3 \textwidth]{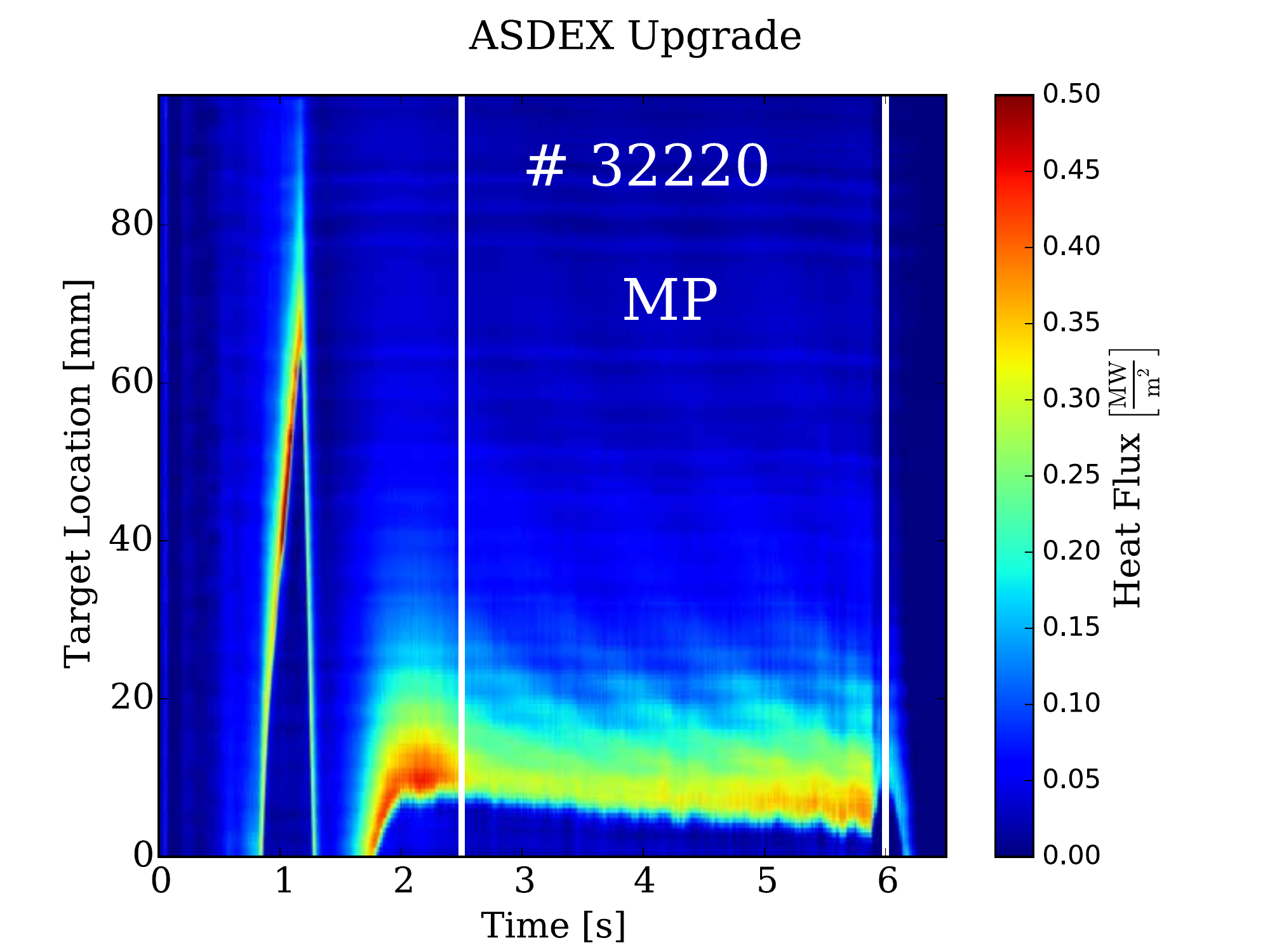}}
\subfigure[\label{fig:2DMid} $\Delta\phi\,=\,0,\,+\pi$]{\includegraphics[width=0.3 \textwidth]{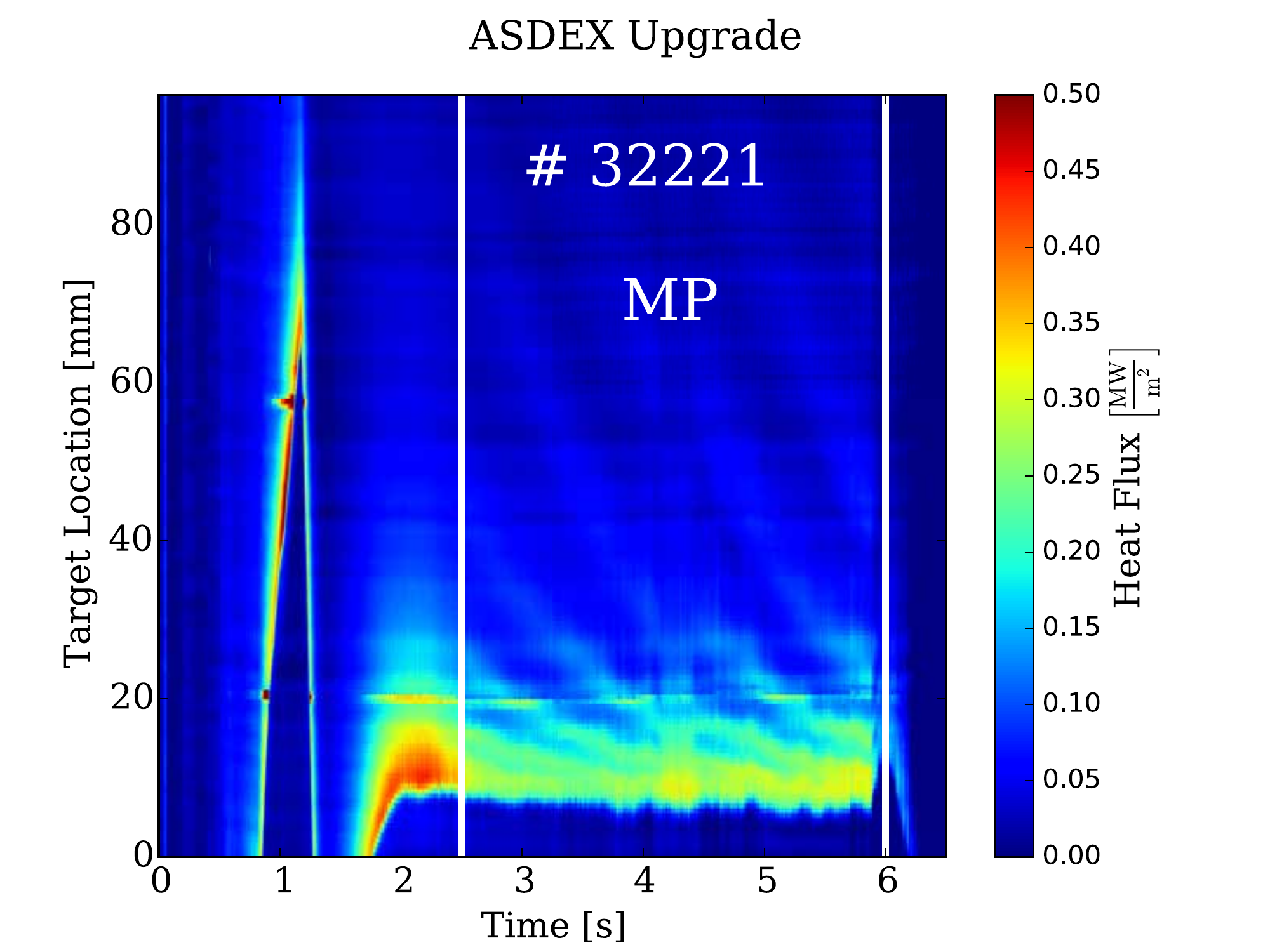}}
\subfigure[\label{fig:2DResonantReduced} Resonant $\Delta\phi\,=\,-\frac{\pi}{2}$ with reduced $I_{\mathrm{coil}}$]{\includegraphics[width=0.3 \textwidth]{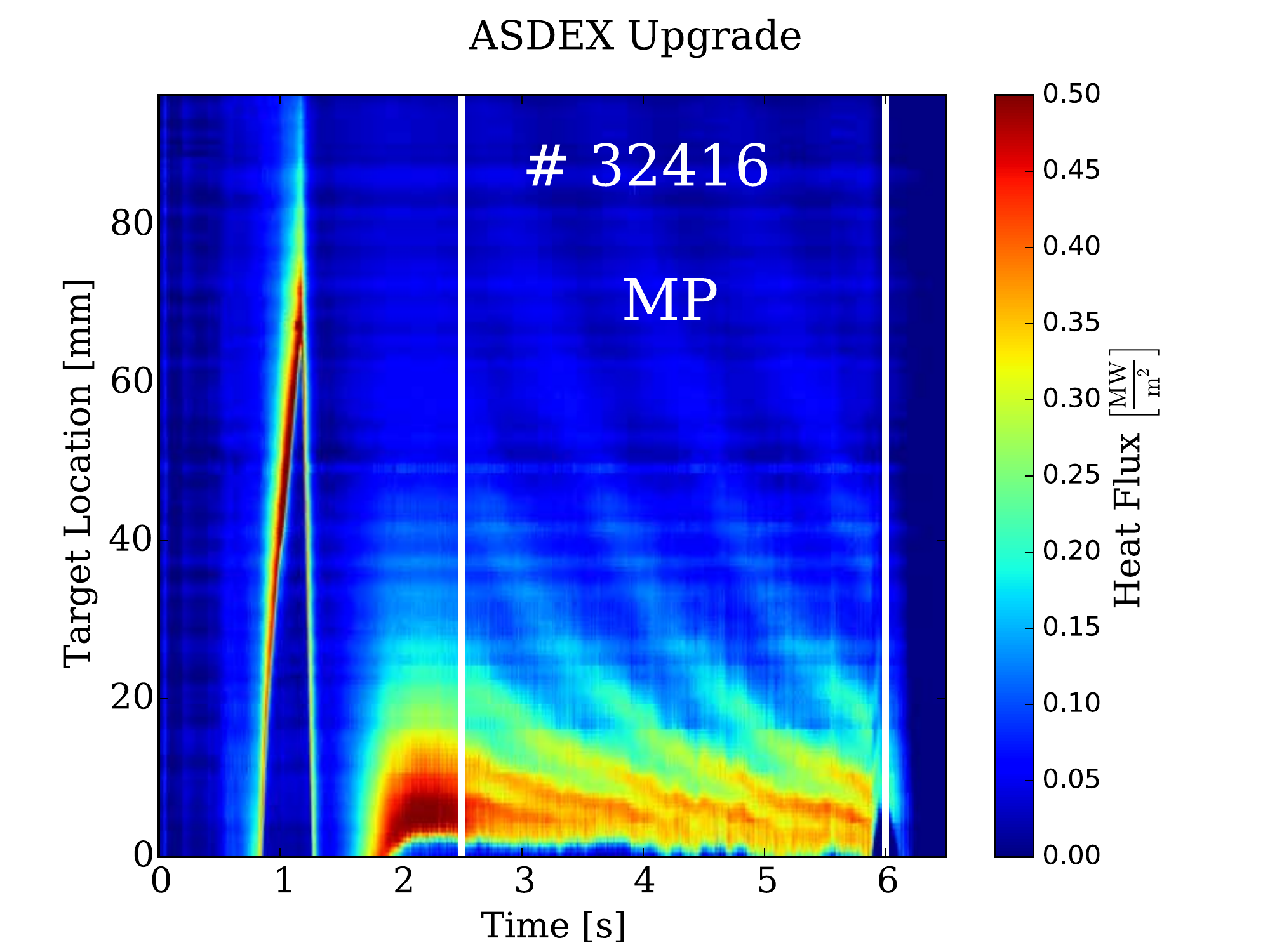}}
\caption{Heat flux time traces for the various \textit{differential phases} between the upper and lower coil currents.
White bars indicate start and end time of the external perturbation, respectively.}
\label{fig:DifferentialPhasesIRTime}
\end{figure}

The largest visual influence of the magnetic perturbation on the heat flux profile evolution is observed in the \textit{resonant} configuration (see \fref{fig:2DResonant}).
It is observed that the strike line position, identified as the sharp rise of the heat flux profile along the target, moves in time when the magnetic perturbation is present.
This is due to imperfections of the attitude control in presence of n\,=\,2 external perturbation, e.g.~\cite{Willensdorfer2016}, possibly enlarged by the presence of internal modes due to the plasma response, for example seen with JOREK simulations~\cite{Orain2016}.
For the further analysis this is taken into account by shifting the heat flux profiles to a fixed strike line position.
The \textit{non-resonant} configuration is shown in~\fref{fig:2DNonResonant}.
The deviation between the axisymmetric heat flux without magnetic perturbation (between 2.0 and 2.5\,s) and with magnetic perturbation is reduced compared to the \textit{resonant} configuration.
However, the helical structure is still present.
The discharges with only the upper and lower coils used are shown in~\fref{fig:2DUpper} and \fref{fig:2DLower}, respectively.
The visual heat flux perturbation is more pronounced than in the \textit{non-resonant} configuration but less pronounced than in the \textit{resonant} configuration.
The phases in between the \textit{resonant} and \textit{non-resonant} configuration are both performed in a single discharge, shown in~\fref{fig:2DMid}.
Neither for a \textit{differential phase} of 0 from 2.5-3.7\,s nor a phase of $\Delta\phi\,=\,+\pi$ between 4.0 and 6.0\,s any difference is seen.
The heat flux profile evolution with the \textit{resonant} configuration, but a reduced current in the coils $I_{\mathrm{coil,red}}\,=\,\frac{1}{\sqrt{2}}~I_{\mathrm{coil}}$, is shown in~\fref{fig:2DResonantReduced}.

\subsection{Toroidally Averaged Heat Flux Profiles} \label{section:averaged}
All global plasma parameters are kept constant in the presented discharges, allowing to average the heat flux profile over one or more full rotation periods.
Four different phases in the rotation of the \textit{resonant} configuration are shown in~\fref{fig:HeatFluxDifferentPosResonant}.
The heat flux profiles are, as mentioned in the previous section, shifted along the target location to match in the strike line position.
The before mentioned hot spot is observed (grey area in~\fref{fig:HeatFluxDifferentPosResonant}) leading to an overestimation of the local heat flux at this position but does not affect the analysis of the fitting to the averaged profile.
The maximum heat flux in all phases is close to the axisymmetric maximum and the maximum of the averaged profile.
A more detailed characterisation of the heat flux distribution is presented in section~\ref{ModellingPeakingProfile} with a comparison to a generic heat flux model.\\
In~\fref{fig:DiffPhaseAveraged} heat flux profiles are shown for the shots shown in~\fref{fig:DifferentialPhasesIRTime}.
The heat flux profiles are normalized to the integrated heat flux for the given profile and to the peak heat flux of the reference profile for \#\,32217.
The normalization is performed because the integrated heat flux in the different shots varies in the order of 10\%.
This is within the regularly observed L-Mode heat flux variation.
A normalization to the peak heat flux for the individual profiles would result in the same figure due to the similarity of the profiles.

\begin{figure}[ht] 
	\centering

\includegraphics[width=0.5\textwidth]{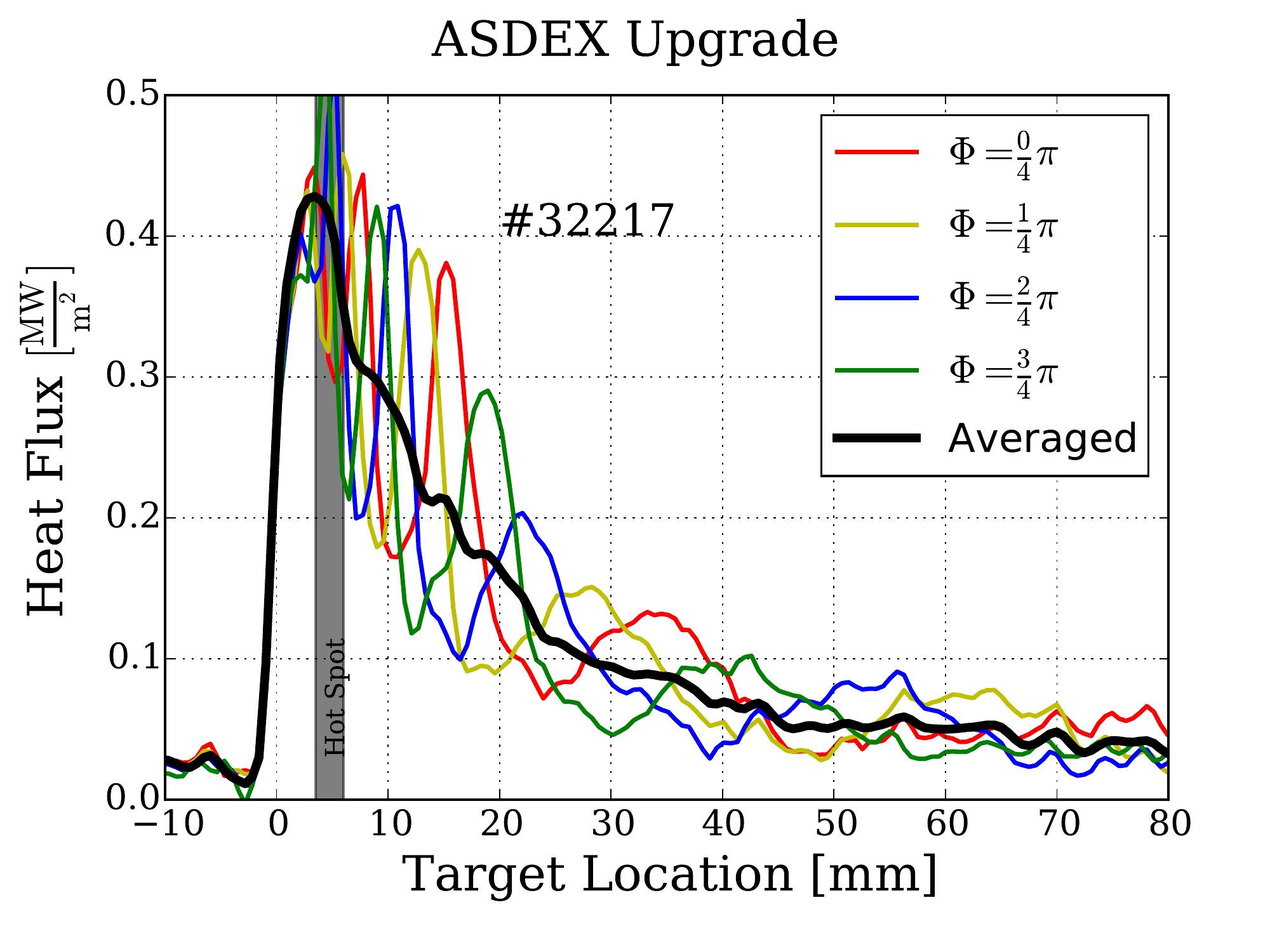}

\caption{Heat flux profiles for different time points in the \textit{resonant} configuration as well as the averaged profile.}
\label{fig:HeatFluxDifferentPosResonant}
\end{figure}

In~\fref{fig:axiWO} heat flux profiles at 2.3\,s are shown.
This is within the reference phase before the magnetic perturbation field is applied at 2.5\,s.
The profiles for all discharges are similar and described by the 1D diffusive model~\eref{eq:diffusiveModel}.
From this it is concluded that the discharges are comparable in terms of edge transport leading to the observed heat flux pattern.
Both transport qualifiers $\lambda_q$ and $S$, obtained by fitting the model to the experimental data, vary only within the fitting uncertainty.
The uncertainty given is the standard deviation for the obtained values.
In~\fref{fig:axiWith} time averaged profiles in the phase with magnetic perturbation are shown.
For the averaging a time window between 3.0\,s and 5.0\,s is chosen.
Except for \#\,32221 where two different \textit{differential phases} are applied and therefore two separate averaged profiles are conducted between 2.7\,s and 3.7\,s for $\Delta \phi \,=\,0$ and between 4.5\,s and 5.5\,s for $\Delta\phi\,=\,+\pi$.
The time windows are chosen to avoid the transient phases during switching of the external magnetic perturbation.
Averaging over 2.0\,s (except for \#\,32221), which corresponds to two periods, ensures that no artificial heat flux variations are present.
The averaged heat flux profiles, in contrast to the single profiles, are described by the 1D diffusive model.
No dependence of the averaged profile on the \textit{differential phase} is observed, in contrast to the different 2D profiles discussed at the beginning of this section.
The parameters describing the heat transport in the scrape-off layer $\lambda_q$ as well as the divertor region $S$ do not show a dependence on the magnetic perturbation.
This is interpreted that there is neither significant additional cross field transport $\chi_{\perp}$ nor significant additional net radial transport along radially deflected field lines due to the change of the radial magnetic field caused by the perturbation coils.
The lobe structure causes a redistribution of heat flux in toroidal direction with an averaged target profile described by the global $\lambda_q$ and $S$.

\begin{figure}[htb!]
	\centering
	
\subfigure[\label{fig:axiWO} Heat flux profiles in the reference time window without magnetic perturbation]{\includegraphics[width=0.45 \textwidth]{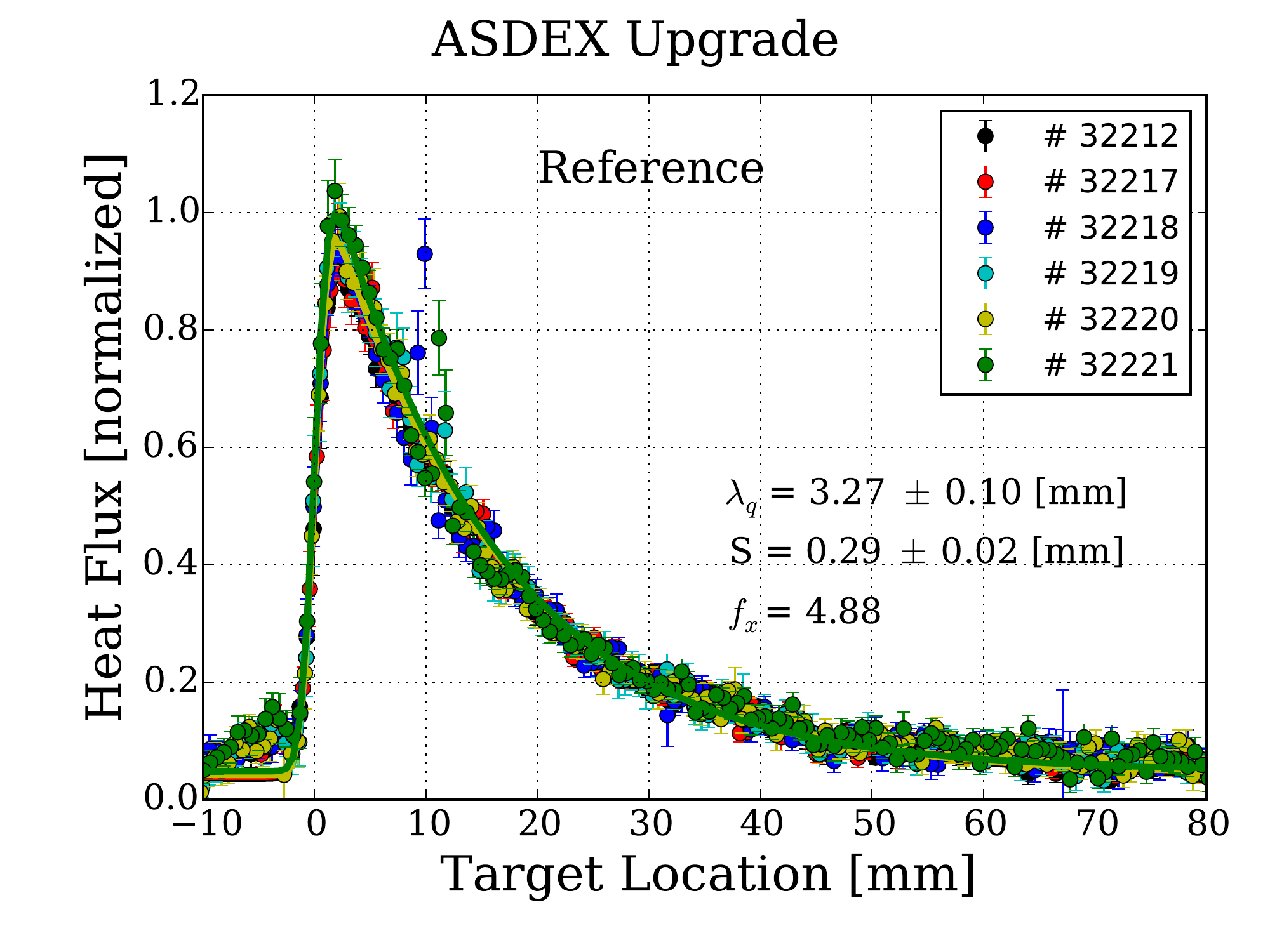}}
\subfigure[\label{fig:axiWith} Time averaged heat flux profiles with different phases between upper and lower coil currents.]{\includegraphics[width=0.45 \textwidth]{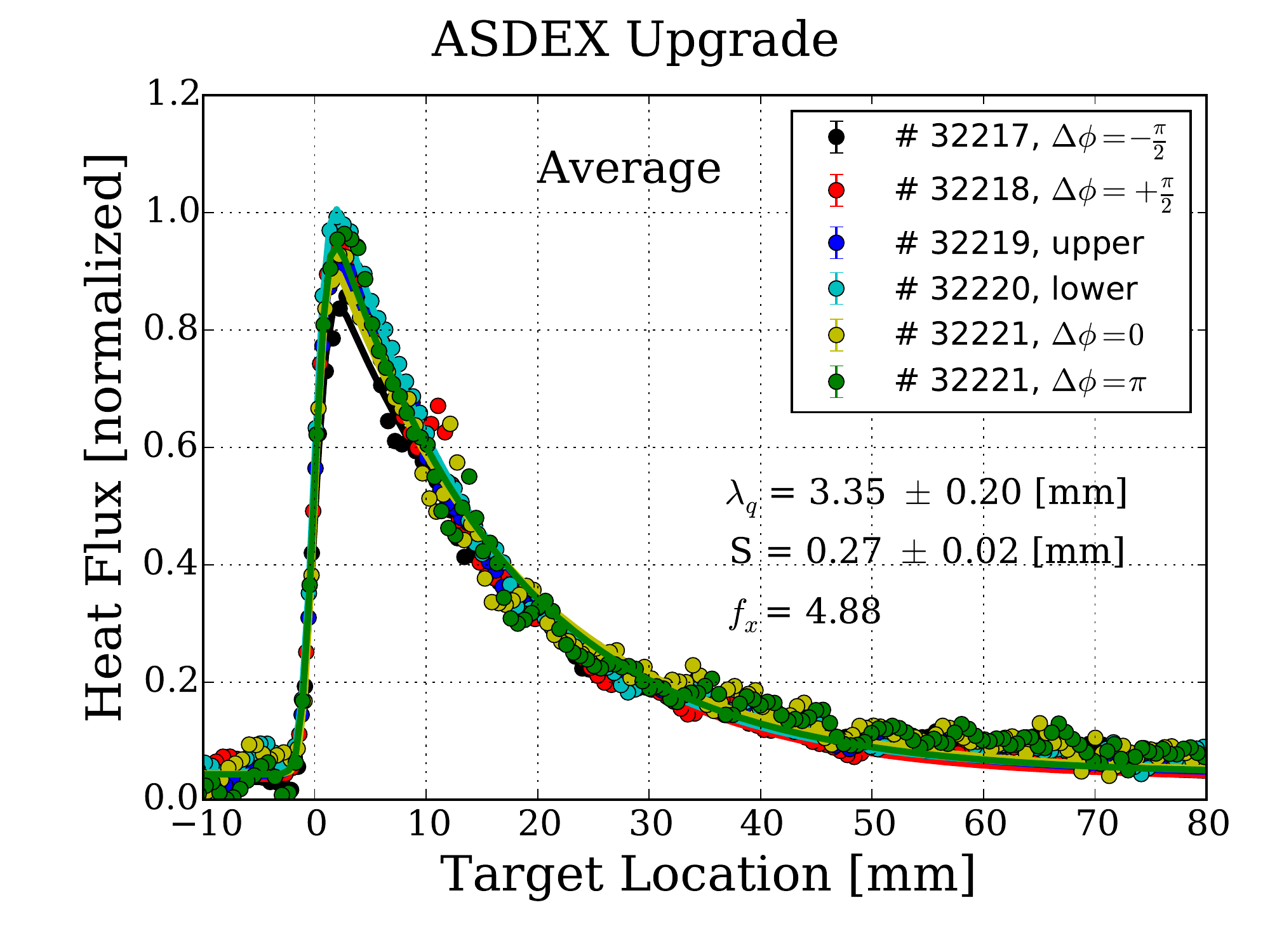}}
\caption{Heat flux profiles with and without magnetic perturbation.}
\label{fig:DiffPhaseAveraged}
\end{figure}

\subsection{Toroidal Heat Flux Variation} \label{section:TimeVariation}
As shown in the previous section, the time averaged profile with rotating magnetic perturbation leads to the same target heat flux profile as the 1D heat flux profile without magnetic perturbation.
Averaging in time with rotating magnetic perturbation is equivalent with averaging in toroidal direction for a static magnetic perturbation or an infinite fast rotation.
Due to the finite rotation frequency in the experiment (1\,Hz) the 2D structure on the target is measured by moving it through the field of view of the IR system.\\
In~\fref{fig:TimeVariation} the heat flux for different target locations in the scrape-off layer, normalized to the averaged heat flux at this position, is shown for the \textit{resonant} and \textit{non-resonant} configuration.
A time window between 3.0\,s and 5.0\,s is chosen containing two complete periods of the rotation.
The target location is expressed in terms of the fitted $\lambda_{q,\mathrm{target}}$ for the averaged heat flux profile.
The black profile (s = 0.21$\cdot \lambda_q$) corresponds to about the peak in the heat flux profile.
Further into the SOL (s \textgreater $\lambda_q$) the IR data becomes more noisy due to the lower signal.
The dots show the single IR measurements, the solid line is smoothed over 125\,ms (100 time points).
The variation between the single time points and the averaged is mainly due to typical heat flux variations in L-Mode and only to a small fraction due to measurement noise.
The heat flux variation for the \textit{resonant} configuration is shown in~\fref{fig:timeVarRes}.
The period in which the heat flux is above the mean value is about the same as the period in which it is below with a nearly sinusoidal structure.
The peak to peak variation varies for the different target locations, with the largest at about 0.67$\cdot \lambda_q$ away from the separatrix with about a factor of 4.
Positions further in the scrape-off layer have less heat flux variation and less averaged heat flux.
Positions closer to the separatrix have less heat flux variation compared to 0.67 $\cdot \lambda_q$ away from the strike line position but with a higher averaged heat flux.\\
This is a direct consequence of the x-point configuration.
The toroidal inclination of the lobes increases towards the strike point which causes a toroidal symmetric profile at the strike line.
This is independent of the heat flux width and set by the unperturbed field configuration.
However, the most critical part for local over-heating along the target of the profile is around $\leq\,\lambda_q$ away from the separatrix, where significant averaged heat flux with a strong variation with the cycle of the magnetic perturbation rotation is observed.\\
The time variation of the heat flux for the \textit{non-resonant} configuration is shown in~\fref{fig:timeVarNon}.
For this configuration the heat flux does not vary significantly in time, as is also seen in the 2D time trace in~\fref{fig:2DNonResonant}.
It has to be noted here, that the structure of the perturbation, although only altering the heat flux marginal, is still observable in the 2D profile.
A comparison between the \textit{toroidal peaking} - maximum value in time in~\fref{fig:TimeVariation} - and the target location is discussed in section~\ref{ModellingPeakingProfile}.

\begin{figure}[htb!]
	\centering
	
\subfigure[\label{fig:timeVarRes} Resonant]{\includegraphics[width=0.45 \textwidth]{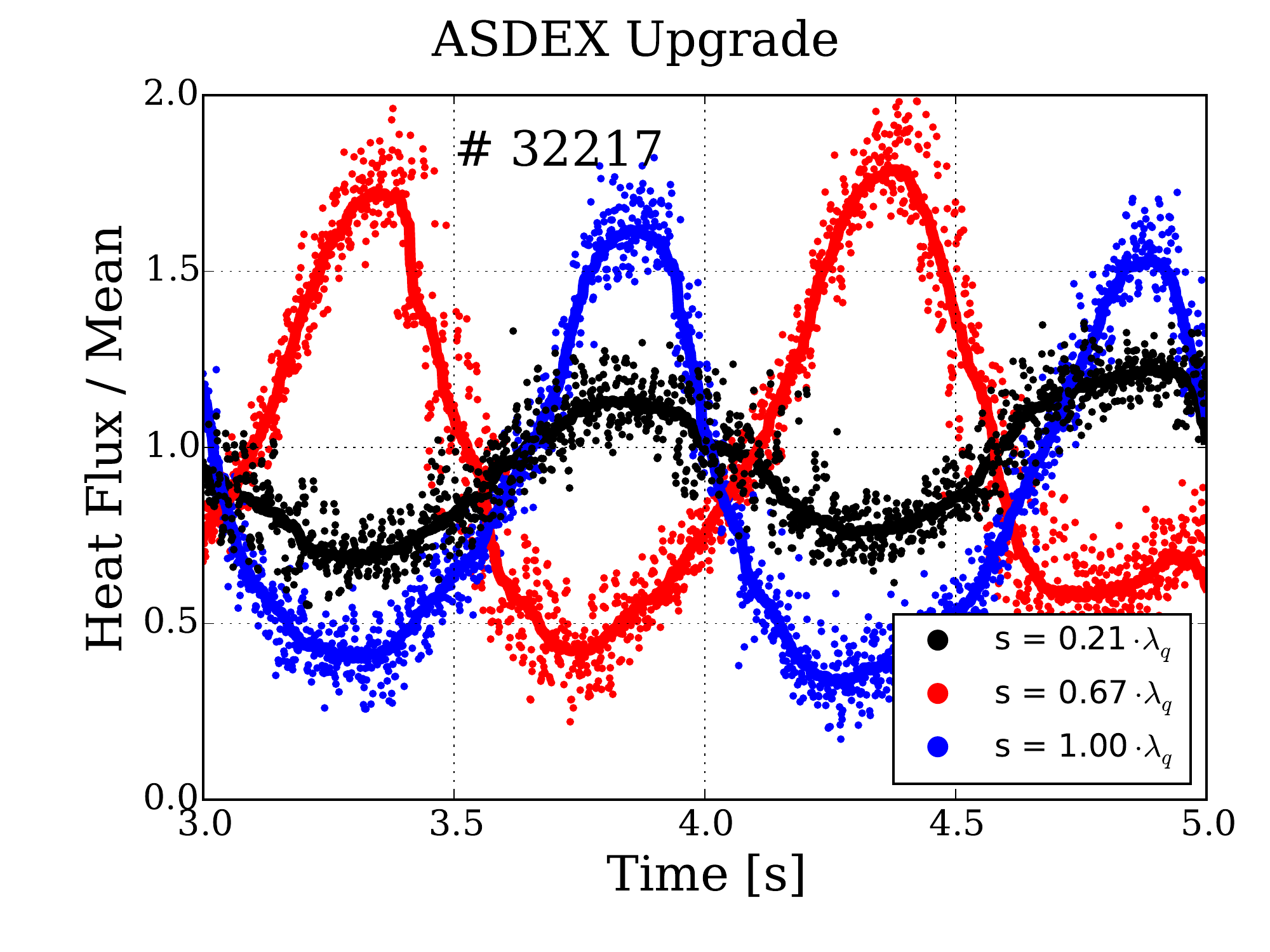}}
\subfigure[\label{fig:timeVarNon} Non-Resonant]{\includegraphics[width=0.45 \textwidth]{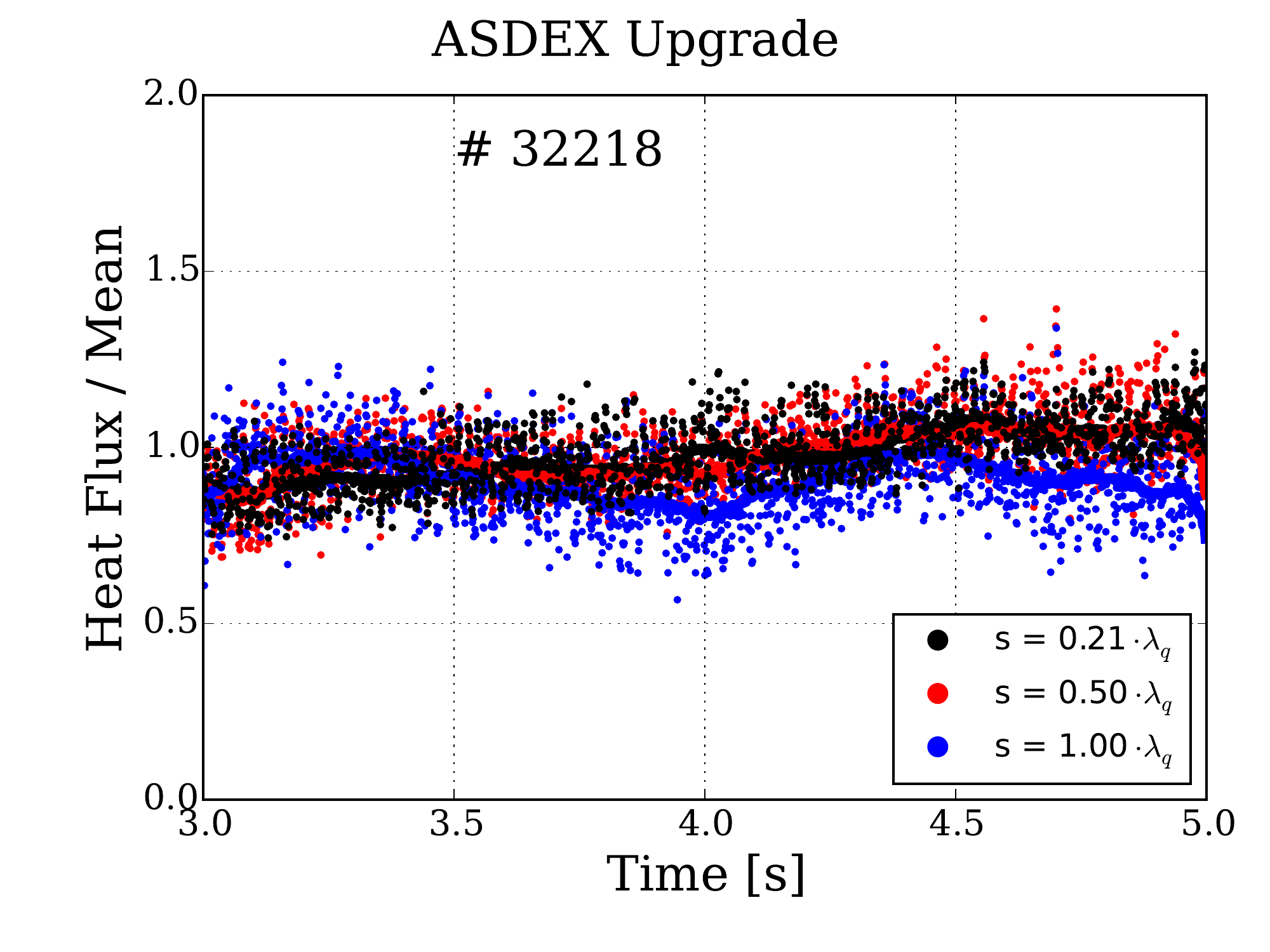}}
\caption{Time variation for different target locations normalized to the average heat flux at this position.}
\label{fig:TimeVariation}
\end{figure}

A variation in time with the rotating magnetic perturbation is also observed by divertor Langmuir probes.
The temperature measured by 4 probes in the scrape-off layer is shown in~\fref{fig:LangmuirTe}.
It has to be noted that the spatial distance between the probes is 20\,mm and there was no strike line sweep in the discharge.
Thus, the spatial resolution is rather low and no proper profile can be constructed.
The about sinusoidal oscillation observed in the heat flux measured by the IR thermography is also observed in the electron temperature.
However, compared to the peak to peak variation of the heat flux of up to a factor of 4, the variation in the electron temperature at the target is in the range of 20\,\%.
The position of the two probes in the scrape-off layer is substantially further away from the strike line position than the positions where the heat flux variation measured by the IR thermography is extracted from.
The probe data shown in black is taken at about the peak heat flux position measured by the IR thermography.
The variation is in both, the electron temperature measured by the Langmuir probes as well as the heat flux measured by the IR thermography, in the same order.
Note here, that the IR thermography data is shifted due to a strike line movement, this is not possible for the Langmuir probes due to the limited spatial resolution.
The time window of 2\,s allows the variation that has a periodicity of 1\,s to be attributed to the MP and not to the strike line movement which seems to have only a minor impact.

\begin{figure}[ht] 
	\centering

\includegraphics[width=0.5\textwidth]{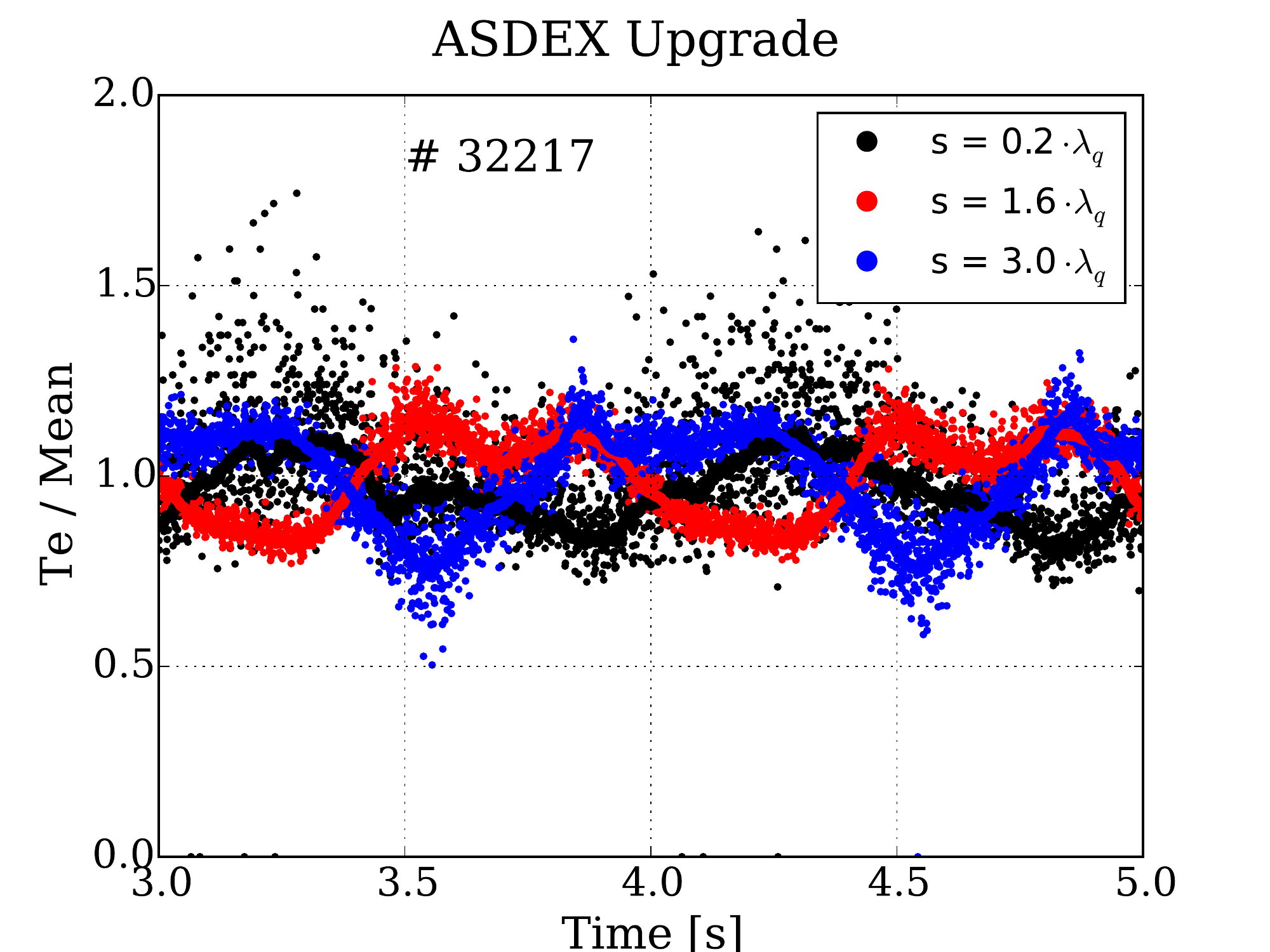}

\caption{Electron temperature variation on the outer divertor target measured with fixed Langmuir probes.}
\label{fig:LangmuirTe}
\end{figure}

\subsection{Effect of Increasing Density}\label{densSteps}
The \textit{resonant} configuration is used to study the effect of magnetic perturbations on $\lambda_q$ and $S$ on the outer divertor target at different densities.
It was shown in previous studies~\cite{Scarabosio2013, Sieglin2013, Sieglin2016} that in L-Mode discharges at ASDEX Upgrade both, $\lambda_q$ and $S$, depend on the electron density at the plasma edge.
The main mechanism to increase $S$ with density is thought to be the reduction of the electron temperature in the divertor region
and, therefore, the reduction of the parallel heat conduction~\cite{Scarabosio2015, Sieglin2016}.\\
A density scan is performed to be able to choose densities that are low enough to have an attached divertor and no significant divertor radiation.
This condidtions result in $T_e\,>$\,10\,eV, whilst still spanning an as large as possible density range.
To study the toroidally averaged profiles, stable conditions are needed.
Three discharges with different density levels are referred to as \textit{low}, \textit{medium} and \textit{high} density corresponding to a line integrated edge density of $n_{e,edge}\,=\,0.8, 1.5, 1.8 \cdot 10^{19} \mathrm{m}^{-2}$, respectively.
The density is measured with the edge channel of the DCN interferometer~\cite{Mlynek2010}.
The terms of low, medium and high density are a choice of convenience for this paper. They do not refer to operation at high density essential for high fusion performance and high radiative scenarios.
The 2D heat flux profiles are shown in~\fref{fig:density2DHeat}, the \textit{medium} density in~\fref{fig:2DMediumDensity} and the \textit{high} in~\fref{fig:2DHighDensity}.
The \textit{low} density reference is shown in~\fref{fig:2DResonant}.
\begin{figure}[htb!]
	\centering

\subfigure[\label{fig:2DMediumDensity} Medium Density, $n_{e,edge}\,=\,1.5 \cdot 10^{19} \mathrm{m}^{-2}$]{\includegraphics[width=0.45 \textwidth]{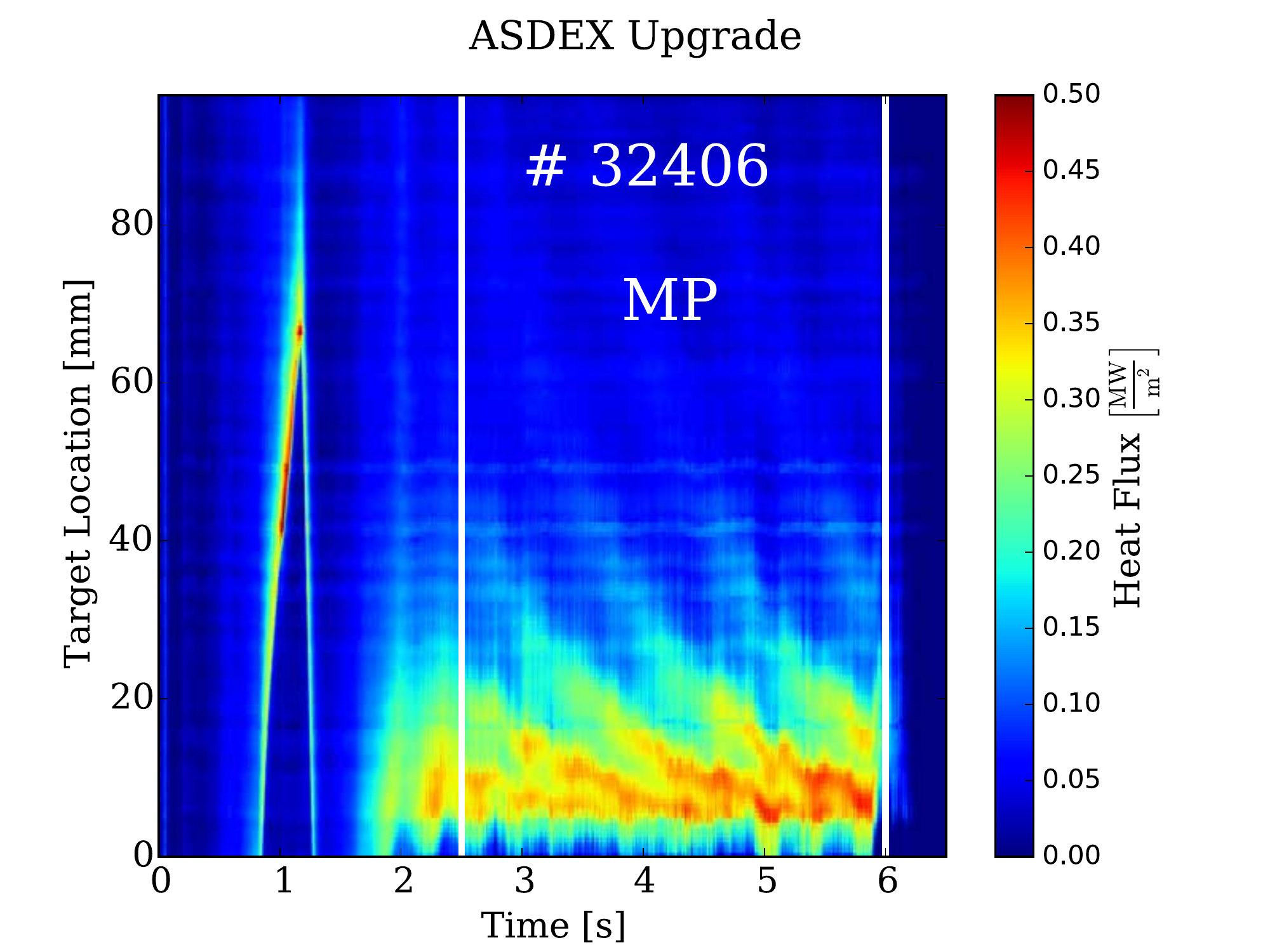}}
\subfigure[\label{fig:2DHighDensity} High Density, $n_{e,edge}\,=\,1.8 \cdot 10^{19} \mathrm{m}^{-2}$]{\includegraphics[width=0.45 \textwidth]{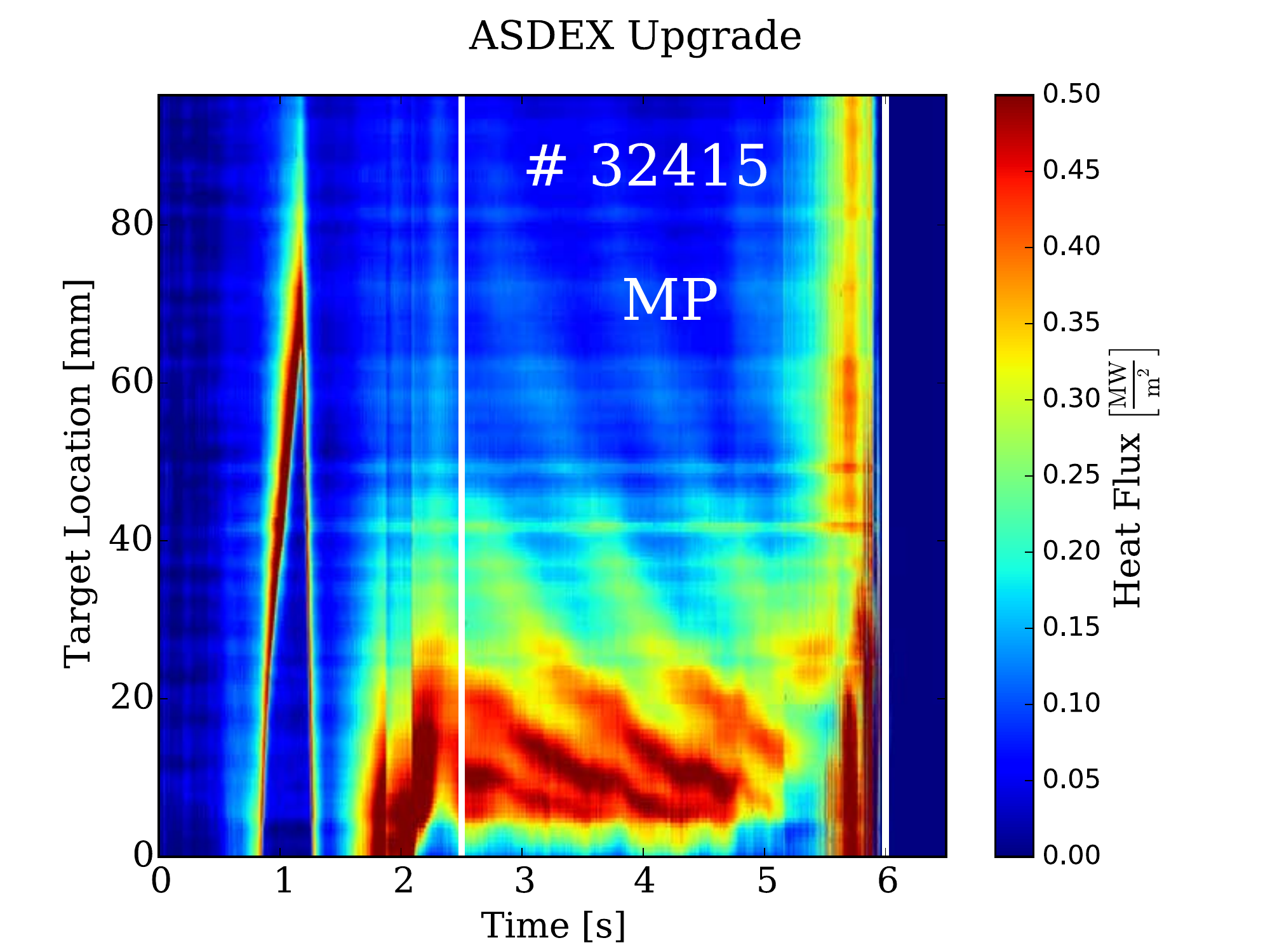}}
\caption{Heat flux time traces for the discharges with higher density.
White bars indicate start and end time of the external perturbation, respectively.}
\label{fig:density2DHeat}
\end{figure}
In all three discharges the 2D structure of the heat flux profile is seen.
Comparing the profiles for the elevated densities with the \textit{low} density reference reveals that increasing the density reduces the heat flux variation.
This is shown in~\fref{fig:densityTimeVar} for all three densities.

\begin{figure}[htb!]
	\centering

\subfigure[\label{fig:timeVarResLow} Low Density, $n_{e,edge}\,=\,0.8 \cdot 10^{19} \mathrm{m}^{-2}$]{\includegraphics[width=0.3 \textwidth]{32217_ToroidalPeaking_TimeTrace.pdf}}
\subfigure[\label{fig:timeVarResMed} Medium Density, $n_{e,edge}\,=\,1.5 \cdot 10^{19} \mathrm{m}^{-2}$]{\includegraphics[width=0.3 \textwidth]{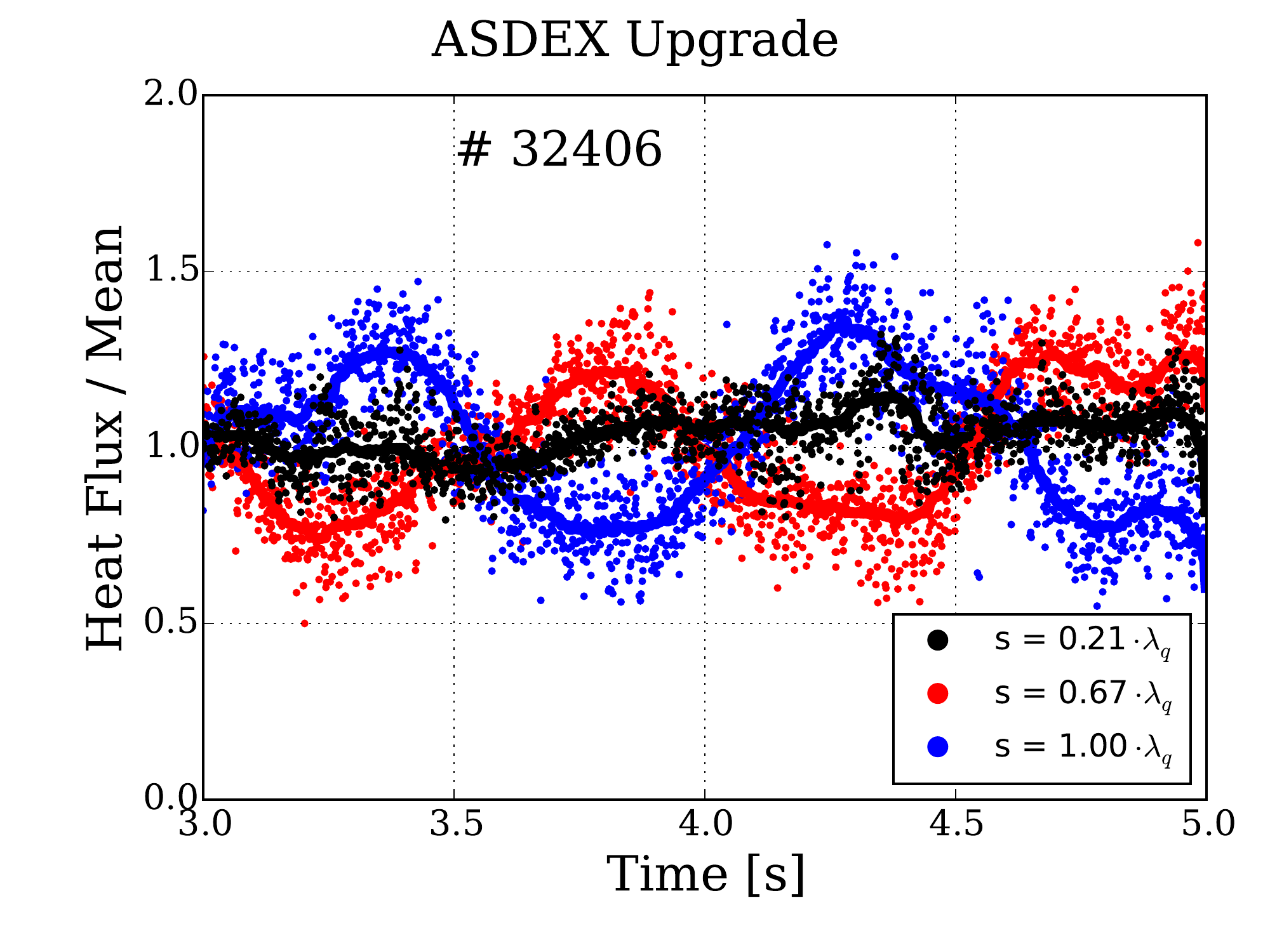}}
\subfigure[\label{fig:timeVarResHigh} High Density, $n_{e,edge}\,=\,1.8 \cdot 10^{19} \mathrm{m}^{-2}$]{\includegraphics[width=0.3 \textwidth]{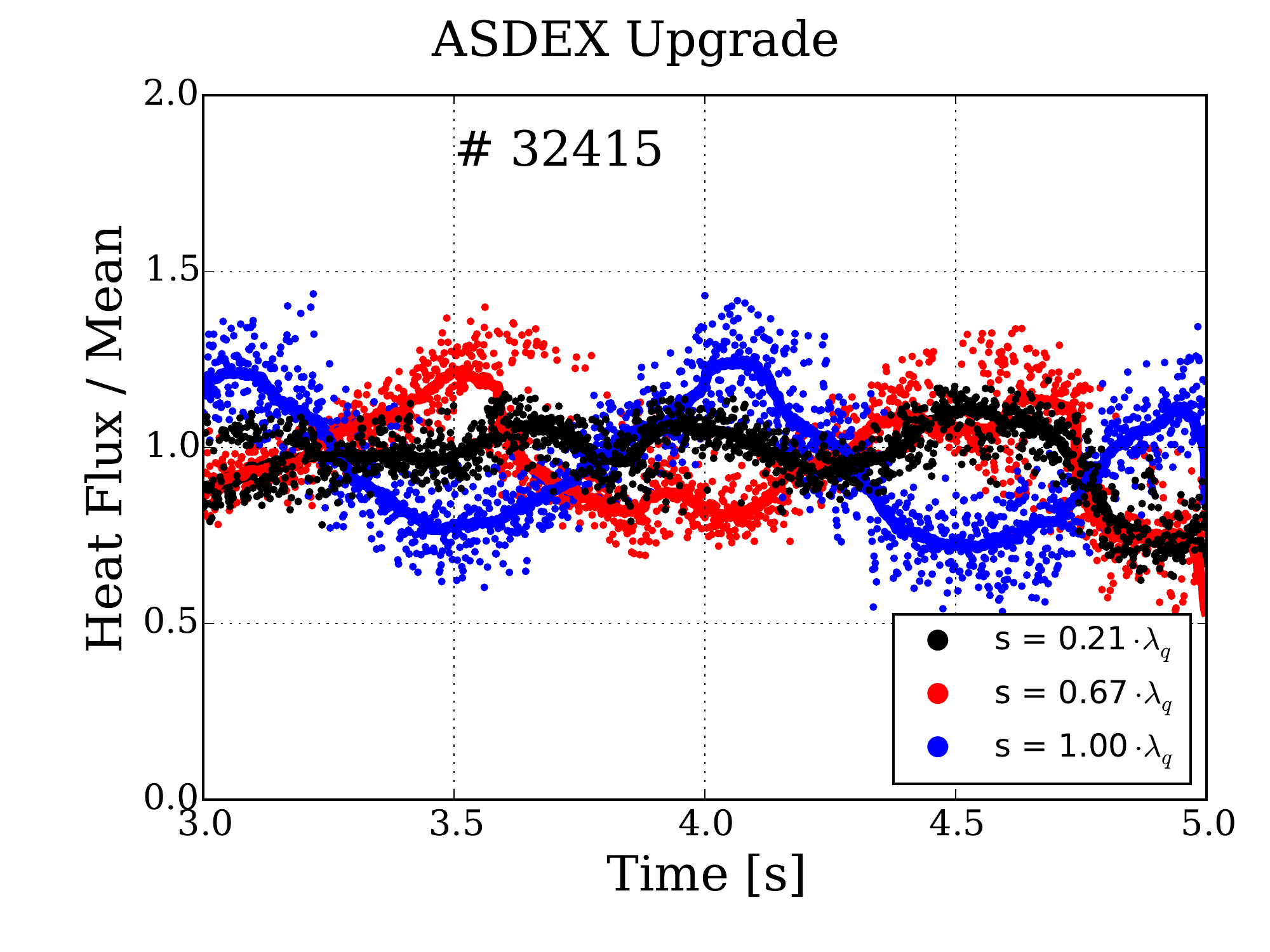}}
\caption{Time variation for different target locations normalized to the average heat flux at this position.}
\label{fig:densityTimeVar}
\end{figure}

The density difference between the \textit{medium} and \textit{high} density discharges is small and no clear difference in the time variation of the heat flux is seen.
The variation in both cases is around 20\,\%.
In~\fref{fig:density1DHeat} the comparison between the 1D profiles for the reference time window (\fref{fig:DensityStepsWithout}) and the toroidally averaged profile with magnetic perturbation in the \textit{resonant} configuration (\fref{fig:DensityStepsWith}) is shown.
The normalization is the same as in section~\ref{section:averaged}.

\begin{figure}[htb!]
	\centering
	
\subfigure[\label{fig:DensityStepsWithout} without magnetic perturbation]{\includegraphics[width=0.45 \textwidth]{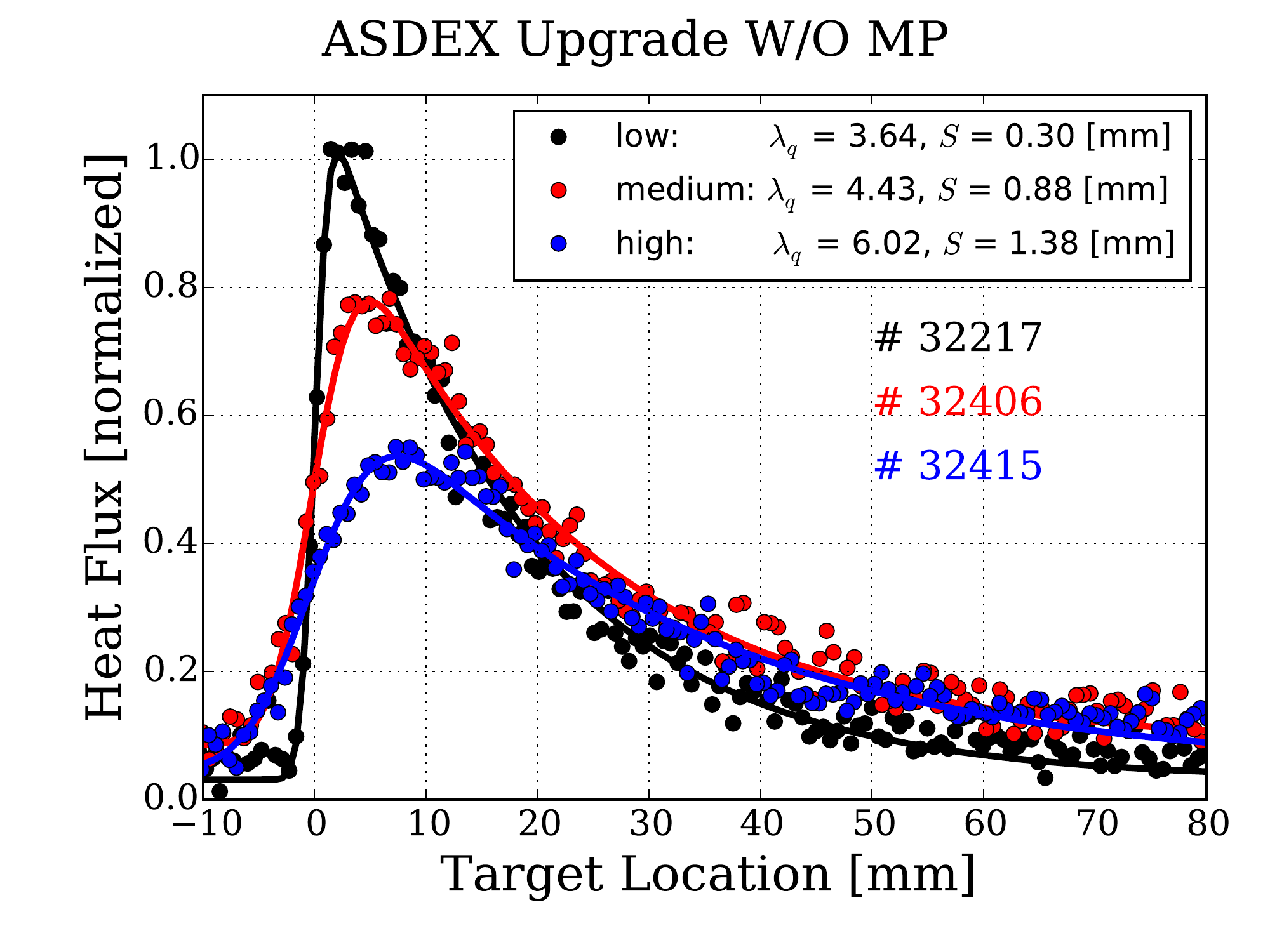}}
\subfigure[\label{fig:DensityStepsWith} with magnetic perturbation]{\includegraphics[width=0.45 \textwidth]{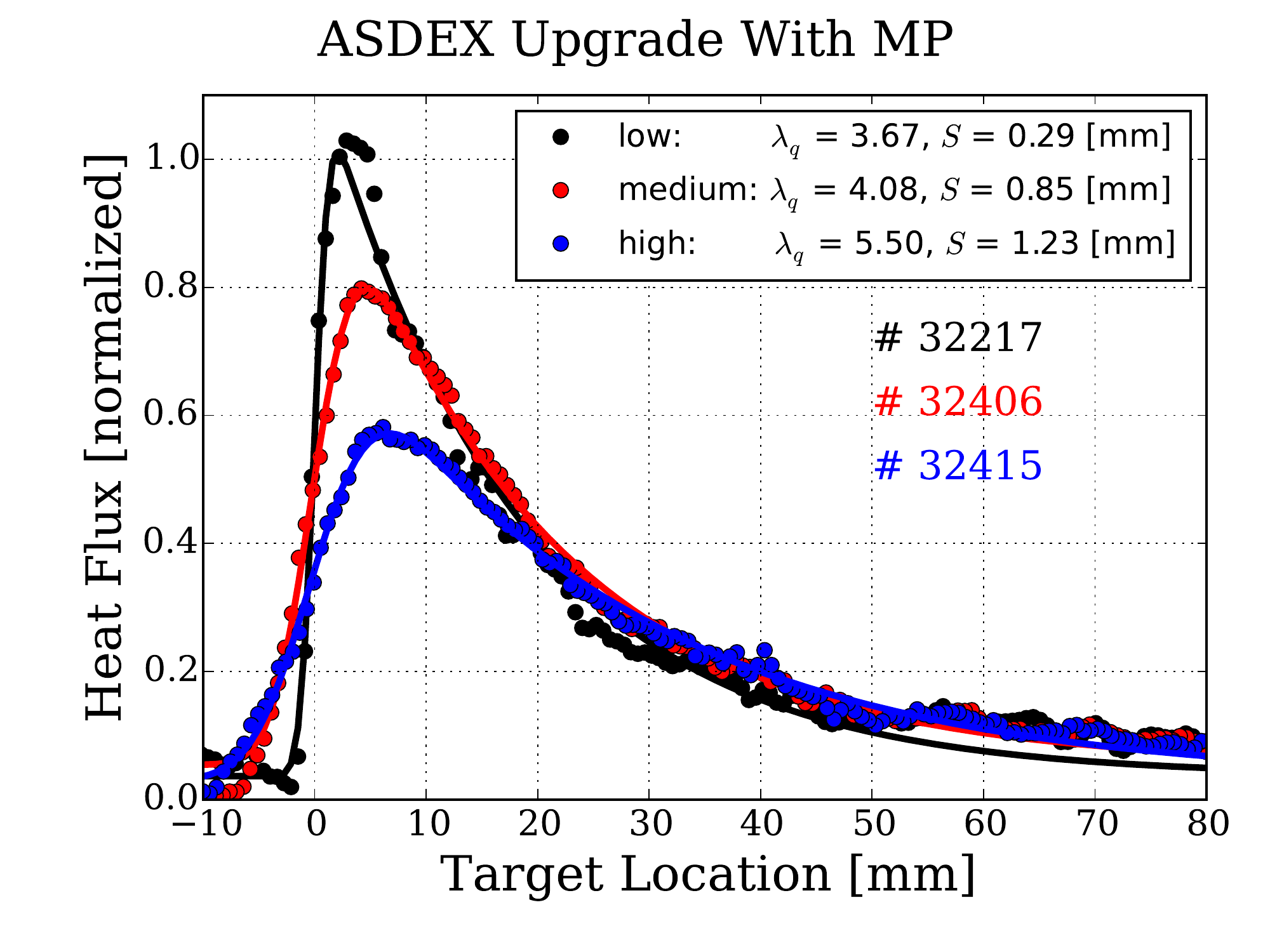}}
\caption{Heat flux profiles with and without magnetic perturbation for the three different densities.}
\label{fig:density1DHeat}
\end{figure}

As mentioned in the beginning of the section, an increase for both, $\lambda_q$ and $S$, is expected for increasing density in ASDEX Upgrade L-Mode.
This is confirmed in the three density steps without magnetic perturbation as the peak heat flux is reduced by a factor of 2.
With increasing density the toroidally averaged profiles are still described by the 1D diffusive model.
The transport qualifiers are similar to the reference values without magnetic perturbation and no significant change with density is observed.

\section{Heat Flux Model}\label{Modelling}
To interpret the experimental results a simple model was developed.
With this model the influence of different quantities, e.g. coil current, divertor broadening $S$ and poloidal spectrum (\textit{resonant} vs. \textit{non-resonant}), is studied.
The intention of this model is to get sufficient agreement with the measured heat flux profiles without treating all the plasma parameters.
This allows to aim for as few as possible (free) input parameters as well as the possibility to change single parameters and study the direct influence they have onto the heat flux distribution.\\
This model is based on the vacuum field approach and a field line tracer, similar approaches are discussed in~\cite{Nguyen1997, Finken1998, Strumberger1996, Eich2000, Cahyna2014}.
The used field line tracer is the 5th order Runge-Kutta GOURDON code~\cite{Gourdon1971, Strumberger1996, Strumberger2000}.
The axisymmetric poloidal flux matrix is calculated using CLISTE~\cite{Schneider2000, McCarthy1999}.
For every step the magnetic field of the perturbation coils is calculated and added to the axisymmetric magnetic field.
The current inside these coils is measured and is corrected taking the conductive passive stabilizing loops (PSL) into account which acts as a low pass filter~\cite{Gruber1993}.
The effective current inside the coils is reduced by 25\% according to FEM calculations at a rotation frequency of 1\,Hz~\cite{Suttrop2009EPS}.
Field lines are traced starting at the outer divertor target and either end at some plasma facing component, e.g. inner divertor, or end up in the confined region and are terminated at a maximum length of \textgreater\,2\,km.
No difference is observed if the field lines are terminated at a length of \textgreater\,200\,m.\\
The following assumptions are made:
\begin{itemize}
\item $R_{sep}(\phi)$ = $A_0 \cdot \sin\left(2\cdot \phi + B_0\right) + A_1 \cdot \sin\left(4\cdot \phi + B_1\right) + R_{sep,axi}$\\
A 2D separatrix at the outer midplane (OMP, z\,=\,0) is defined using the major radius of the axisymmetric separatrix $R_{sep,axi}$ as a mean value (mean($R_{sep}(\phi)$) = $R_{sep,axi}$) and a toroidal $\phi$ sinusoidal component ($\phi\,\epsilon\,[0,2\pi)$).
The periodicity of the sinusoidal is given by the dominant mode number n of the applied magnetic perturbation, which is kept constant at n\,=\,2 for these experiments.
The amplitude of the deformation is fitted using an arbitrary but fixed field line length approaching the separatrix in the unperturbed case.
In the presented results this was fixed to a range between 120-125\,m.
This separatrix is used to define the poloidal flux coordinate $\rho_{pol}(R,\phi)$ also as a 2D quantity.
\item $q_{||}$ = const. along the field line.\\
No perpendicular heat transport. The perpendicular information is covered with the temperature fall-off length. The cross field transport in the divertor region is simplified by convolving the target profile with a Gaussian. 
\item $T_e$($\rho_{pol}$), $q_{||}$($\rho_{pol}$) are flux quantities and have an exponential fall-off length.\\
The electron temperature fall-off length $\lambda_{Te}$ is calculated using the two point model~\ref{eq:2Point} and the power-fall-off length $\lambda_q$.
The two-point model is in good agreement for ASDEX Upgrade L-Mode discharges in the density range from 2 to $6\cdot10^{19} \mathrm{m}^{-2}$ using the power fall-off length measured at the outer divertor target and the temperature fall-off length measured with the Thomson scattering diagnostics~\cite{Faitsch2015, Sun2015}.
\item The local field line angle at the target $\alpha$ is calculated for tiles without tilting using.\\
The tilting of the divertor targets is needed to prevent leading edges. A second minor change in the field line angle is the flat tile surface, not following the toroidal direction of the vessel.
The measurement position is not changed within the presented study and thus the effect of the tilting is only a constant attenuation.
Both effects together lead to a variation of the heat flux amplitude in toroidal direction of about 20\,\% and an additional, but less pronounced, difference in direction of the target location.
The angle is calculated with the magnetic perturbation present, although it does not change the angle significantly.\\

\item $L_{\mathrm{OMP}}$ = constant.\\
The length between the OMP and the outer divertor target is approximated to be a constant.
This is done in order to have both an exponential temperature decay and an exponential heat flux decay in the OMP for an axisymmetric configuration.
This reduces the number of free parameters to the separatrix temperature.
\item $T_{e\mathrm{,target}}$\,=\,0\,eV.\\
Using the same approximation as in the derivation for~\eref{eq:2Point}.
\end{itemize}
The following numerical values are used for all calculations:
\begin{itemize}
\item $L_{\mathrm{OMP}}$ = 25\,m.
\item $\lambda_q$\,=\,3.67\,mm, $S = 0.3$\,mm (deduced from measurement at \#\,32217 @ 2.3\,s).
\item $\kappa_0$ = 2000\,$\frac{\mathrm{W}}{\mathrm{m} \cdot \left(\mathrm{eV}\right)^{\frac{7}{2}}}$~\cite{Stangeby2000, Kallenbach2001}
\item $T_{e,sep} = 46, 45, 46$\,eV for the axisymmetric, \textit{resonant}, \textit{non-resonant} case, respectively. 
The separatrix temperature is set to match the peak heat flux from the IR measurement.
This leads to reasonable values for a low power L-Mode in ASDEX Upgrade.
\end{itemize}
The parallel heat flux $q_{||}$ is calculated using the two point model with Spitzer-H\"arm electron conduction:
\begin{eqnarray}
q_{||} = -\kappa_0 T^{\frac{5}{2}} \frac{\mathrm{d}T}{\mathrm{d}x}
\end{eqnarray}
with $x$ the coordinate along the field line and $T$ the electron temperature~\cite{Stangeby2000}.
Integrating along $x$ leads to a relation between the upstream temperature $T_u$ and the target temperature $T_t$ with the connection length $L_{\mathrm{OMP}}$
\begin{eqnarray}
T_u = \left(T_t^{\frac{7}{2}} + \frac{7 q_{||} L_{\mathrm{OMP}}}{2 \kappa_0}\right)^{\frac{2}{7}}
\end{eqnarray}
Neglecting the target electron temperature $T_t$ ($T_u\,\gg\,T_t$):
\begin{eqnarray}
T_u \approx \left(\frac{7 q_{||} L_{\mathrm{OMP}}}{2 \kappa_0}\right)^{\frac{2}{7}}
\end{eqnarray}
leading to a ratio between upstream temperature fall-off length~$\lambda_{T_e}$ and power fall-off length~$\lambda_q$ of 
\begin{eqnarray} \label{eq:2Point}
\frac{\lambda_{T_e}}{\lambda_q} = \frac{7}{2}
\end{eqnarray}
and a parallel heat flux $q_{||}$ of 
\begin{eqnarray}
q_{||} \approx \frac{2}{7} \frac{\kappa_0 T_u^{\frac{7}{2}}}{L_{OMP}}
\end{eqnarray}
The heat flux perpendicular to the target is calculated using the pitch angle $\alpha$ giving:
\begin{eqnarray} \label{eq:perpendicular}
q_{\perp} = \sin\left(\alpha\right) \cdot q_{||}
\end{eqnarray}
which is in the order of $\sin\left(\alpha\right) = \frac{1}{20}$ for ASDEX Upgrade and the present divertor configuration Div\,III.

The field line tracer allows the calculation of the parallel heat flux at a given target position (before the convolution with the Gaussian):
\begin{eqnarray} \label{eq:parallel}
q_{||, target}\left(s,\phi\right) = f\left(T_u\right) = f\left(R_{\mathrm{OMP}}, \phi_{\mathrm{OMP}}\right)
\end{eqnarray}
with the before mentioned assumptions and numerical constants.
The heat flux profile is interpolated in order to get a uniform distribution along the target location s.
This heat flux is then convolved with a Gaussian function g with a width $S \cdot f_x$:
\begin{eqnarray}
q_{\perp}^{*} = (q_{\perp}*g)(s)
\end{eqnarray}
with \eref{eq:parallel}, \eref{eq:perpendicular} and
\begin{eqnarray}
g(s) ={\frac {1}{S \cdot f_x {\sqrt {\pi }}}}e^{-\left({\frac {s}{S \cdot f_x}}\right)^{2}}
\end{eqnarray}
the Gaussian as used in~\eref{eq:diffusiveModel}.

\FloatBarrier
\section{Comparison: Heat Flux Model and Experiment}\label{ModellingResults}
In this section the heat flux profiles obtained from the presented model in section~\ref{Modelling} are compared to the experimental heat fluxes measured with the IR system.
\subsection{2D Heat Flux Structure}
The 2D structure of the heat flux profile is examined in this section.
In order to compare the measured heat flux with the model, the time variation from~\fref{fig:2DResonant} is transferred into a toroidal distribution.
This is justified by the constant background plasma parameters.
The strike line position is corrected as mentioned in section~\ref{ExperimentResults}.
The 2D structure for the \textit{resonant} configuration is shown in~\fref{fig:2DResonantAngle}.
Note that here the scale in toroidal direction and along the divertor target are a factor of 100 different, with the circumference of ASDEX Upgrade is in the order of 10\,m and the extent along the target is in the order of 5\,cm.
The toroidal angle of $\phi\,=\,0$ is arbitrary and set to be the heat flux at $t\,=\,3\,$s.

\begin{figure}[ht] 
	\centering

\includegraphics[width=0.5\textwidth]{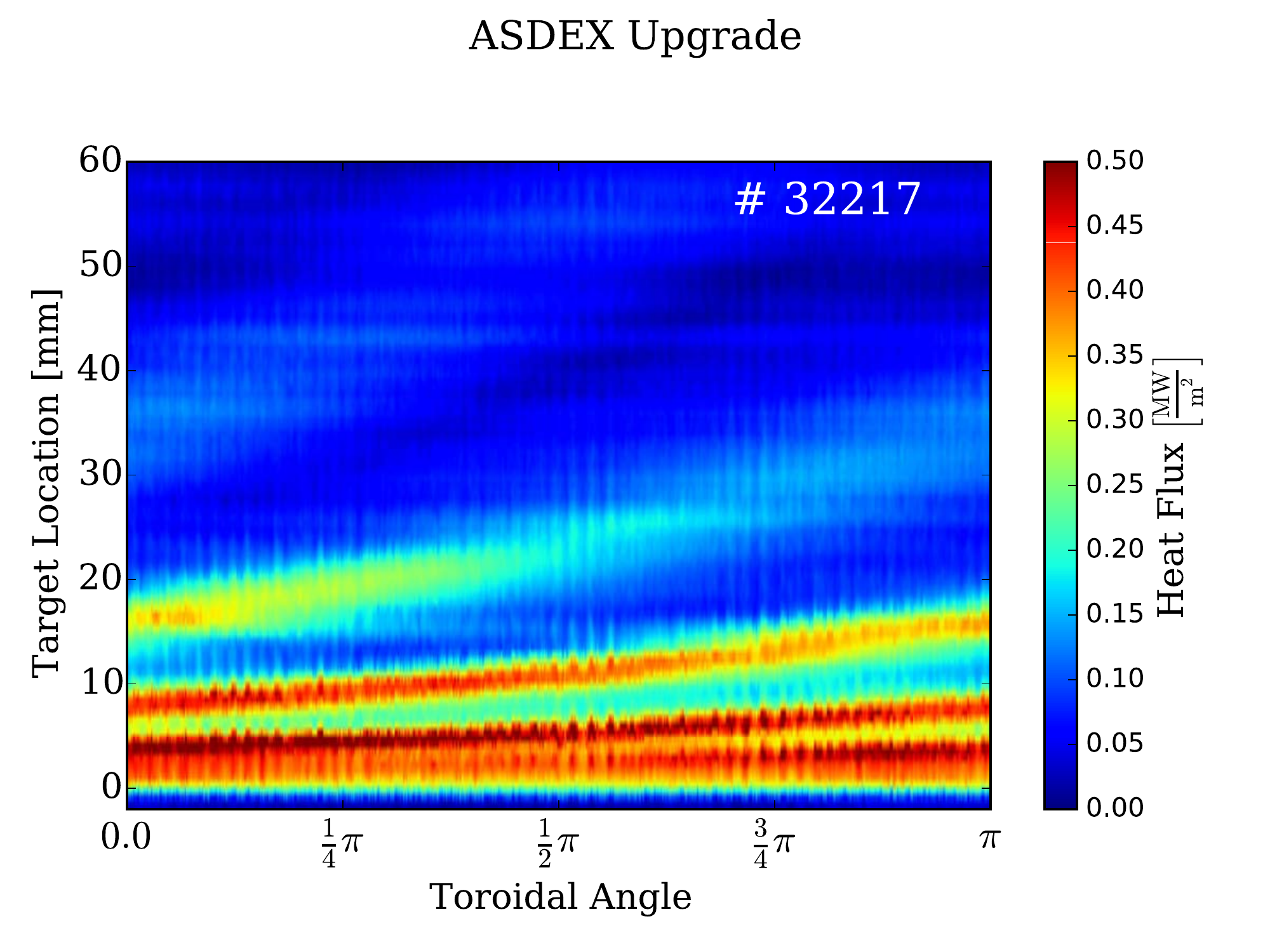}

\caption{2D Heat Flux for the \textit{resonant} configuration in \#\,32217 deduced from IR measurements.}
\label{fig:2DResonantAngle}
\end{figure}

\begin{figure}[ht] 
	\centering

\includegraphics[width=0.5\textwidth]{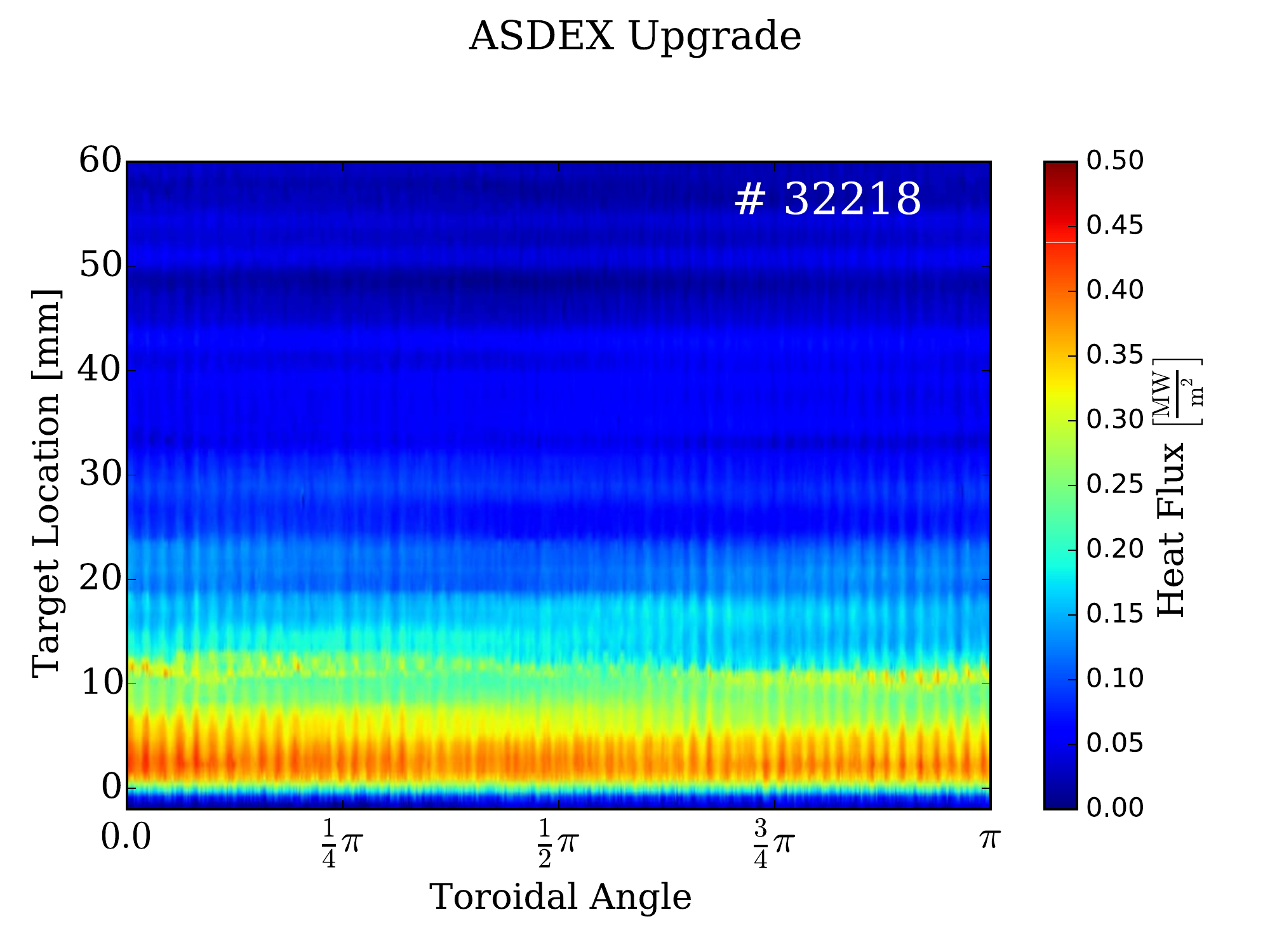}

\caption{2D Heat Flux for the \textit{non-resonant} configuration in \#\,32218 deduced from IR measurements.}
\label{fig:2DNonResonantAngle}
\end{figure}

\subsubsection{Influence of the Conductive Wall onto the Perturbation Field}
The \textit{resonant} configuration with the rotation in the opposite direction is performed to measure the phase delay due to the conductive passive stabilizing loops (PSL) nearby, acting as a low pass~\cite{Suttrop2009EPS, Rott2009}.
From ref~\cite{Suttrop2009EPS} it is known that the phase delay should be about $\frac{1}{24} \pi$ (15$^{\circ}$) for a rotation frequency of 1\,Hz.
Thus, the difference between the two discharges with opposite rotation directions should be $\frac{1}{12} \pi$ (30$^{\circ}$).\\
A comparison between both rotation directions is shown in~\fref{fig:LobePositionRotationDir}.
The dots represent the local maxima $s_m$ in direction of the target location s, $q(s_m-1) < q(s_m)>  q(s_m+1)$.
The agreement obtained without a phase shift for the profile close to the former separatrix position can't be improved by adding a phase shift to one of the profiles.
The distribution close to the former separatrix is nearly toroidal due to the x-point geometry.
The change in the target location along the toroidal direction is less than 5\,mm per $\pi$ up to one $\lambda_q$ away from the former separatrix position.
A phase shift close to the former separatrix is not resolvable with the IR system due to the limited spatial resolution.
Further into the scrape-off layer the change in the target location along the toroidal direction becomes larger.
The determination of the local maxima gets more uncertain with less arriving heat flux.
A phase shift in the expected range of $\frac{1}{12} \pi$ is not resolvable with the heat flux data from the IR system.\\
The low pass filter not only leads to a phase shift but also to an attenuation of the amplitude.
The amplitude is reduced by about 25\,\% compared to a static field according to ref~\cite{Suttrop2009EPS}.
This attenuation is accounted for in the calculation of the magnetic field for the modelling.

\begin{figure}[ht] 
	\centering

\includegraphics[width=0.5\textwidth]{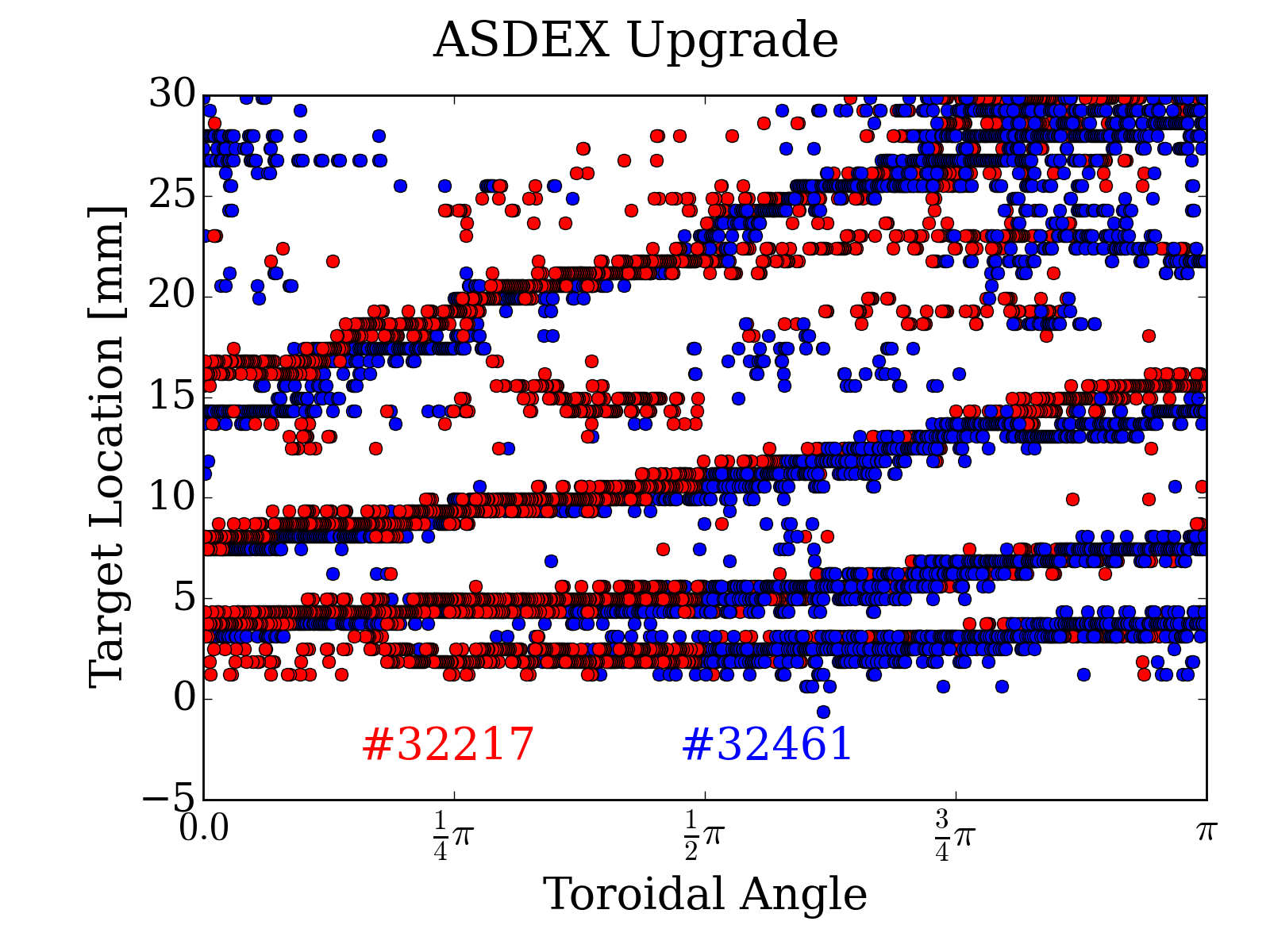}

\caption{Lobe position for both rotation directions without phase shift.}
\label{fig:LobePositionRotationDir}
\end{figure}

The positions of the local maxima in the modelled heat flux distribution as well as in the experiment for the \textit{resonant} configuration are shown in~\fref{fig:LobePosition}.
The modelled data is calculated at 3.5\,s and shifted according to the phase shift due to the PSL and the offset of the IR position.
The position of the local maxima is in agreement within the uncertainty.
However, as already discussed in the previous section, the phase information in the measured data is limited due to the nearly toroidal direction close to the former separatrix due to the x-point shear.

\begin{figure}[ht] 
	\centering

\includegraphics[width=0.5\textwidth]{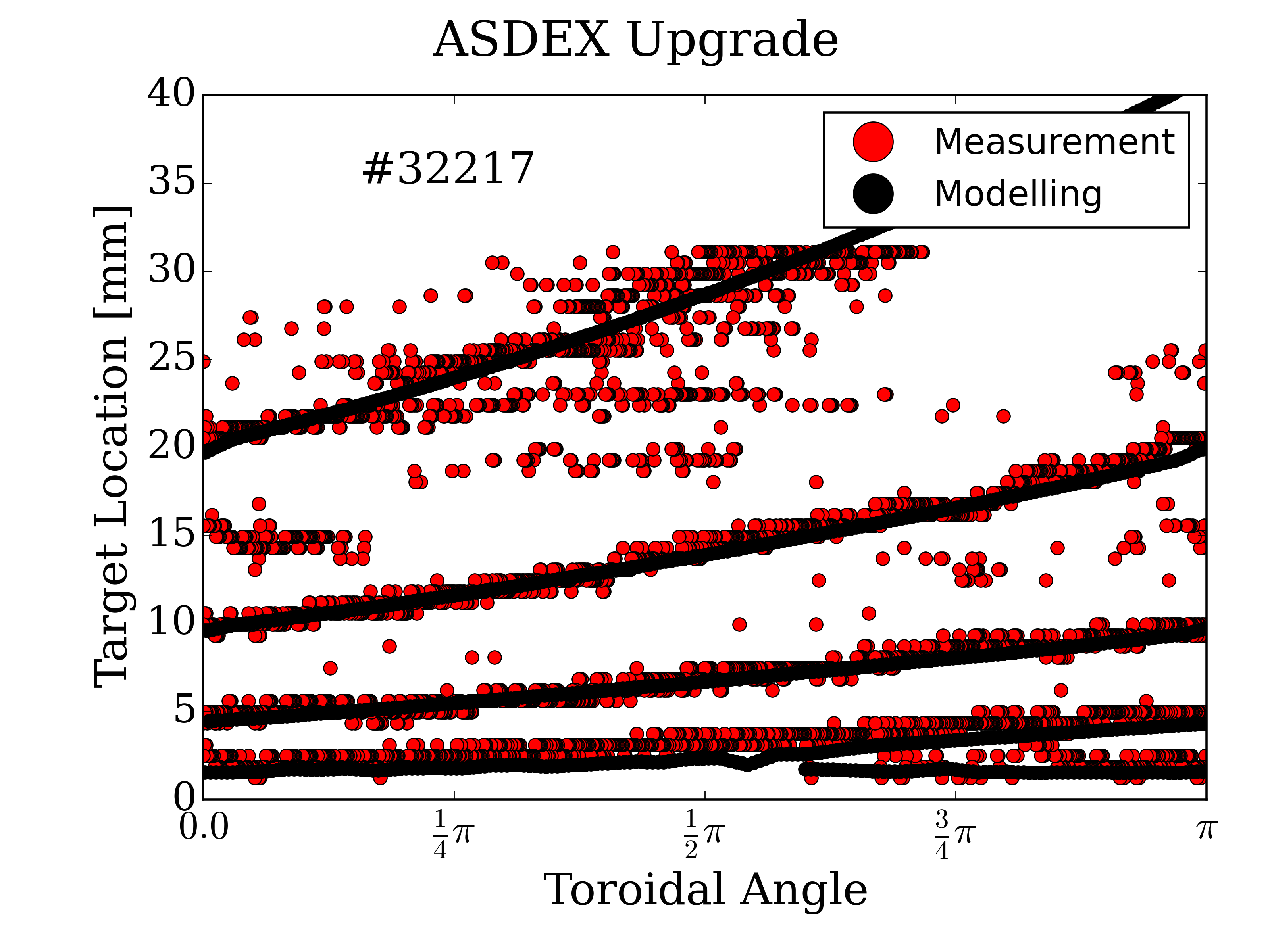}

\caption{Lobe position for modelling (black) and experiment (red dots) for the \textit{resonant} configuration \#\,32217.}
\label{fig:LobePosition}
\end{figure}

The local maxima for three different coil configurations is shown in~\fref{fig:LobePositionCurrent}.
It is observed that the relative position of the maxima is independent of the coil current setup.
The local maxima represent two fixed upstream toroidal positions (magnetic perturbation with a toroidal mode number n\,=\,2) at the OMP.
The toroidal angle is fixed by the absolute phase of the magnetic perturbation, the position at the target is - after shifting to the same absolute phase - independent of the magnetic perturbation.
This holds as long as the perturbation is small as well as no significant non ideal plasma response shifts the absolute phase of the perturbation, e.g.~\cite{Liu2011, Orain2016}.
 
\begin{figure}[ht] 
	\centering

\includegraphics[width=0.5\textwidth]{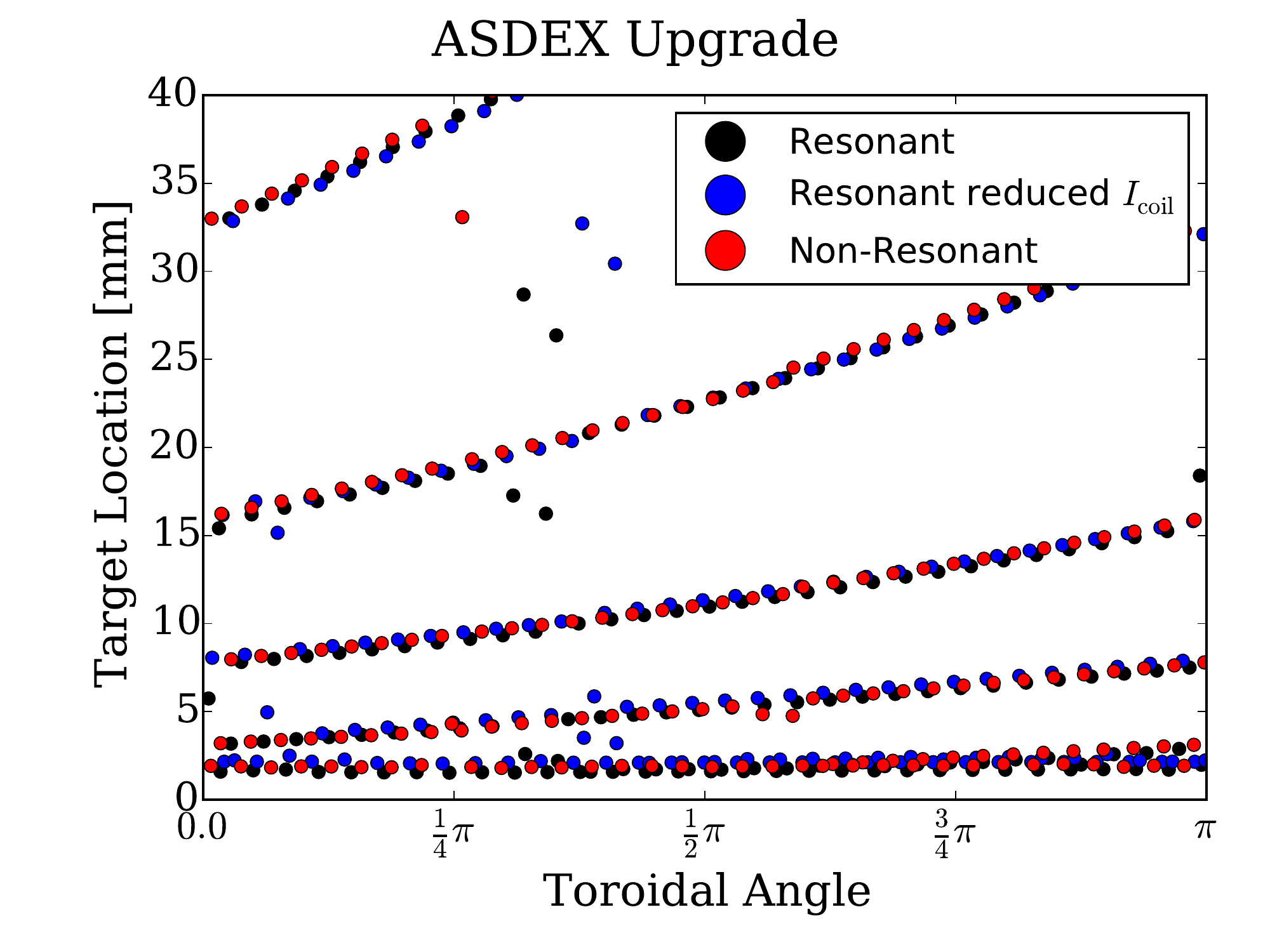}

\caption{Lobe position for the \textit{resonant} configuration with two different coil currents $I_{\mathrm{coil}}$ and for the \textit{non-resonant} configuration.}
\label{fig:LobePositionCurrent}
\end{figure}

\FloatBarrier
\subsection{Transformation of Time into Toroidal Angle with Rotating Magnetic Perturbation}
In the experimental results we transferred from the time evolution of the heat flux at a fixed toroidal location to a toroidal distribution for a fixed time.
In this section the differences between a toroidal distribution and the variation of the coil currents in order to rotate the magnetic perturbation field are discussed.
In the presented model both approaches can be compared.
One is the \textit{fixed time}, starting field lines at different toroidal positions at the target and computing the 2D heat flux structure for the given fixed magnetic coil currents.
The other is the \textit{fixed position}, starting the field lines at a fixed toroidal position at the target and changing the currents in the magnetic perturbation coils in the same way as in the experiment.
For an infinite number of coils the result would be the same.
For a low number of coils per row, as it is the case for the experiments (8 in the case of ASDEX Upgrade, 9 foreseen for ITER~\cite{Daly2013}), the representation of the sinusoidal perturbation varies for different absolute phases.
For an n\,=\,2 perturbation with 8 toroidal coils the difference is negligible.
The 2D heat flux is shown in~\fref{fig:ModellCompFixTimeLoc2D} for both approaches.
The corrugation of the separatrix is not taken into account here.
The variation in toroidal direction (or time transferred to a toroidal angle) at fixed target locations is shown in\fref{fig:ModellCompFixTimeLocVar}.

\begin{figure}[htb!]
	\centering
	
\subfigure[\label{fig:2DTimeFix} Fixed time]{\includegraphics[width=0.3 \textwidth]{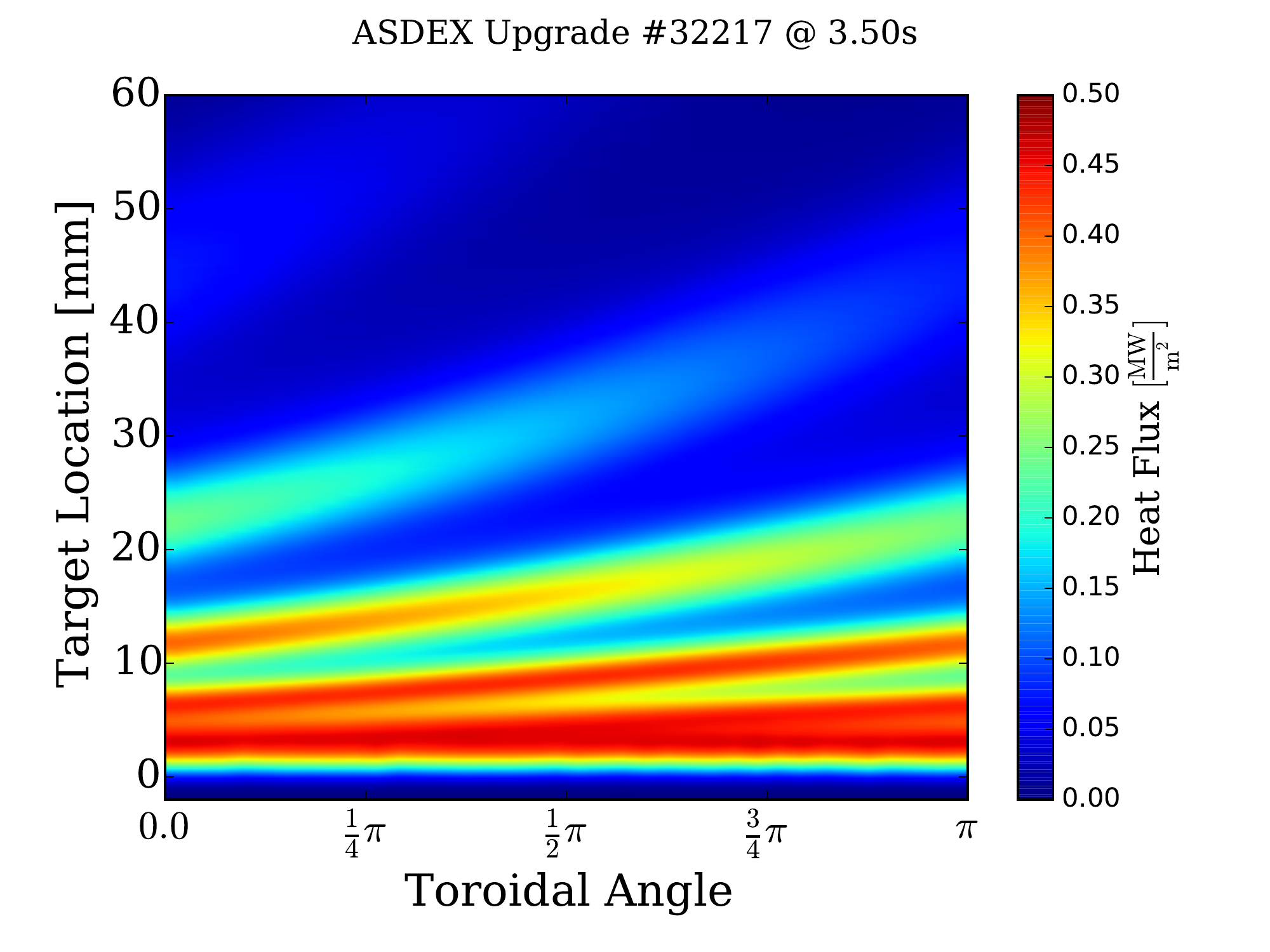}}
\subfigure[\label{fig:2DTimeVar} Fixed position]{\includegraphics[width=0.3 \textwidth]{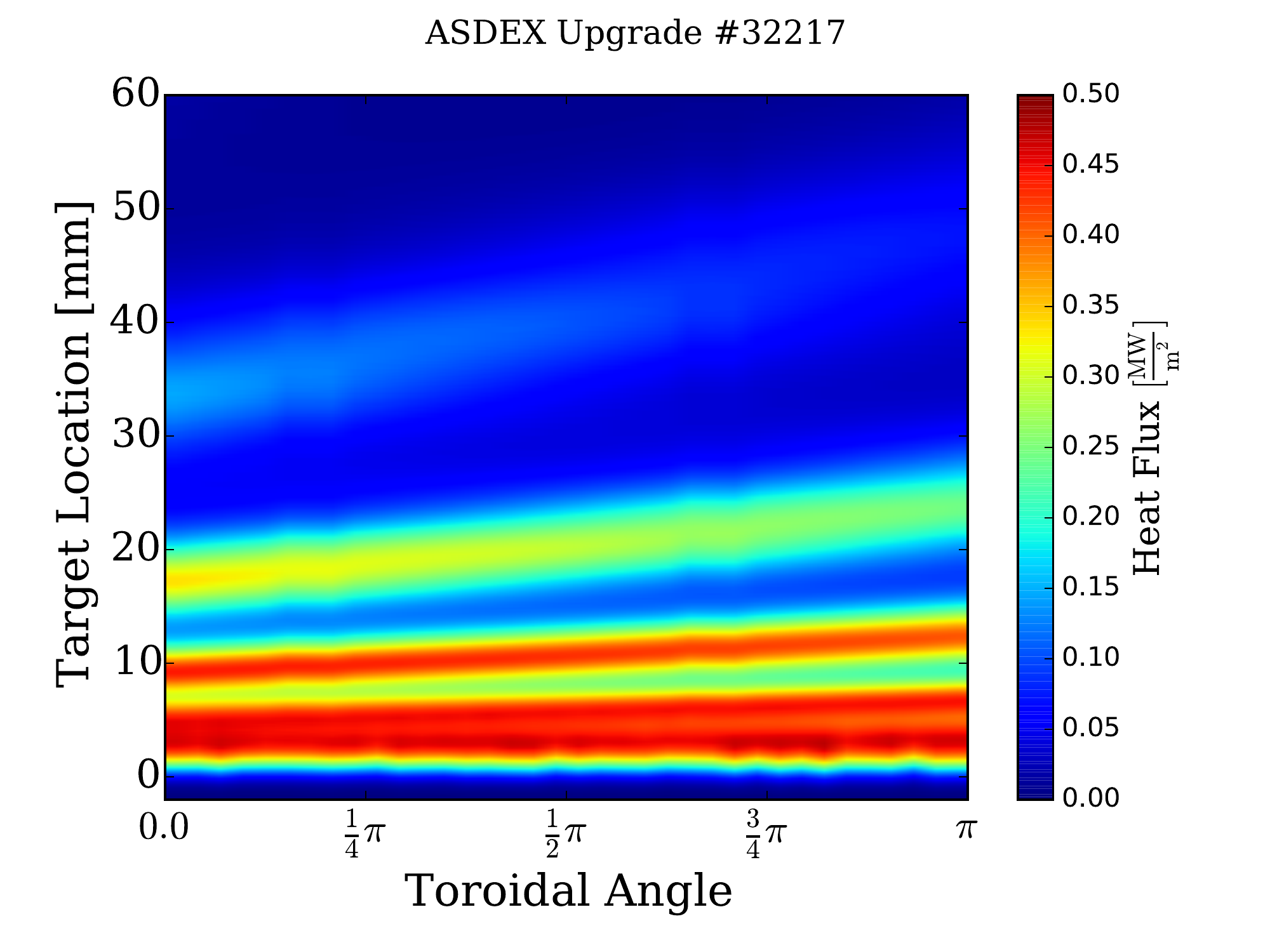}}
\caption{2D heat flux profile for fixed time and fixed position, both without taking the separatrix corrugation into account.}
\label{fig:ModellCompFixTimeLoc2D}
\end{figure}

\begin{figure}[htb!]
	\centering
	
\subfigure[\label{fig:torVarTimeFix} Fixed Time]{\includegraphics[width=0.3 \textwidth]{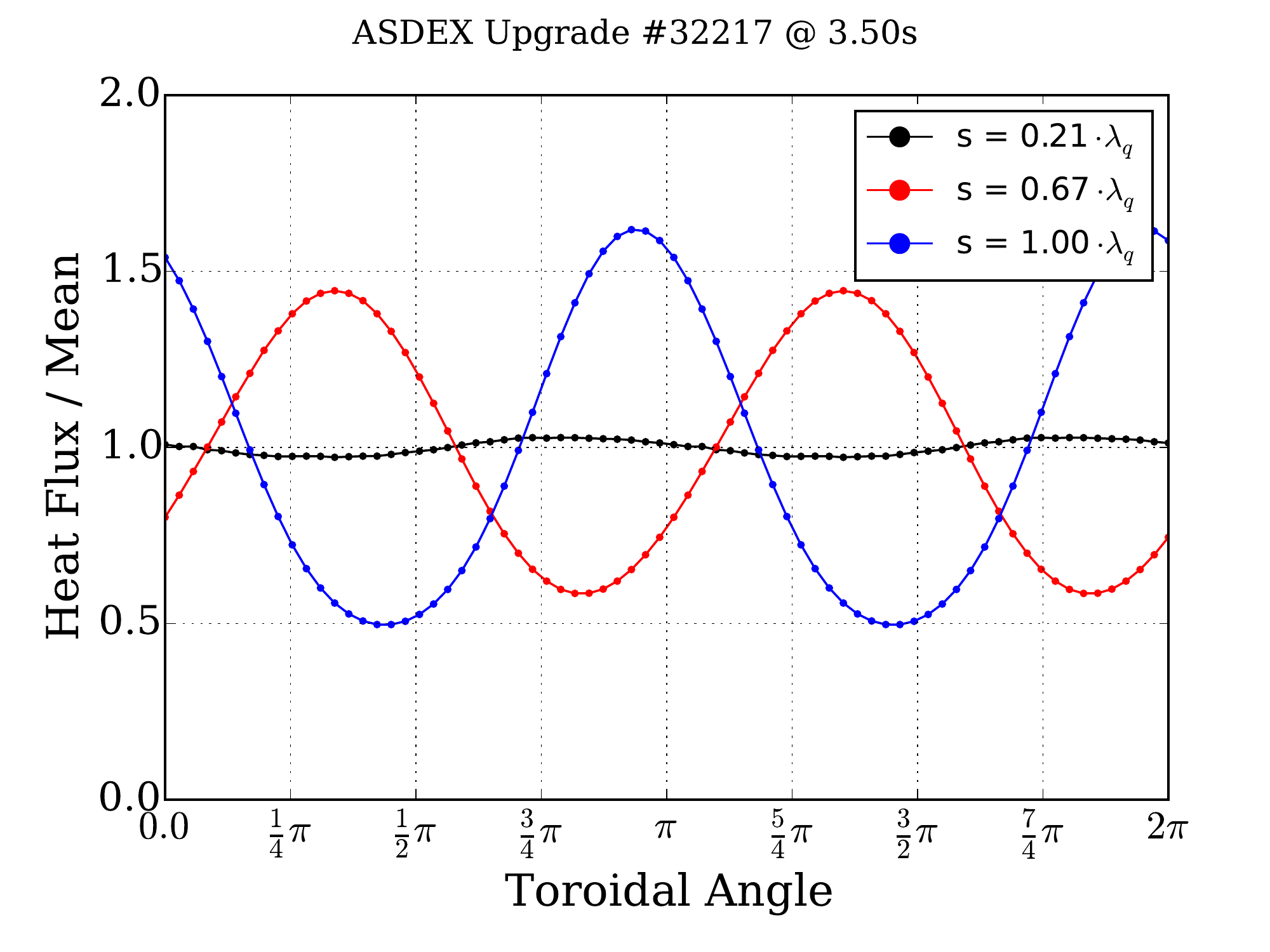}}
\subfigure[\label{fig:torVarTimeVar} Fixed Position]{\includegraphics[width=0.3 \textwidth]{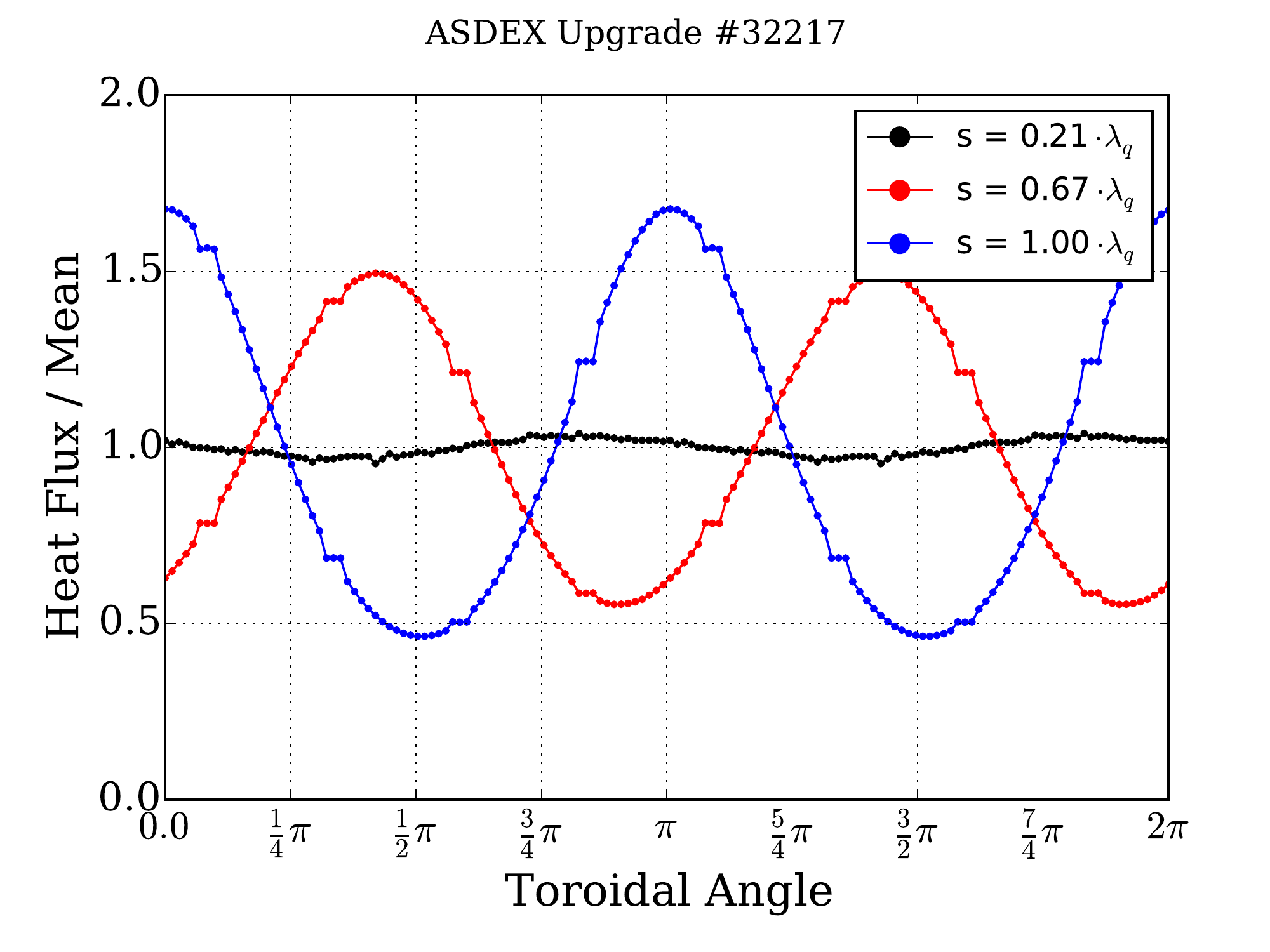}}
\caption{Toroidal (time) variation of the heat flux for fixed time and fixed position, both without taking the separatrix corrugation into account.}
\label{fig:ModellCompFixTimeLocVar}
\end{figure}
\FloatBarrier
\subsection{Radial Displacement of the Plasma Boundary}
With the application of the magnetic perturbation the plasma boundary is radially corrugated, e.g.~\cite{Willensdorfer2016}.
As discussed in the beginning of this section, this deformation can be treated with the presented model.
The model is based on field lines, making it convenient to handle the displacement as the change of the length of field lines at the OMP (\fref{fig:ConnectionLengthOMPFit}).
The black line indicates the used non-axisymmetric separatrix.

\begin{figure}[ht] 
	\centering

\includegraphics[width=0.5\textwidth]{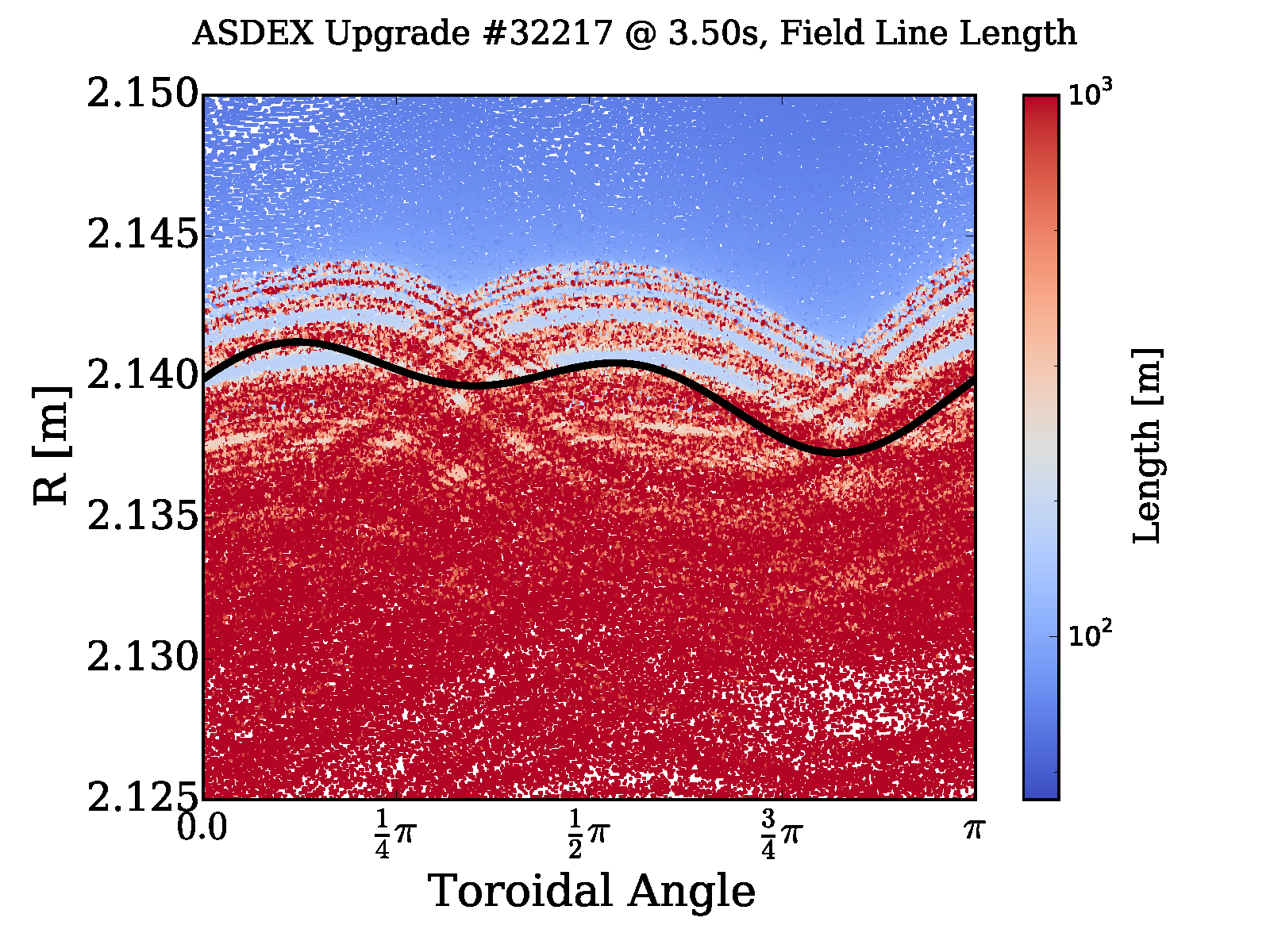}

\caption{Field line length at the OMP with the separatrix obtained as presented in section~\ref{Modelling}.}
\label{fig:ConnectionLengthOMPFit}
\end{figure}

The correction of the radial displacement at the OMP leads to an increase of the heat flux variation in the \textit{resonant} configuration and a decrease in the \textit{non-resonant} configuration in the model.
The 2D heat flux is compared in~\fref{fig:2DCompSep}, different radial positions and the averaged profile in~\fref{fig:1DCompSep} as well as the variation at some target positions in~\fref{fig:TimeDepentendCompSep} for the \textit{resonant} configuration.
\begin{figure}[htb!]
	\centering
	
\subfigure[\label{fig:Modell2DFixed} Without separatrix corrugation]{\includegraphics[width=0.45 \textwidth]{Modelling_32217_350s_fixedSep_2DHeatFlux.pdf}}
\subfigure[\label{fig:Modell2DCorr} With separatrix corrugation $R_{sep}(\Phi)$]{\includegraphics[width=0.45 \textwidth]{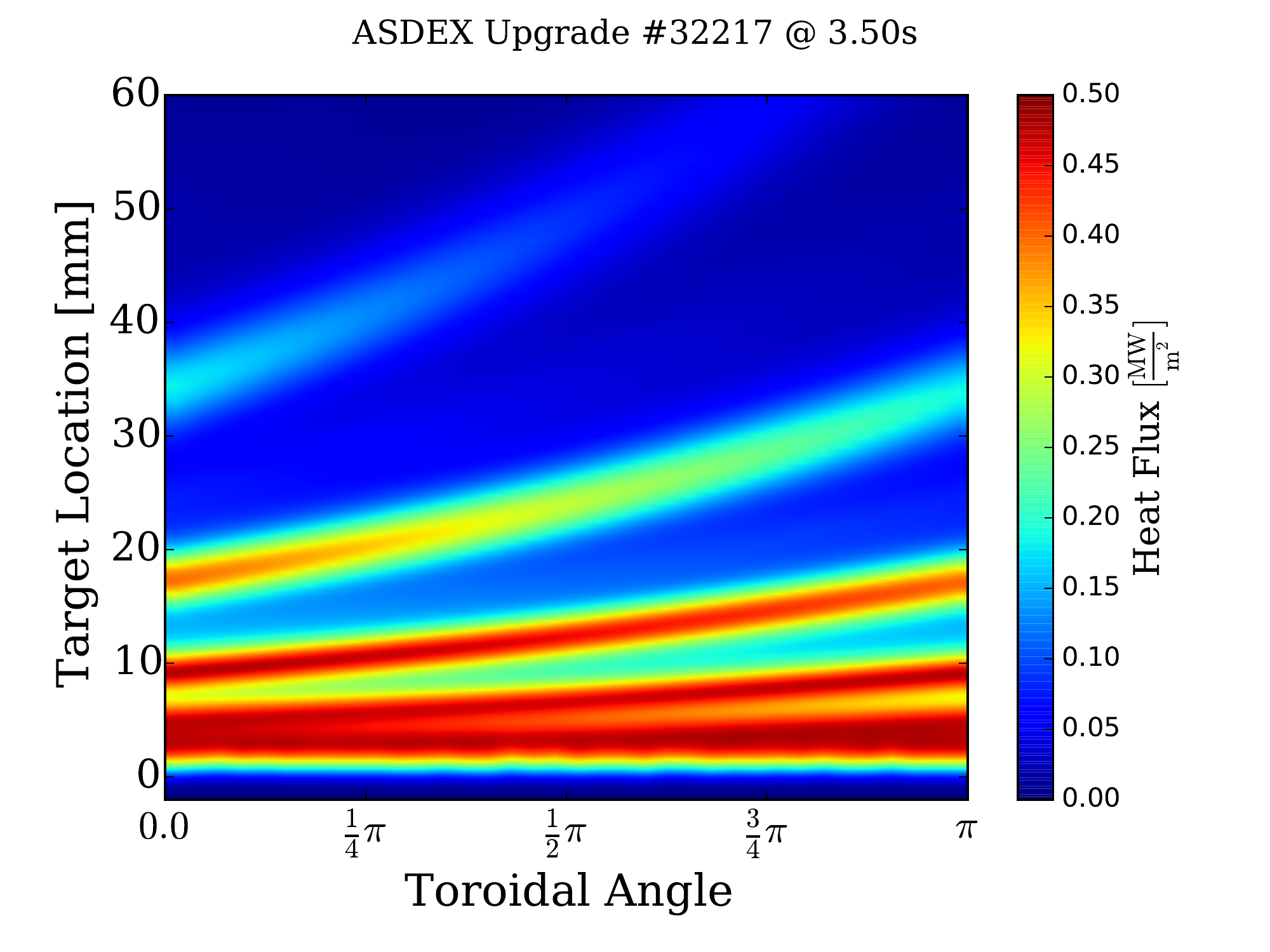}}
\caption{2D heat flux profile for the \textit{resonant} configuration.}
\label{fig:2DCompSep}
\end{figure}

\begin{figure}[htb!]
	\centering
	
\subfigure[\label{fig:ModellAveragedFixed} Without separatrix corrugation]{\includegraphics[width=0.45 \textwidth]{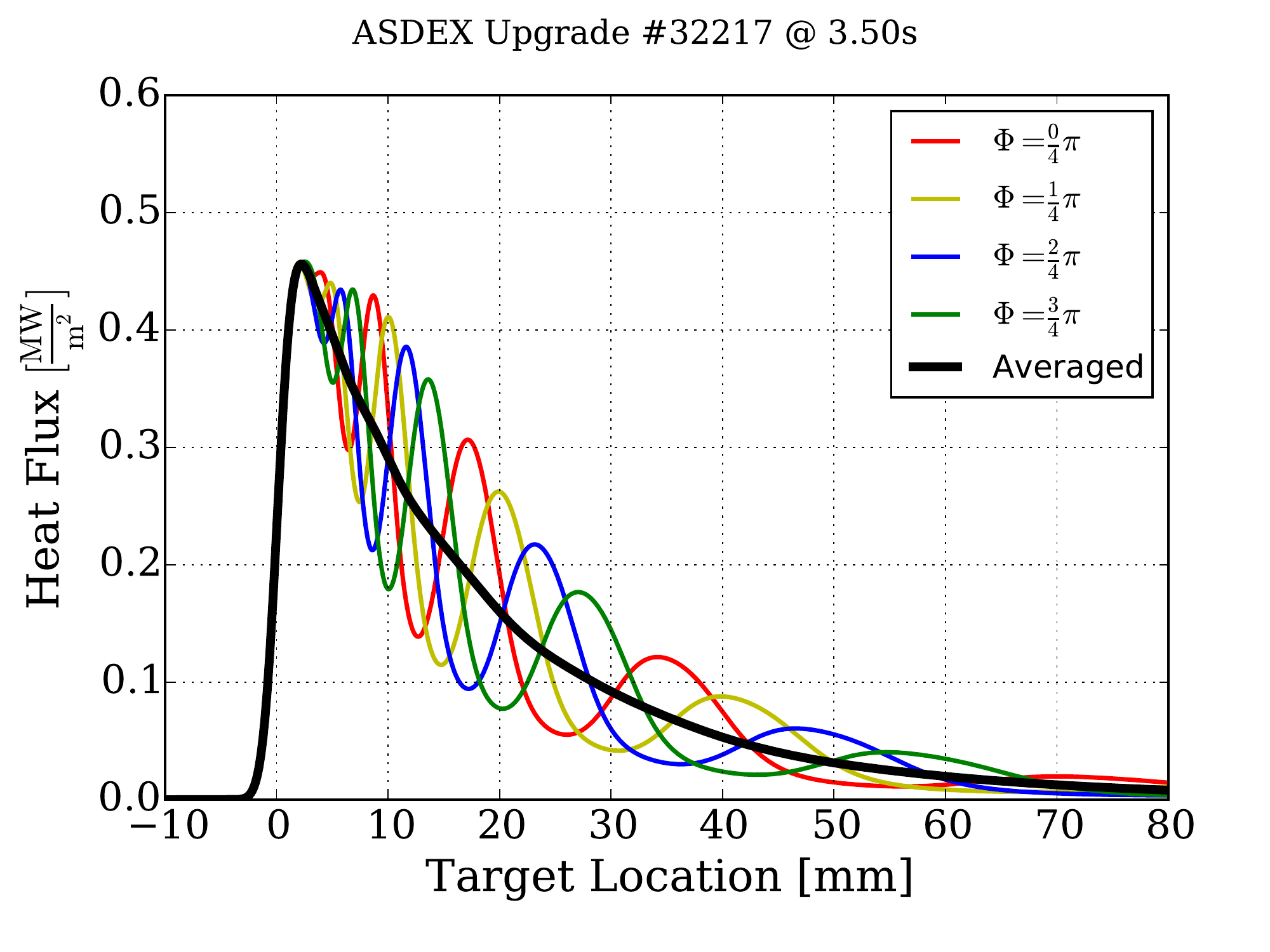}}
\subfigure[\label{fig:ModellAveragedCorr} With separatrix corrugation $R_{sep}(\Phi)$]{\includegraphics[width=0.45 \textwidth]{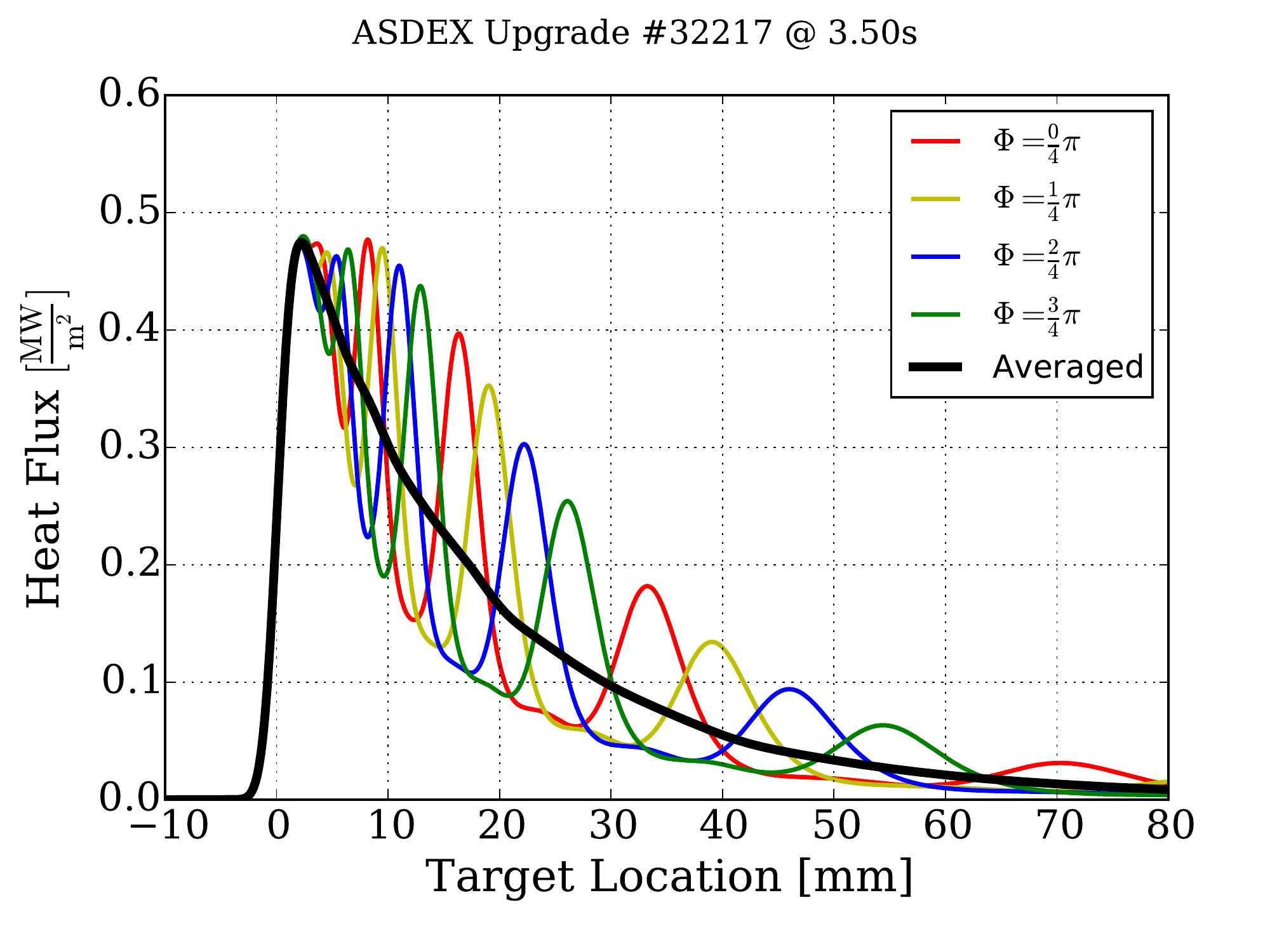}}
\caption{Averaged heat flux profiles for the \textit{resonant} configuration.}
\label{fig:1DCompSep}
\end{figure}

\begin{figure}[htb!]
	\centering
	
\subfigure[\label{fig:ModellTimeVariationFixed} Without separatrix corrugation]{\includegraphics[width=0.3 \textwidth]{Modelling_32217_350s_fixedSep_ToroidalPeaking_3Positions.pdf}}
\subfigure[\label{fig:ModellTimeVariationCorr} With separatrix corrugation $R_{sep}(\Phi)$]{\includegraphics[width=0.3 \textwidth]{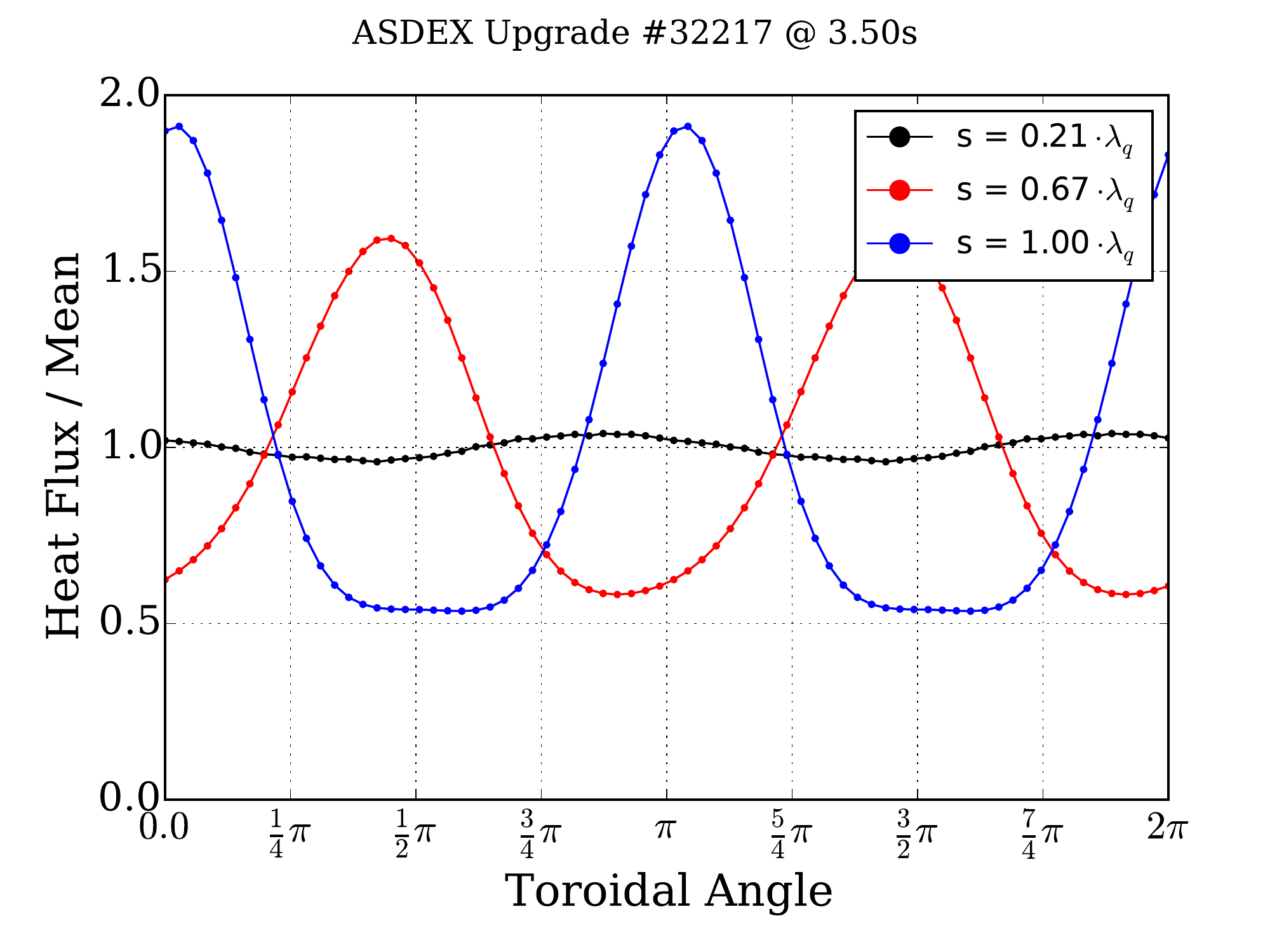}}
\subfigure[\label{fig:ExperimentTimeVariation} Measurement]{\includegraphics[width=0.3 \textwidth]{32217_ToroidalPeaking_TimeTrace.pdf}}
\caption{Toroidal (time) variation of the heat flux for the \textit{resonant} configuration.}
\label{fig:TimeDepentendCompSep}
\end{figure}
The 2D structure does not change significantly.
However, the single profiles show a larger variation when taking the corrugation into account.
The maximum heat flux at a given toroidal position is not changed and is at the position of the axisymmetric, or toroidally averaged, maximum where the variation is negligible.
The variation at a given target location changes from a nearly sinusoidal for the complete profile to a more triangular shaped variation far away from the separatrix (blue lines in~\fref{fig:TimeDepentendCompSep}).\\
The same plots for the \textit{non-resonant} configuration are shown in figures~\ref{fig:2DCompSepNon},~\ref{fig:1DCompSepNon} and~\ref{fig:TimeDepentendCompSepNon}.
Without taking the displacement into account the modelled profiles do not differ when varying the \textit{differential phase} (\textit{resonant}, \textit{non-resonant}).
The heat flux variation significantly varies with the correction of the corrugation in the \textit{non-resonant} configuration.
\begin{figure}[htb!]
	\centering
	
\subfigure[\label{fig:Modell2DFixedNon} Without separatrix corrugation]{\includegraphics[width=0.45 \textwidth]{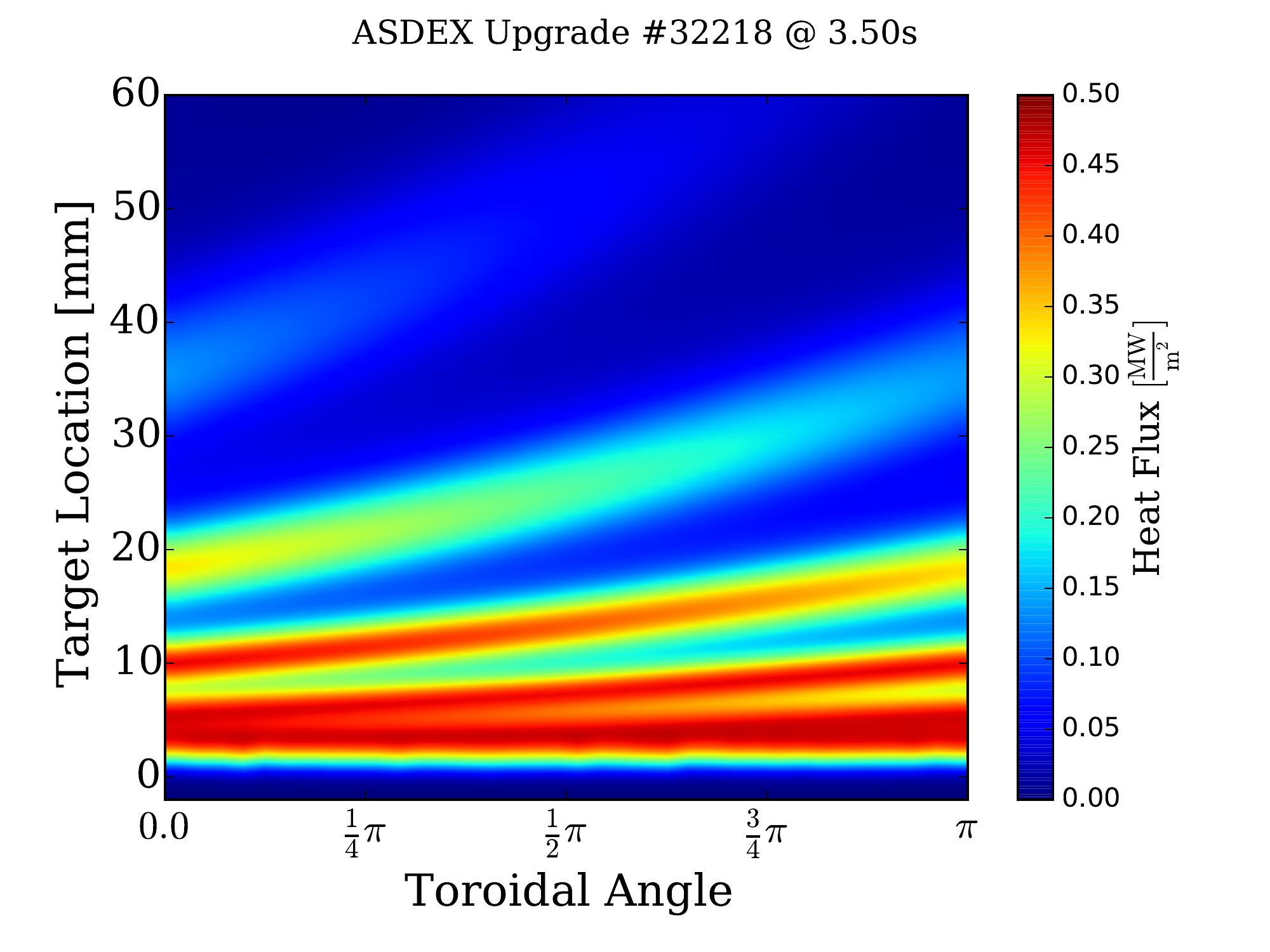}}
\subfigure[\label{fig:Modell2DCorrNon} With separatrix corrugation $R_{sep}(\Phi)$]{\includegraphics[width=0.45 \textwidth]{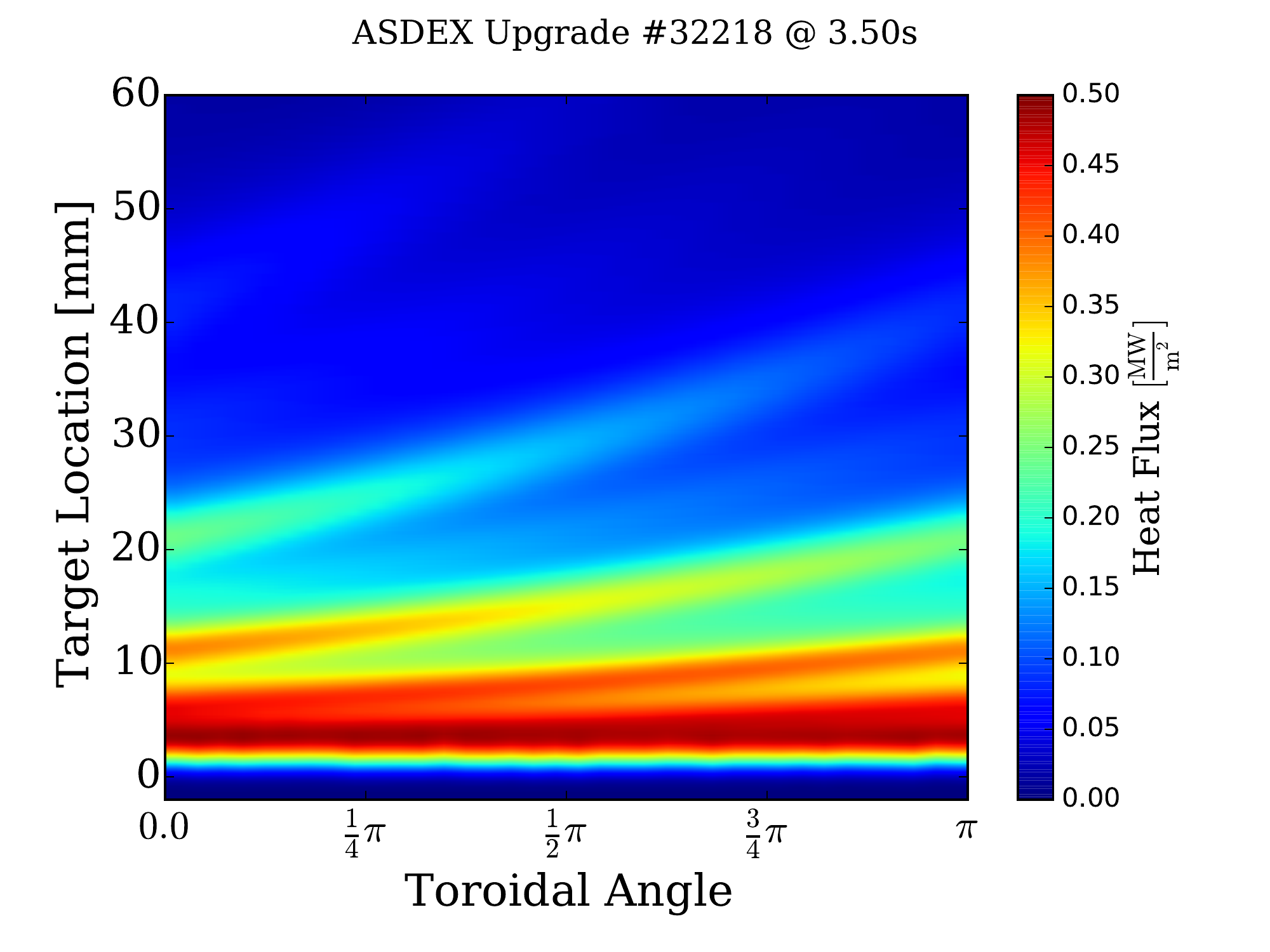}}
\caption{2D heat flux profile for the \textit{non-resonant} configuration.}
\label{fig:2DCompSepNon}
\end{figure}

\begin{figure}[htb!]
	\centering
	
\subfigure[\label{fig:ModellAveragedFixedNon} Without separatrix corrugation]{\includegraphics[width=0.45 \textwidth]{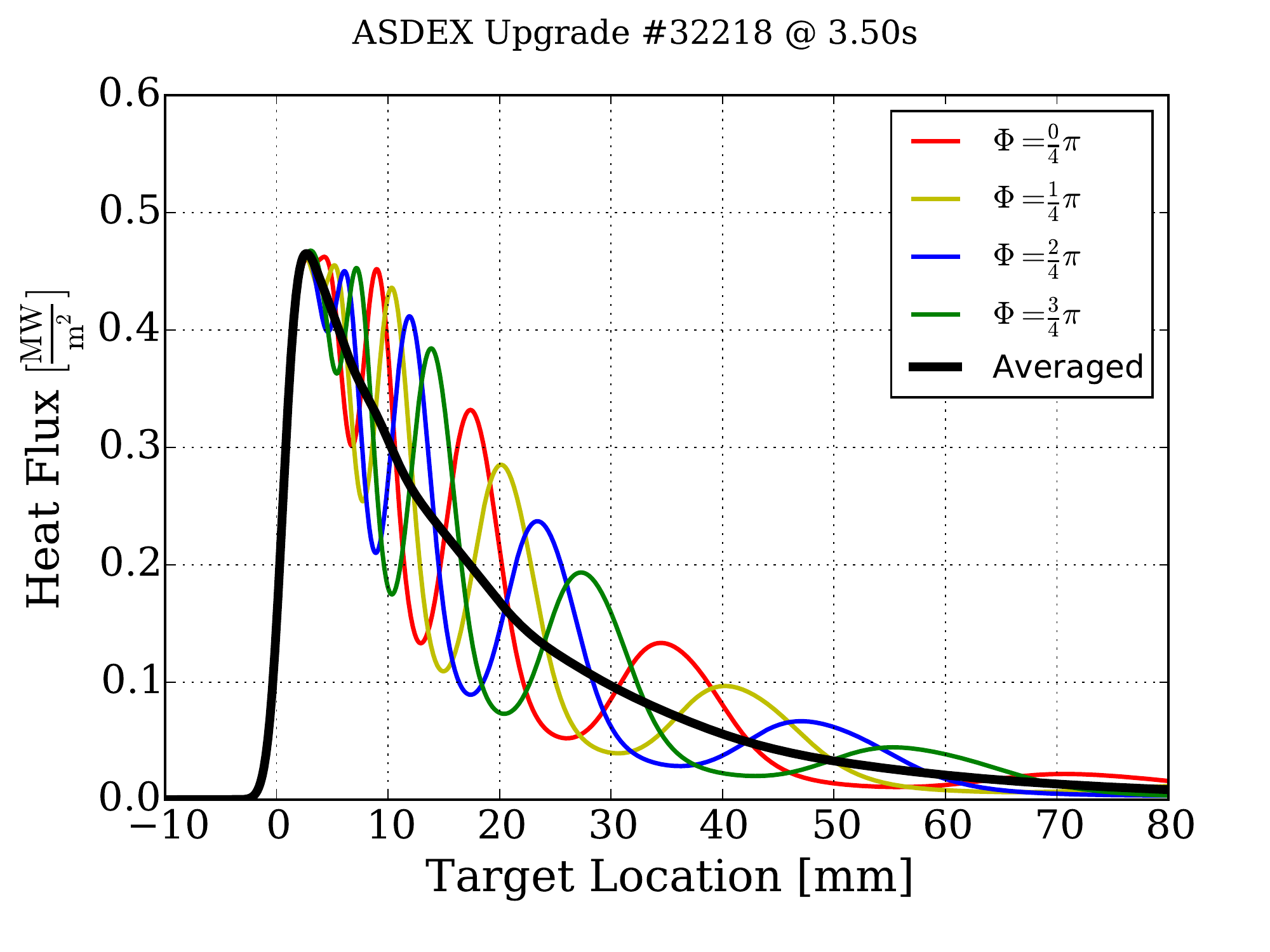}}
\subfigure[\label{fig:ModellAveragedCorrNon} With separatrix corrugation $R_{sep}(\Phi)$]{\includegraphics[width=0.45 \textwidth]{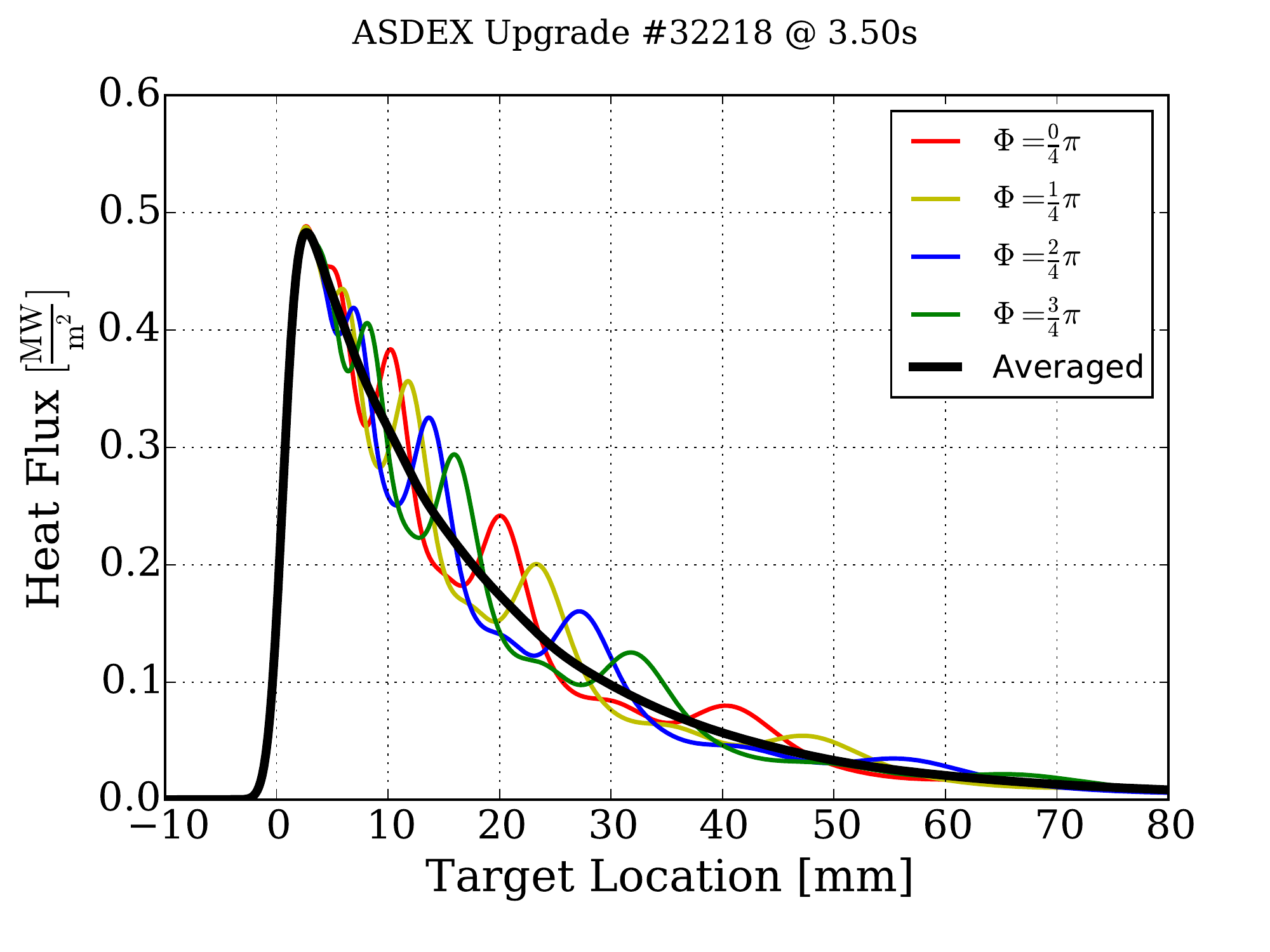}}
\caption{Averaged heat flux profiles for the \textit{non-resonant} configuration.}
\label{fig:1DCompSepNon}
\end{figure}

\begin{figure}[htb!]
	\centering
	
\subfigure[\label{fig:ModellTimeVariationFixedNon} Without separatrix corrugation]{\includegraphics[width=0.3 \textwidth]{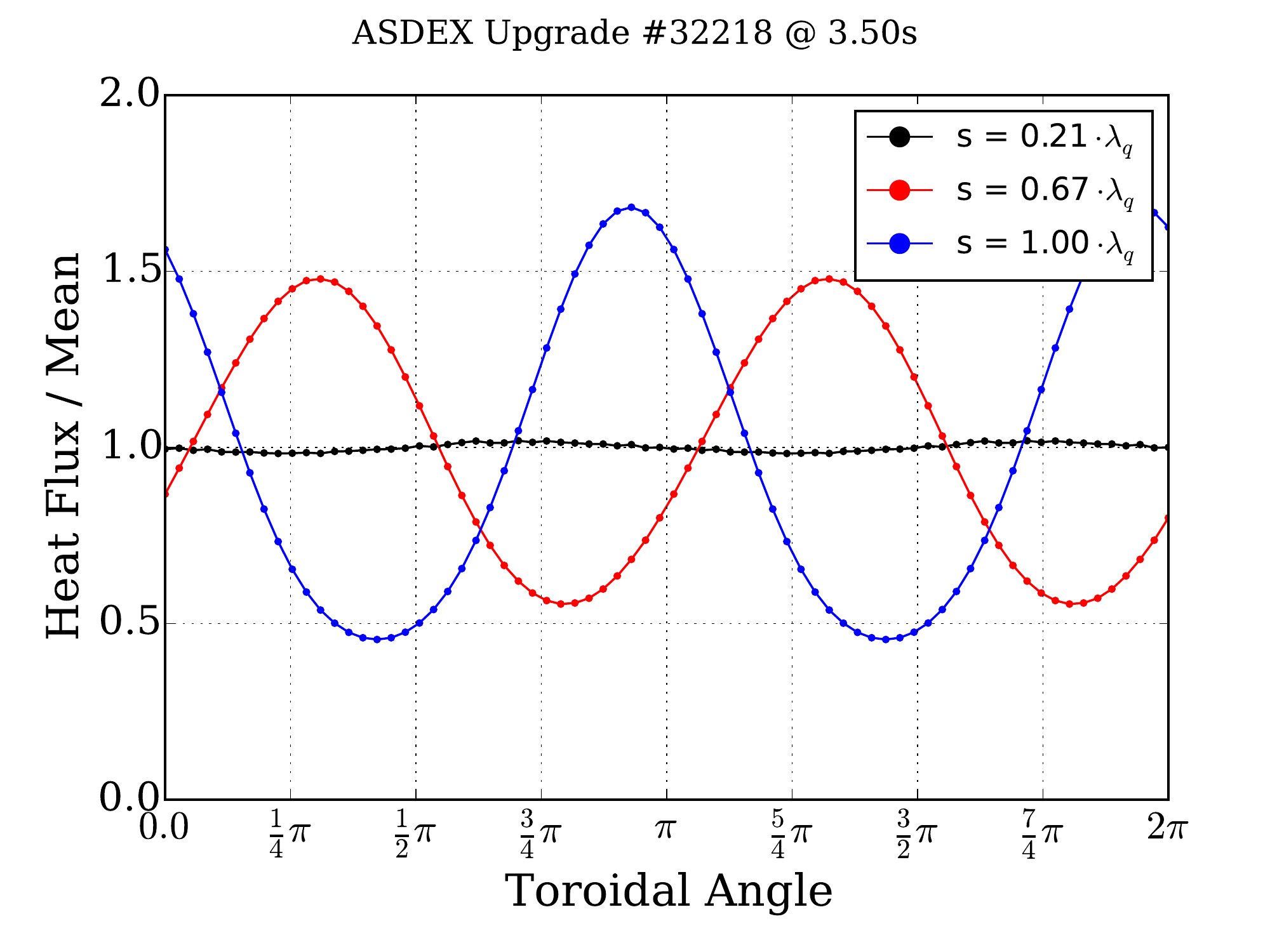}}
\subfigure[\label{fig:ModellTimeVariationCorrNon} With separatrix corrugation $R_{sep}(\Phi)$]{\includegraphics[width=0.3 \textwidth]{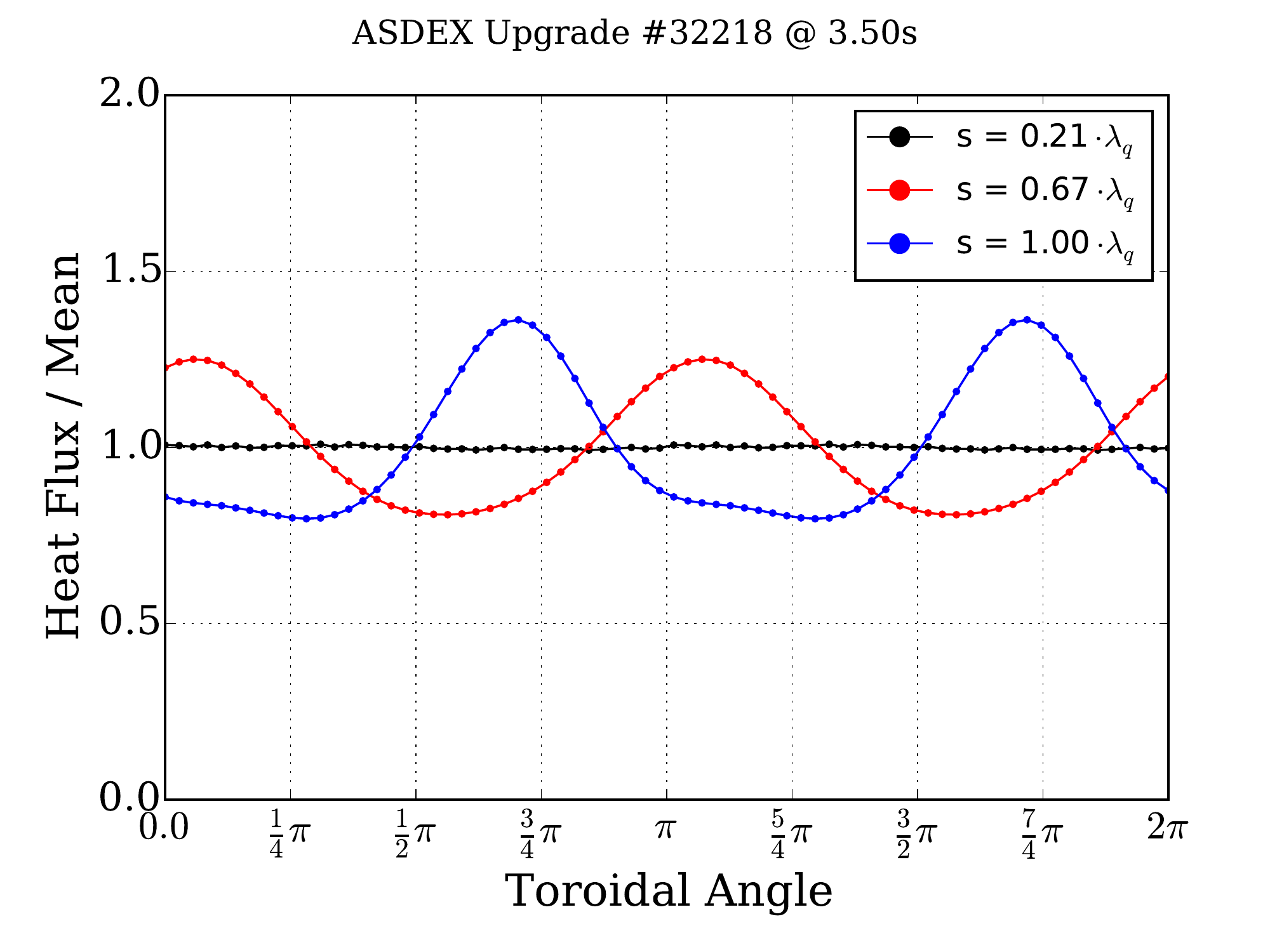}}
\subfigure[\label{fig:ExperimentTimeVariationNon} Measurement]{\includegraphics[width=0.3 \textwidth]{32218_ToroidalPeaking_TimeTrace.pdf}}
\caption{Toroidal (time) variation of the heat flux for the \textit{non-resonant} configuration.}
\label{fig:TimeDepentendCompSepNon}
\end{figure}
In the experiment, a clear difference is observed with the variation of the \textit{differential phase}.
In the model this is only reproduced by taking the plasma boundary displacement into account, although, the \textit{non-resonant} case in the model still exhibits stronger variations than observed in the experiment.

\FloatBarrier
\subsection{Description of Toroidal Heat Flux Peaking}\label{ModellingPeakingProfile}
The \textit{toroidal peaking} is defined as the maximum value along the toroidal direction at a given target location normalized to the mean value at this target location.
The \textit{toroidal peaking} along the divertor target is shown in~\fref{fig:ModellingPeakingToroidalvsTarget}.
\begin{figure}[htb!]
	\centering
	
\subfigure[\label{fig:MeasurementPeakingTarget} Measurement]{\includegraphics[width=0.45 \textwidth]{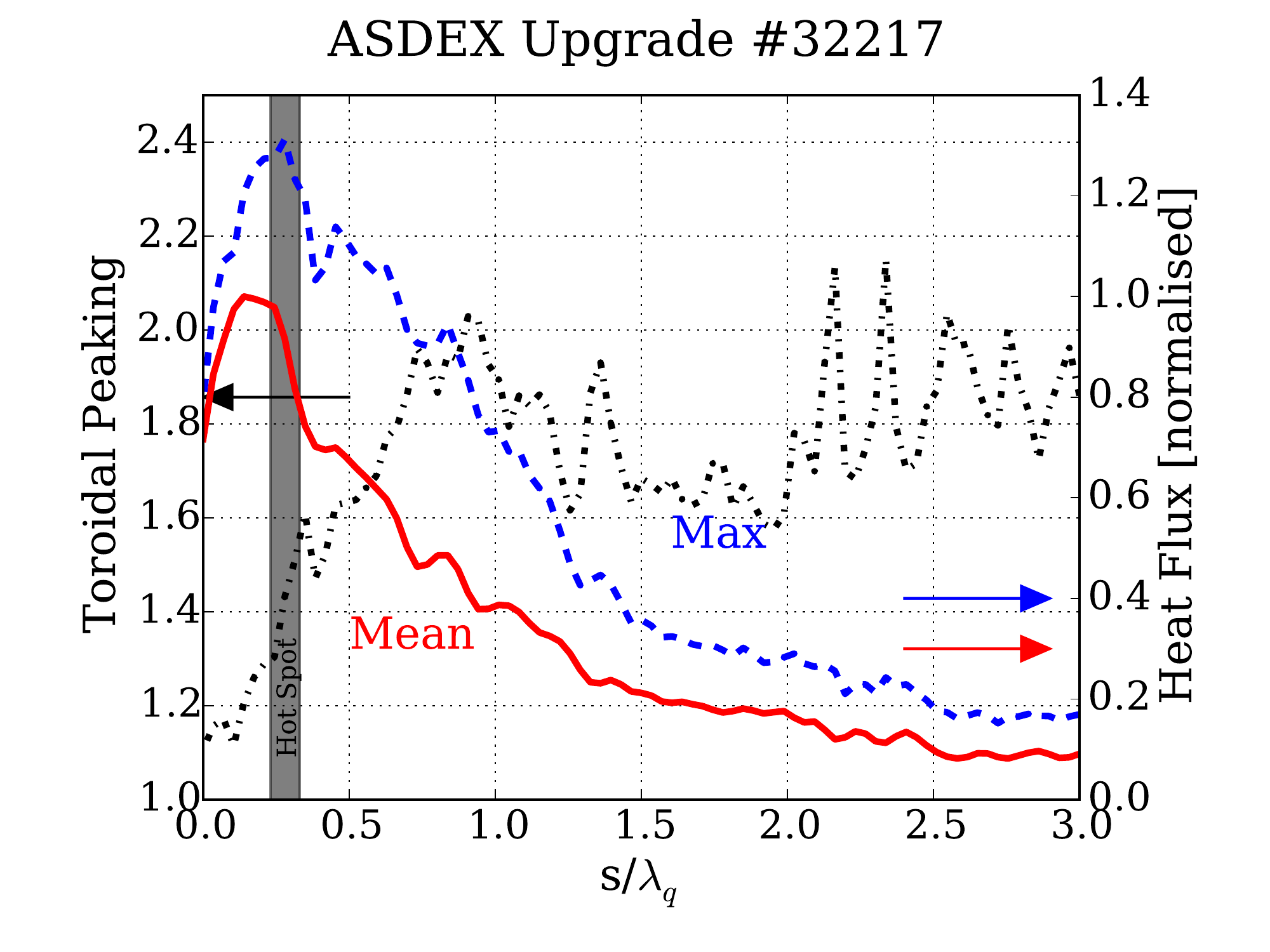}}
\subfigure[\label{fig:ModelPeakingTarget} Modelling]{\includegraphics[width=0.45 \textwidth]{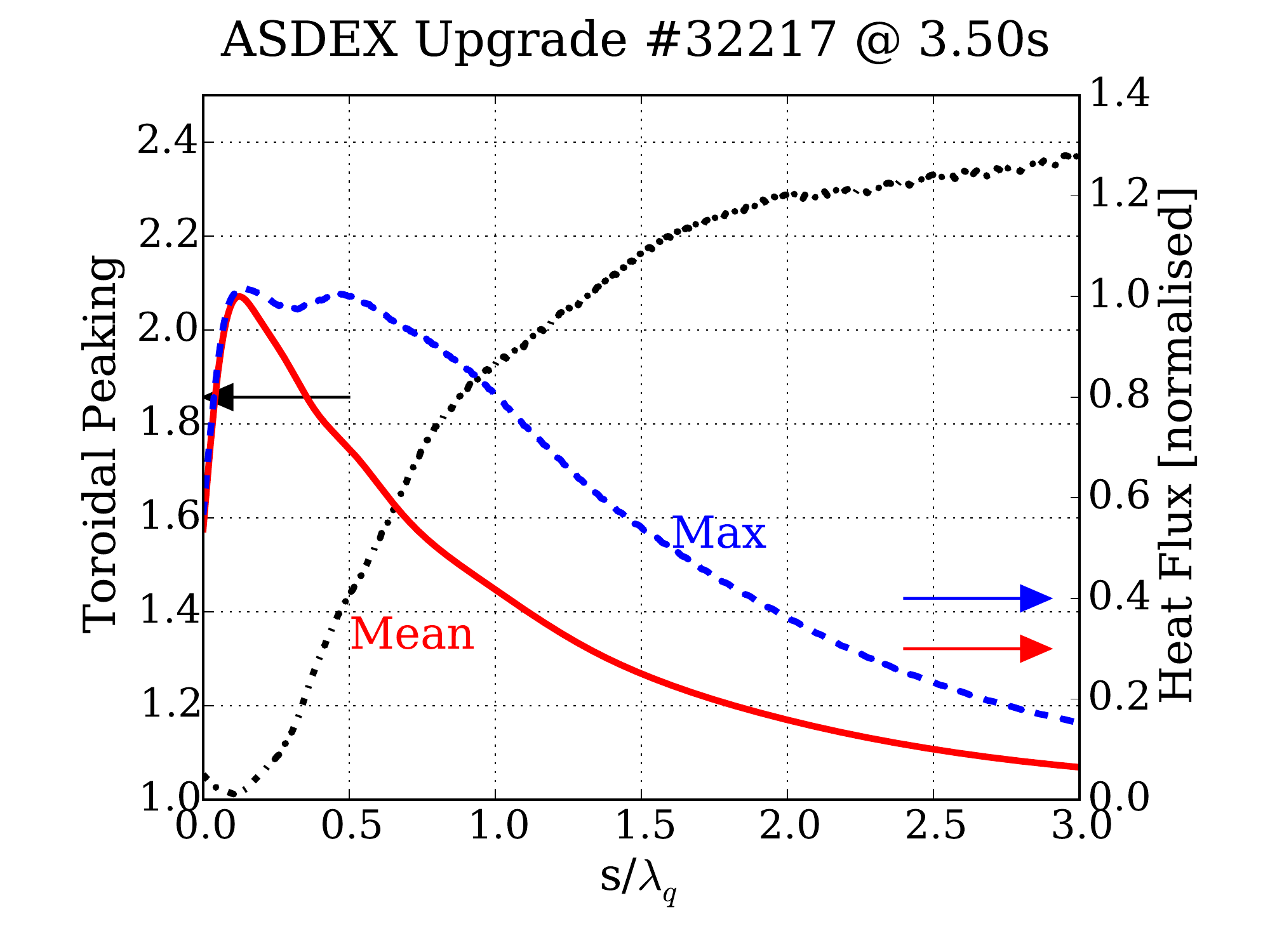}}
\caption{Impact of the \textit{toroidal peaking} along the target location.}
\label{fig:ModellingPeakingToroidalvsTarget}
\end{figure}
The toroidally averaged heat flux profile (mean) is shown in red, the toroidal maximum (max) in blue with the normalization described in section~\ref{section:averaged}.
The max profile reveals the extend along the target which is above a certain heat flux.
The \textit{toroidal peaking}, shown in black.
The measurement shown in~\fref{fig:MeasurementPeakingTarget} shows some deviations from the model~\fref{fig:ModelPeakingTarget}.
The maximum value in the model is never exceeding the peak value of the averaged profile whereas in the experiment the maximum is in the order of 20\,\% larger than the averaged peak value.
However, a hot spot in the measurement close to the peak position leads to an overestimation.
The \textit{toroidal peaking} in the experiment is always about 20\,\% larger, mainly due to the L-Mode heat flux variation of the heat flux, see section~\ref{section:TimeVariation}.
In the modelled data the \textit{toroidal peaking} monotonically increases with increasing distance from the strike line position.
In the measurement the \textit{toroidal peaking} seems to saturate at a distance of 1 $\lambda_q$.
However, since the incident (mean) heat flux decreases with distance to the strike line, the uncertainty becomes larger and the interpretation of this saturation has to be treated with caution.
The proposed explanation is the divertor broadening $S$.
The 1D diffusive model (shown in~\eref{eq:diffusiveModel}) has a single parameter for the diffusion in the divertor region, leading to an effective parameter $S$ for the complete profile.
It was shown that the parameter depends on the electron temperature in the divertor volume, sufficiently characterized by the target electron temperature for attached conditions~\cite{Sieglin2016, Scarabosio2015}. As a result this leads to a variation along the divertor target as the temperature is not constant.
An increased divertor broadening in the far SOL explains the reduction in the \textit{toroidal peaking}.

\FloatBarrier
\subsubsection{Influence of the Magnetic Perturbation Coil Current}
The influence of the magnetic perturbation strength for the \textit{resonant} and \textit{non-resonant} configuration on the variation in toroidal direction is shown in~\fref{fig:toroidalPeakingCurrent}.
\begin{figure}[ht] 
	\centering

\includegraphics[width=0.5\textwidth]{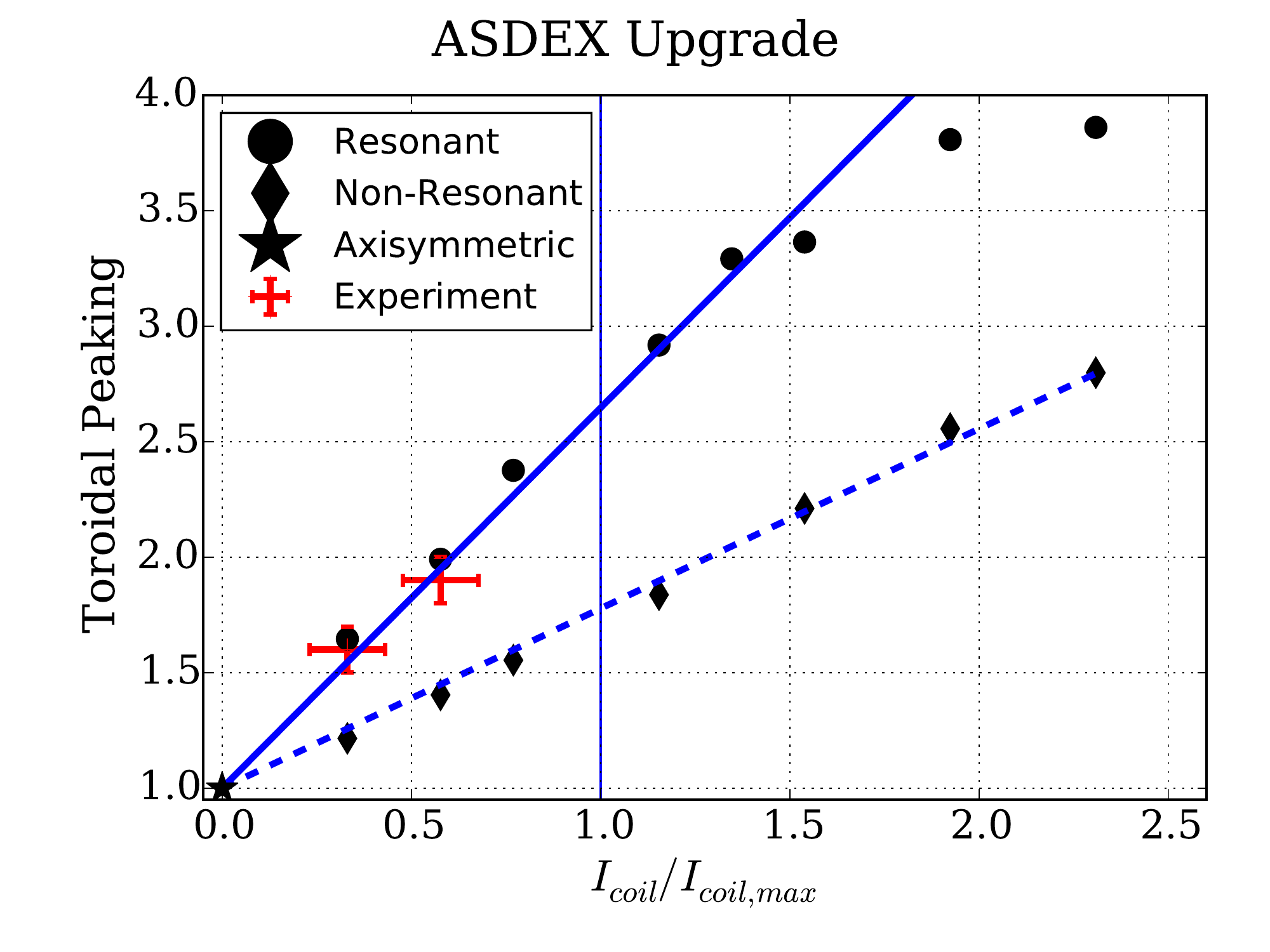}

\caption{\textit{Toroidal peaking} for different coil currents at the target position 1 $\lambda_q$ away from the strike line. The current is normalized to the maximum current of the ASDEX Upgrade coils.}
\label{fig:toroidalPeakingCurrent}
\end{figure}
A target location of one $\lambda_q$ away from the strike line is chosen in the figure.
For zero coil current an axisymmetric heat flux is expected, having a \textit{toroidal peaking} of 1.0.
The coil current in the figure is normalized to the maximum coil current of the ASDEX Upgrade coils (1.3\,kA x 5 turns).
The nominal current used in the presented study is 5\,kAt, the PSL attenuates 25\,\% of the magnetic field.
The maximum achievable normalized coil current in the experiment is thus $\approx$\,0.6.
In order to rotate the field with a constant amplitude the effective magnetic perturbation is a square-root reduced for the same coil currents (not all coils have the maximum current at the same time).
This is not reflected in the normalization, all data that is shown is from discharges with rotated field.
The perturbation strength with a static field can be about 1.9 times larger with the same current limit of 5\,kAt.\\
The \textit{toroidal peaking} increases nearly linear with increasing current for a given \textit{differential phase}.
The slope is different for the two configurations, the \textit{resonant} configuration has a larger \textit{toroidal peaking} with about a factor two difference in the slope.
It is seen, that for too high currents in the model (more than 2.0 times the maximum current possible in the experiment) this linear dependence is lost.
This is due to the large influence of the perturbation and might be an artifact of this model.
The red dots in the figure show the \textit{toroidal peaking} for the \textit{resonant} case observed in the experiment.
The agreement with the model is remarkable, keeping in mind the simplifications made for this model.
The experimental data for the \textit{non-resonant} case does not allow to make the same comparison, which is due to the low expected variation and the influence of noise in the measurement.
In~\fref{fig:ExperimentTimeVariationNon} it is seen that the experimental \textit{toroidal peaking} at a normalized current of 0.6 is well below the 1.5 observed in the model.
\FloatBarrier
\subsubsection{Influence of the Divertor Broadening $S$}
In section~\ref{densSteps} it is shown that the \textit{toroidal peaking} decreases with increasing density.
From the axisymmetric reference as well as the toroidally averaged profiles it is observed that both transport qualifiers, $\lambda_q$ and $S$, increase with increasing density.
In the presented model the divertor broadening $S$ is applied after the heat flux calculations and thus can be varied independently.
\begin{figure}[ht] 
	\centering

\includegraphics[width=0.5\textwidth]{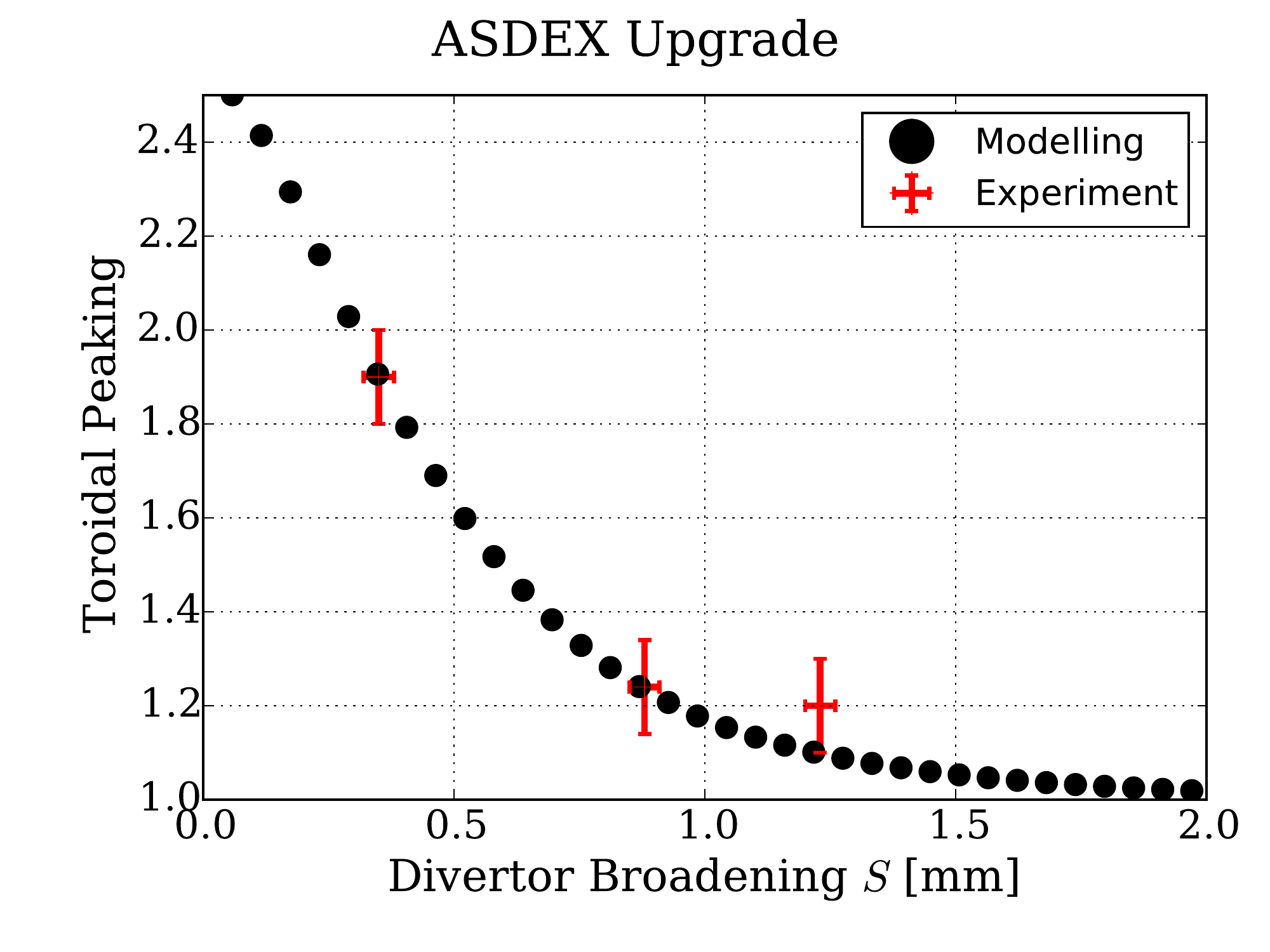}

\caption{\textit{Toroidal Peaking} in dependence of the divertor broadening $S$ one $\lambda_q$ away from the strike line.}
\label{fig:ModellingBroadening}
\end{figure}
The \textit{toroidal peaking} decreases with increasing divertor broadening $S$.
The decrease seen in the experiment with increasing density is explained by the increase of the divertor broadening $S$.
The measured \textit{toroidal peaking} at the fitted divertor broadening $S$ for the three different density steps are plotted in red.
The power fall-off length $\lambda_q$ is kept constant in the modelling, although it changes in the experiment with density.
\FloatBarrier

\section{Summary and Conclusions} \label{Conclusion}
The effect of external magnetic perturbation on the divertor heat load is studied in ASDEX Upgrade L-Mode.
ASDEX Upgrade is able to measure the 2D heat flux profile with n\,=\,2 and various poloidal phasing due to the versatile power supply of the saddle coils.
A clear change of the divertor heat flux with the poloidal phasing (\textit{resonant} vs. \textit{non-resonant}) is observed that is in agreement with modelling using the vacuum field approach and field line tracing from the outer divertor target to the outer midplane combined with the two point model.\\
The time averaged heat flux profiles are similar to the axisymmetric reference profiles without MP leading to the same transport qualifiers, power fall-off length $\lambda_q$ and divertor broadening $S$, for both the measurements and the model for all \textit{differential phases}.
No change in the heat transport is observed.\\
The peak heat flux is unchanged and at the same location for all toroidal phases with magnetic perturbation.
This is the same location than in the reference phase without magnetic perturbation.
Although the toroidal averaged heat flux is unchanged, the application of the magnetic perturbation has an affect onto the local heat flux.\\
The \textit{toroidal peaking} is largest for the \textit{resonant} configuration and at lowest density with up to a factor of 2 locally increased heat flux.
The variation decreases with shifting the \textit{differential phase} away from the \textit{resonant} configuration and is for the \textit{non-resonant} configuration within the typical heat flux variation in L-Mode.
Increasing the density increases the divertor broadening $S$ for the outer target of ASDEX Upgrade.
Increasing the density leads to a reduced \textit{toroidal peaking} and a nearly axisymmetric profile in still attached conditions for the discharge parameters used in the presented study.
The reduction of the \textit{toroidal peaking} is explained in the model with the increase of the divertor broadening $S$ solely and is in quantitative agreement with the measurements.
The model suggests a linear increase of the \textit{toroidal peaking} with perturbation strength, obtained by an increase of the current in the saddle coils.\\
The overall agreement between the measurements and the model leads to the conclusion that in these L-Mode conditions plasma response is not a dominant factor for the heat transport.
However, in the foreseen H-Mode regime for ITER this might change.
The influence of this response onto the heat transport in the scrape-off layer is up to now unknown and outside of the scope of the presented study.
The 2D pattern might be less or more extended.
Less due to the more narrow power fall-off length $\lambda_q$ and possible shielding or more due to field amplification from the plasma.
The effect of the divertor broadening $S$ onto the \textit{toroidal peaking} should be similar in H-Mode compared to the presented results in L-Mode.

\section*{Acknowledgements}
This work has been carried out within the framework of the EUROfusion Consortium and has received funding from the Euratom research and training programme 2014-2018 under grant agreement No 633053. The views and opinions expressed herein do not necessarily reflect those of the European Commission.

\section*{References}

\bibliographystyle{nf}
\bibliography{library}

\end{document}